\documentclass[pdflatex,sn-aps]{sn-jnl}%

\usepackage{graphicx,array}%
\usepackage{multirow}%
\usepackage{amsmath,amssymb,amsfonts,bm}%
\usepackage{amsthm}%
\usepackage{mathrsfs}%
\usepackage[title]{appendix}%
\usepackage{textcomp}%
\usepackage{manyfoot}%
\usepackage{booktabs}%
\usepackage{titlesec}
\usepackage{tabularray}
\usepackage{algorithm}%
\usepackage{algorithmicx}%
\usepackage{algpseudocode}%
\usepackage{listings}%
\usepackage{comment}
\usepackage[percent]{overpic}
\usepackage{graphicx}
\usepackage[percent]{overpic}

\newsavebox{\rightfigbox}

\usepackage{braket}%

\usepackage{tabularray}
\UseTblrLibrary{booktabs}
\usepackage{booktabs}%
\usepackage{multirow}%
\usepackage{siunitx}%
\usepackage[table]{xcolor} %

\usepackage{bibunits}
\defaultbibliographystyle{sn-aps}
\defaultbibliography{ref-insulators}

\newcommand{\suppref}[1]{Supplementary Note~\ref{#1}}
\makeatletter
\newcommand{\beginsupplementary}{%
  \clearpage
  
  \setcounter{page}{1}
  \renewcommand{\thepage}{S\arabic{page}}

  \section*{Supplementary Information}
  \addcontentsline{toc}{section}{Supplementary Information}

  \setcounter{section}{0}
  \setcounter{subsection}{0}
  \setcounter{subsubsection}{0}
  \setcounter{figure}{0}
  \setcounter{table}{0}
  \setcounter{equation}{0}

  \renewcommand{\thesection}{\arabic{section}}
  \renewcommand{\thesubsection}{\arabic{section}.\arabic{subsection}}
  \renewcommand{\thesubsubsection}{\arabic{section}.\arabic{subsection}.\arabic{subsubsection}}

  \def\supp@sectionname{section}
  \renewcommand{\@seccntformat}[1]{%
    \def\supp@tmp{##1}%
    \ifx\supp@tmp\supp@sectionname
      Supplementary Note~\thesection\quad
    \else
      \csname the##1\endcsname\quad
    \fi
  }

  \renewcommand{\theequation}{S\arabic{equation}}

  \renewcommand{\thefigure}{S\arabic{figure}}
  \renewcommand{\thetable}{S\arabic{table}}
  \renewcommand{\fnum@figure}{Fig. \thefigure}
  \renewcommand{\fnum@table}{Table \thetable}

  \@ifundefined{theHsection}{}{%
    \renewcommand{\theHsection}{supp.\arabic{section}}%
    \renewcommand{\theHsubsection}{supp.\arabic{section}.\arabic{subsection}}%
    \renewcommand{\theHfigure}{suppfig.\arabic{figure}}%
    \renewcommand{\theHtable}{supptab.\arabic{table}}%
    \renewcommand{\theHequation}{suppeq.\arabic{equation}}%
  }%
}
\makeatother

\definecolor{mirrorEven}{RGB}{220,235,250} 
\definecolor{mirrorOdd}{RGB}{255,235,205}

\theoremstyle{thmstyleone}%

\theoremstyle{thmstyletwo}%

\theoremstyle{thmstylethree}%

\begin{document}
\begin{bibunit}[sn-aps]

\title{Design Principles for Quasi-Isotropic Exchange in Rare-Earth Quantum Magnets}

\author*[1]{\fnm{Kotaro} \sur{Shimizu}}\email{shimizu@ap.t.u-tokyo.ac.jp}

\author[2,3]{\fnm{Esteban Agustin} \sur{Ghioldi}}%

\author[3]{\fnm{Filip} \sur{Ronning}}%

\author[4,5]{\fnm{Cristian D.} \sur{Batista}}%

\affil*[1]{\orgdiv{Department of Applied Physics}, \orgname{The University of Tokyo}, \orgaddress{\city{Bunkyo-ku}, \postcode{113-8656}, \state{Tokyo}, \country{Japan}}}

\affil[2]{\orgdiv{Theoretical Division}, \orgname{Los Alamos National Laboratory}, \orgaddress{\city{Los Alamos}, \postcode{87545}, \state{NM}, \country{USA}}}

\affil[3]{\orgdiv{Institute for Materials Science}, \orgname{Los Alamos National Laboratory}, \orgaddress{\city{Los Alamos}, \postcode{87545}, \state{NM}, \country{USA}}}

\affil[4]{\orgdiv{Quantum Condensed Matter Division and Shull-Wollan Center}, \orgname{Oak Ridge National Laboratory}, \orgaddress{\city{Oak Ridge}, \postcode{37831}, \state{TN}, \country{USA}}}

\affil[5]{\orgdiv{Department of Physics and Astronomy}, \orgname{University of Tennessee}, \orgaddress{\city{Knoxville}, \postcode{37996}, \state{TN}, \country{USA}}}

\abstract{
Rare-earth quantum materials provide a promising platform for emergent phenomena ranging from quantum spin liquids with long-range entanglement to topological magnetic textures. However, the strong spin--orbit coupling that stabilizes their low-energy pseudospin degrees of freedom also tends to generate strongly anisotropic exchange interactions, complicating the realization of quasi-isotropic Heisenberg magnetism.
Here we investigate the microscopic origin of superexchange in $\mathrm{Ce}^{3+}$- and $\mathrm{Yb}^{3+}$-based insulators with edge-sharing octahedral geometry. Using degenerate perturbation theory for a multiorbital Hubbard model, we show that isotropic exchange originates predominantly from virtual hopping within the ground-state Kramers doublet, whereas anisotropic interactions arise primarily from processes involving excited multiplets. This leads to a simple orbital design principle: quasi-isotropic exchange is promoted when the ground-state doublet has a strong maximal-angular-momentum character with respect to the quantization axis perpendicular to the superexchange plane spanned by rare-earth and ligand ions.
We demonstrate this mechanism for both ideal and distorted geometries and show that it is broadly consistent with experimentally studied Yb-based insulators. Our results establish a practical framework for engineering quasi-isotropic interactions in rare-earth quantum materials.
}

\maketitle

\section{Introduction \label{sec:introduction}}

The search for exotic states of quantum matter has become a central theme in modern condensed matter physics, driven by the prospect of realizing phases characterized by long-range quantum entanglement and non-trivial topology~\cite{WenXG1991,Balents2010review}. A particularly fruitful guiding principle in this effort is the identification of systems whose low-energy physics is governed by \emph{quasi-isotropic Heisenberg interactions}. Such interactions provide a well-understood and versatile platform in which a wide variety of emergent phenomena can arise, ranging from quantum spin liquids (QSLs) in frustrated lattices to topological magnetic textures such as skyrmion crystals.

In geometrically frustrated lattices, such as triangular and kagome geometries, quasi-isotropic antiferromagnetic interactions can suppress conventional magnetic ordering and stabilize QSL ground states~\cite{Anderson1973,Savary_2016review,broholm2019quantum}, characterized by fractionalized excitations and long-range entanglement, with potential applications in topological quantum computation~\cite{Kitaev2003,Nayak2008}. Quasi-isotropic interactions are also central to systems with long-wavelength spiral order, where weak perturbations--such as chiral interactions~\cite{Bogdanov1989,Bogdanov1994,Roessler2006,Yi2009,Nagaosa2013} or magnetic anisotropies~\cite{Butenko2010,Wilson2014,Lin2016,Leonov2016,Hayami2016,Leonov2017}--can stabilize skyrmion crystal phases. This dual relevance underscores quasi-isotropy as a guiding principle for realizing both quantum-disordered and topological magnetic states.

Rare-earth-based materials provide a remarkably flexible platform for realizing quantum spin models across a wide range of lattice geometries. The crystal structure determines the spatial arrangement of magnetic ions, enabling the realization of triangular, kagome, honeycomb, and three-dimensional frustrated lattices. At the same time, the local electronic configuration of the rare-earth ions selects the effective low-energy degrees of freedom, ranging from large-$S$ moments in $L=0$ ions to effective spin-$1/2$ Kramers doublets~\cite{Lea1962}. Finally, the form of the magnetic interactions can be tuned through the interplay of spin-orbit coupling, crystal electric fields, and ligand-mediated hopping processes~\cite{Jang2019,Jang2020,Motome2020,Jang2024}. This combination of structural, local, and interaction-level tunability makes rare-earth compounds an ideal setting for engineering and exploring a broad class of quantum spin Hamiltonians. 

This flexibility, however, is accompanied by important microscopic constraints that govern the form of the exchange interactions. Due to the strong localization of $4f$ orbitals, direct $f$-$f$ hopping is strongly suppressed, and magnetic exchange is mediated predominantly by ligand-assisted processes. As a result, achieving quasi-isotropic interactions in these systems is generally nontrivial. A notable exception arises for ions such as Gd$^{3+}$ and Eu$^{2+}$, where the $4f^7$ configuration is half-filled, leading to a vanishing orbital angular momentum ($L=0$) and, consequently, nearly isotropic magnetic interactions. This property underlies the prominent role of Gd- and Eu-based compounds among known hosts of magnetic skyrmion crystals~\cite{Tokura2020,Kurumaji2019,Hirschberger2019,Kakihana2019,Khanh2020,Takagi2022}. In contrast, most rare-earth insulators exhibit strongly anisotropic exchange interactions due to the combined effects of spin-orbit coupling and crystal electric field splitting, which entangle spin and orbital degrees of freedom and generate bond-dependent interactions.

This observation highlights an important limitation of the $L=0$ ions: while they provide nearly ideal realizations of isotropic Heisenberg interactions, their large spin ($S=7/2$) suppresses quantum fluctuations and is therefore unfavorable for stabilizing quantum spin liquid phases. By contrast, quantum spin liquids are typically realized in systems with effective spin-$1/2$ degrees of freedom, where quantum fluctuations are maximized. In rare-earth materials, such two-level systems naturally emerge from Kramers doublets of ions such as Ce$^{3+}$ and Yb$^{3+}$, whose low-energy physics can be described by effective pseudospin-$1/2$ moments. These ions, however, generally exhibit strong anisotropic exchange interactions due to spin-orbit coupling and crystal electric field effects. This tension raises a central question: can one simultaneously realize spin-$1/2$ degrees of freedom and quasi-isotropic Heisenberg interactions in rare-earth systems?

\begin{figure}[h]
\centering
\includegraphics[width=1.0\textwidth]{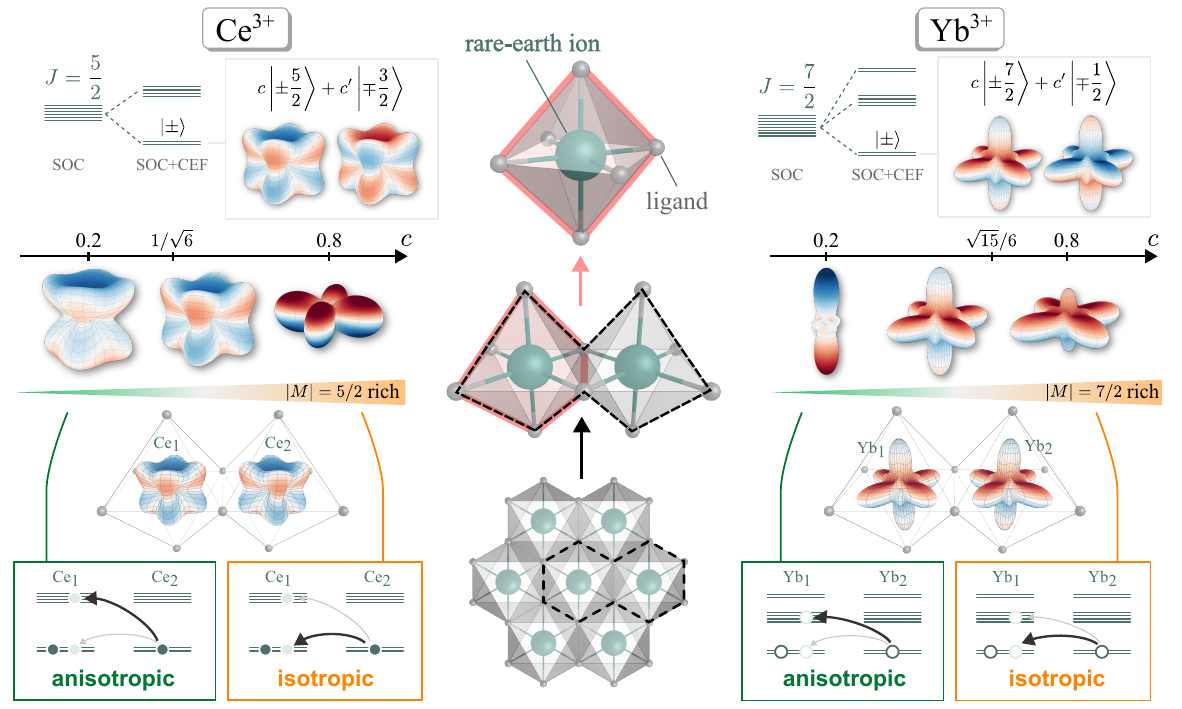}
\caption{
\textbf{Schematic illustration of the crystal-field splitting, ground-state wave function, and the qualitative trend of magnetic interactions in Yb- and Ce-based insulators.}
The left and right columns show Ce- and Yb-based cases, respectively. 
The upper panels illustrate the single-ion level structure: spin--orbit coupling selects the lowest total-angular-momentum manifold, $J=5/2$ for $\mathrm{Ce}^{3+}$ and $J=7/2$ for $\mathrm{Yb}^{3+}$, and the crystal electric field further splits this manifold into Kramers doublets. 
Within the ground multiplet, the doublets are parametrized as $\ket{\pm}=c\ket{\pm 5/2}+c'\ket{\mp 3/2}$ for Ce and $\ket{\pm}=c\ket{\pm 7/2}+c'\ket{\mp 1/2}$ for Yb, where $\ket{M}$ denotes the total-angular-momentum projection state. 
The shapes and colors of the wavefunctions represent the charge and magnetic distributions, respectively; electron and hole charge distributions are shown for $\mathrm{Ce}^{3+}$ and $\mathrm{Yb}^{3+}$. 
The middle panels show how the wavefunction evolves as the amplitude of the maximal-projection component is varied; as $c$ approaches unity, the wavefunction becomes increasingly flattened. 
The lower panels summarize the corresponding trend in magnetic interactions.
When the wavefunction is dominated by the $|M|=J$ component, magnetic exchange is governed primarily by intra-doublet hopping processes between the ground-state doublets of neighboring ions, resulting in nearly isotropic interactions. Reducing the $|M|=J$ character enhances out-of-doublet hopping processes involving the excited crystal-field multiplets of neighboring ions, which generates increasingly anisotropic exchange interactions.
}
\label{fig:schematic}
\end{figure}

In this work, we investigate the microscopic origin of magnetic interactions in Ce- and Yb-based compounds and identify a mechanism to enhance their isotropic character, as schematically illustrated in Fig.~\ref{fig:schematic}. We employ degenerate perturbation theory for $\mathrm{Ce}^{3+}$ and $\mathrm{Yb}^{3+}$ ions, whose low-energy degrees of freedom are described by Kramers doublets with total angular momentum $J=\frac{5}{2}$ and $\frac{7}{2}$, respectively. By systematically disentangling the perturbative contributions to the exchange interactions into isotropic (Heisenberg-like) and anisotropic components, we show that the isotropic interaction can be enhanced by increasing the hopping processes that connect the ground-state (GS) Kramers doublets (intra-doublet processes). 
We then demonstrate that a dominant contribution from the maximal angular-momentum projection---$M=\pm5/2$ for $\mathrm{Ce}$ and $M=\pm7/2$ for $\mathrm{Yb}$---naturally favors isotropic exchange because these states localize the orbital weight near the hopping plane, enhancing the intra-doublet process. 

Moreover, we show that this orbital criterion is broadly consistent with experimentally reported quasi-isotropic Yb-based compounds.
Finally, based on the microscopic understanding, we provide an explanation for why Yb is generally more favorable than Ce for realizing quasi-isotropic exchange, owing to differences in total angular momentum, electron-versus-hole character, and virtual intermediate states.

\section{Isotropic Exchange from a Perturbative Perspective}
\label{sec:perturbation}

\begin{figure}[h]
\centering
\includegraphics[width=1.0\textwidth]{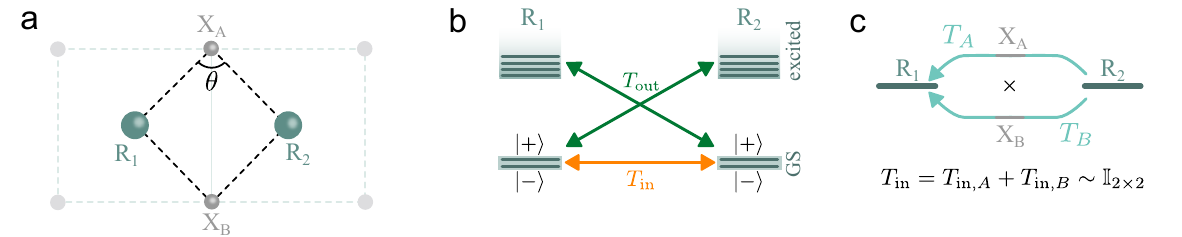}
\caption{
\textbf{Schematic illustration of the origin of isotropic and anisotropic exchange interactions.}
(\textbf{a}) Minimal cluster considered in the degenerate perturbation analysis, consisting of two rare-earth ions, $\mathrm{R}_1$ and $\mathrm{R}_2$, bridged by two ligand ions, $\mathrm{X_A}$ and $\mathrm{X_B}$.
(\textbf{b}) Multiorbital structure of the rare-earth ions and the two classes of effective hopping relevant to the exchange interaction: The intra-doublet processes $T_{\rm in}$, which connect the GS Kramers doublets, and the out-of-doublet processes $T_{\rm out}$, which involve excited multiplets.
(\textbf{c}) Destructive interference between the two ligand-mediated paths under inversion symmetry, which renders $T_{\rm in}$ pseudospin conserving.
}
\label{fig:perturbation}
\end{figure}

Figure~\ref{fig:perturbation} distills the perturbative picture that underlies our design principle for enhancing isotropic exchange in rare-earth oxides. Rather than beginning from the full crystal structure, we first isolate the minimal motif that controls the superexchange process: two rare-earth ions connected through two ligands, specifically relevant to the edge-shared octahedron case, as shown in Fig.~\ref{fig:perturbation}\textbf{a}. This construction allows us to separate the generic ingredients of the mechanism from materials-specific details and, in turn, to identify which virtual processes favor isotropic exchange and which generate anisotropy.

For the two ionic configurations of interest, $\mathrm{Ce}^{3+}$ and $\mathrm{Yb}^{3+}$, the local $4f$ manifolds correspond to a single electron ($4f^1$) and a single hole ($4f^{13}$), respectively. Strong spin-orbit coupling, followed by the weaker crystal electric field (CEF), selects a low-energy Kramers doublet on each site~\cite{Lea1962}; these doublets define the pseudospin-$1/2$ degrees of freedom relevant to the magnetic interaction. Having identified the appropriate low-energy variables, we describe the two-ion problem by a multi-orbital Hubbard model and derive the exchange interaction by degenerate perturbation theory~\cite{VSMironov1996,Palii2005,Ghioldi2024,Pourovskii2025-5d,Pourovskii2025-Ce-Yb}, as detailed in Methods and \suppref{sm:perturbation}.

Tracing out the ligand degrees of freedom then leads naturally to an effective hopping between the rare-earth ions. For the present discussion, the most important distinction is whether this effective hopping remains within the GS doublet manifold or instead accesses excited multiplets. Following Ref.~\cite{Ghioldi2024}, we therefore separate the virtual processes into the intra-doublet channel, $T_{\rm in}$, and the out-of-doublet channel, $T_{\rm out}$, as illustrated in Fig.~\ref{fig:perturbation}\textbf{b}. This classification already suggests the central physical idea: if the effective hopping is dominated by $T_{\rm in}$, the exchange problem reduces, at the level of the projected doublets, to the familiar single-orbital situation in which spin-conserving hopping produces Heisenberg-like exchange.

This expectation becomes rigorous once the bond symmetries are taken into account. In the presence of inversion symmetry together with time-reversal symmetry, the intra-doublet hopping is constrained to be pseudospin conserving, so that its contribution to the effective interaction is isotropic~\cite{Ghioldi2024}. This symmetry implies the constraint $T_{\rm in}=(i\sigma_y)T_{\rm in}^{\mathsf{T}}(-i\sigma_y)$, where $\sigma_y$ acts on the GS Kramers doublet subspace, leading to the pseudospin-conserving structure $T_{\rm in}\propto \mathbb{I}_{2\times2}$. This constraint can be understood as the consequence of destructive interference between the two ligand-mediated paths, each of which can flip pseudospin, as sketched in Fig.~\ref{fig:perturbation}\textbf{c}. The anisotropic terms must therefore originate primarily from processes involving excited multiplets, namely from $T_{\rm out}$.

This observation yields the guiding principle used throughout this paper: \emph{the exchange interaction becomes more isotropic as the relative weight of the out-of-doublet channel is reduced}. Put differently, the microscopic problem of designing isotropic exchange can be reformulated as the problem of enhancing $T_{\rm in}$ relative to $T_{\rm out}$. The detailed classification of the fourth-order processes %
is deferred to the \suppref{sm:perturbation}. With this organizing principle in place, we now discuss the quantum chemical conditions to realize such an isotropic situation.

\section{Designing Isotropic Superexchange}
\label{sec:design}

\begin{figure}[h]
\centering
\includegraphics[width=1.0\textwidth]{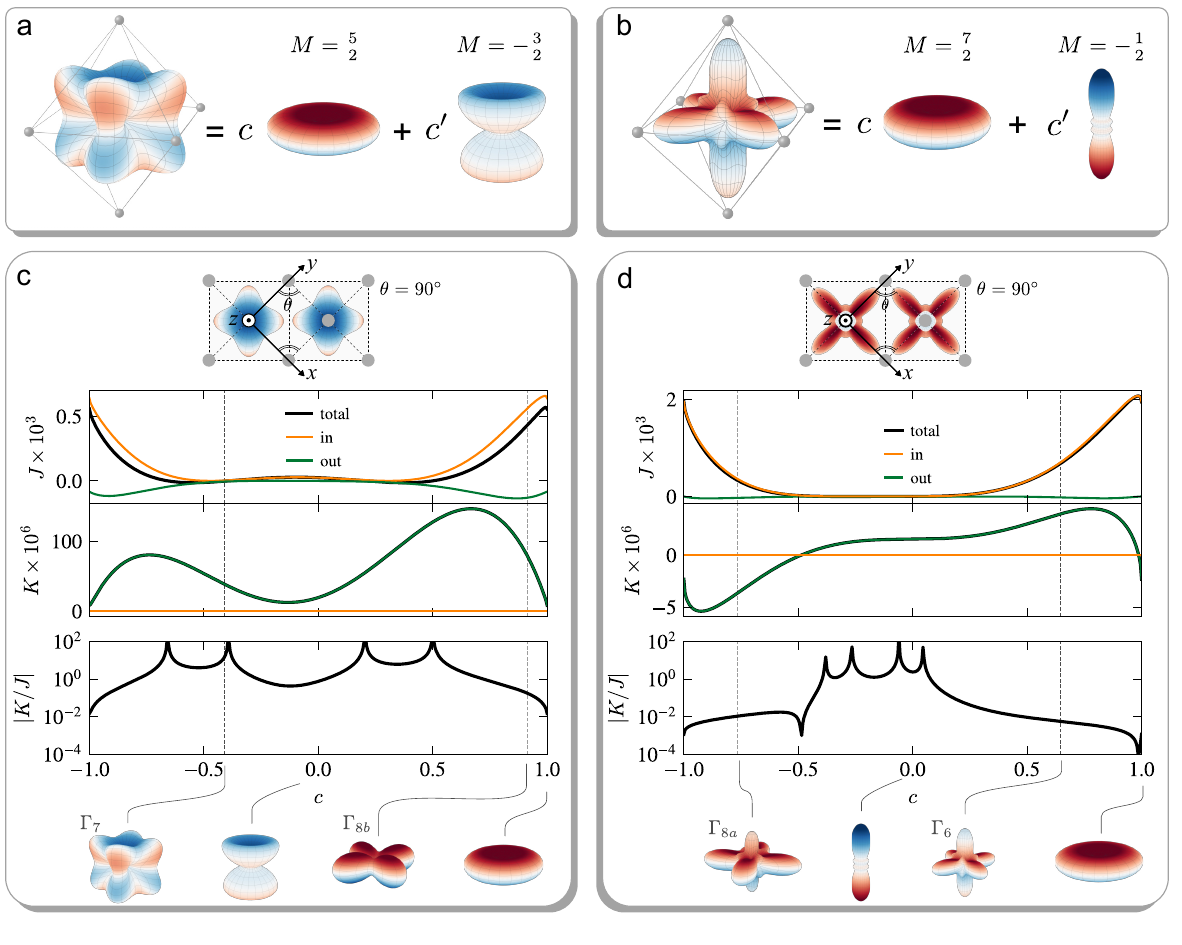}
\caption{
\textbf{Orbital shape of the ground-state wavefunction and its impact on isotropic exchange.}
Visualization of one component of the ground-state (GS) Kramers doublet for $\mathrm{Ce}^{3+}$ (\textbf{a}) and $\mathrm{Yb}^{3+}$ (\textbf{b}). 
The distance from the center to the surface represents the electron (hole) charge distribution in \textbf{a} (\textbf{b}), while the color represents the magnetic charge density~\cite{Kusunose2008} (see \suppref{sm:cef_review}). 
The gray circles indicate the ligand positions. 
Panels \textbf{c} and \textbf{d} show the evolution of the Heisenberg-like interaction $J$ and the anisotropic interaction $K$ for Ce and Yb, respectively, as the orbital shape is varied through the amplitude $c$ of maximum angular-momentum projection, $c$, of the Kramers doublets in Eq.~\eqref{eq:KD_c_Ce} and Eq.~\eqref{eq:KD_c_Yb}. All energies are given in eV and scale as $(t_{pf\sigma}/1\,\mathrm{eV})^4$ with the rare-earth--ligand $\sigma$-bond hopping amplitude, $t_{pf\sigma}$.
The total value is shown in black, and contributions from the intra- and out-of-doublet processes are represented by orange and green, respectively. 
Vertical dashed lines mark the values of $c$ corresponding to ideal octahedral crystal-field eigenstates: $\Gamma_7$ and $\Gamma_{8b}$ in \textbf{c}, and $\Gamma_6$ and $\Gamma_{8a}$ in \textbf{d}~\cite{Lea1962}. 
The corresponding wavefunctions, together with the limiting cases $c=0$ and $c=1$, are shown below the plots.
}
\label{fig:wfn_J}
\end{figure}

The perturbative analysis of Sec.~\ref{sec:perturbation} establishes the central microscopic criterion for quasi-isotropic superexchange: the intra-doublet hopping channel must dominate over the out-of-doublet one. 
The remaining task is to convert this criterion into a practical rule for materials design. 
To do so, we now reformulate the problem in terms of the real-space geometry of the crystal-field GS wavefunction, which directly controls the overlap with the ligand orbitals and hence the relative magnitudes of the two effective hopping channels.

Because the intra-doublet hopping is mediated by ligand $p$ orbitals, its amplitude is controlled primarily by the angular distribution of the rare-earth wavefunction relative to the superexchange plane, where the rare-earth $f$ orbitals hybridize with the ligand $p$ orbitals. In particular, orbitals whose weight is concentrated near the hopping plane enhance the intra-doublet process. To illustrate the connection between orbital shape and exchange interactions, we choose the local quantization axis such that $\hat{\mathbf{z}}$ is normal to the rare-earth--ligand hopping plane. Hereafter, we refer to this local reference frame as the \emph{hopping frame}.
Although the detailed form of the wavefunction depends on the local ligand environment and the electronic structure of each material, the essential mechanism is particularly transparent in the cubic case, shown in Figs.~\ref{fig:wfn_J}\textbf{a} and \ref{fig:wfn_J}\textbf{b}.  
Fourfold rotational symmetry constrains the mixing of $| J, M \rangle$ states such that only pairs of states with $\Delta M =4n$ ($n$ is an integer) can hybridize. 
Under this constraint, the electron and hole charge distributions tend to avoid and align with the negatively charged ligand directions, respectively. 
As a result, the ground-state wavefunctions of $\mathrm{Ce}^{3+}$ and $\mathrm{Yb}^{3+}$ are described by linear combinations of a planar $M=\frac{5}{2}$ component and a vertically extended $M=-\frac{3}{2}$ component within the $J=5/2$ multiplet, and of a planar $M=\frac{7}{2}$ component and a vertically elongated $M=-\frac{1}{2}$ component within the $J=7/2$ multiplet, respectively.

This cubic example immediately suggests a design rule: \textit{The intra-doublet channel is enhanced when the ground-state doublet contains a large weight of the maximal angular-momentum component $|M|=J$.}
As indicated by the wavefunction decomposition, orbitals with larger $|M|$ exhibit charge distributions that extend more strongly within the hopping plane, thereby increasing the hybridization with the ligand orbitals that mediate effective hopping between neighboring rare-earth ions. 
Increasing the $|M|=J$ amplitude in the GS therefore enhances the intra-doublet hopping while reducing the relative importance of the out-of-doublet channel, driving the exchange interaction toward the isotropic limit. 
This is the microscopic content of the rule illustrated in Fig.~\ref{fig:schematic}: a $|M|=J$-rich doublet promotes isotropic exchange because it selectively amplifies the intra-doublet channel, which is the primary source of isotropy.

We now test this orbital criterion explicitly, first for the ideal edge-sharing octahedral geometry and then in the presence of symmetry-lowering distortions. 
Taking the octahedral ground-state wavefunction as a reference, we parametrize the Kramers doublets as
\begin{align}
    \ket{\pm}
    =
    c\Ket{J=\frac{5}{2},M=\pm\frac{5}{2}}
    +c'\Ket{J=\frac{5}{2},M=\mp\frac{3}{2}},
    \label{eq:KD_c_Ce}
\end{align}
for $\mathrm{Ce}^{3+}$, and
\begin{align}
    \ket{\pm}
    =
    c\Ket{J=\frac{7}{2},M=\pm\frac{7}{2}}
    +c'\Ket{J=\frac{7}{2},M=\mp\frac{1}{2}},
    \label{eq:KD_c_Yb}
\end{align}
for $\mathrm{Yb}^{3+}$, with $c'=\sqrt{1-c^2}$. 
For $\mathrm{Ce}^{3+}$, the choice $c=-1/\sqrt{6}$ recovers the octahedral ground-state doublet $\Gamma_7$, whereas for $\mathrm{Yb}^{3+}$, $c=\sqrt{15}/6$ corresponds to $\Gamma_6$. 
The decomposition of these states in the basis specified by the orbital angular momentum $m_{\ell}$ and spin $m_s$, which clarifies the different orbital character of Ce and Yb, is given in Methods and \suppref{sm:angular_character}.
Using these states, we evaluate the exchange interactions within a perturbative framework as a function of the maximal-angular-momentum amplitude $c$. 
Details of the calculations are provided in the Methods.

The $\mathrm{RE}_1$--$\mathrm{X_{A(B)}}$--$\mathrm{RE}_2$ bond angle is fixed to $\theta=90^\circ$ (see also Fig.~\ref{fig:perturbation}\textbf{a}), for which symmetry constrains the exchange matrix to the diagonal form $\mathsf{J}_{12}=\mathrm{diag}(J,J,J+K)$. 
Figures~\ref{fig:wfn_J}\textbf{c} and \ref{fig:wfn_J}\textbf{d} show the $c$ dependence of the isotropic interaction $J$ (top), the anisotropic interaction $K$ (middle), and their ratio $|K/J|$ (bottom) for $\mathrm{Ce}$ and $\mathrm{Yb}$, respectively. Further decompositions into different perturbation channels are given in \suppref{sm:exchange_decomposition}.
The black curves show the total values, while the orange and green curves denote their decomposition into the intra- and out-of-doublet contributions, respectively. 
The wavefunctions shown at the bottom of  Figs.~\ref{fig:wfn_J}\textbf{c} and \ref{fig:wfn_J}\textbf{d} illustrate the corresponding evolution of the GS doublet; in particular, $c=\pm1$ corresponds to a pure $\ket{M=\pm J}$ state, whose orbital shape is most extended within the hopping plane.

For both Ce and Yb, the isotropic interaction $J$ is governed predominantly by the intra-doublet process, whereas the anisotropic interaction $K$ arises entirely from the out-of-doublet channel, directly confirming the microscopic disentanglement established in Sec.~\ref{sec:perturbation}.
In addition, $J$ and $K$ exhibit qualitatively distinct behaviors; $J$ is enhanced near $c\sim\pm1$, where the orbital extends in the hopping plane, while $K$ is strongly suppressed. 
By contrast, near $c\sim0$ the orbital is elongated along the direction perpendicular to the plane, reducing the ligand-mediated hopping and thus the overall exchange scale. In this regime, both $J$ and $K$ become small and may change sign, producing accidental sharp spikes and dips in $|K/J|$, which are sensitive to microscopic details, as further illustrated below by the $\theta$ dependence. 
Most importantly, the strong reduction of $K$ along with the enhancement of $J$ in $c \sim \pm 1$ leads to the stable suppression of the relative anisotropy $|K/J|$. These results provide direct support for the microscopic rule proposed above: enhancing the maximal-$|M|$ component drives the system toward more isotropic exchange by favoring the intra-doublet hopping channel.

Although the quasi-isotropic condition is satisfied in the $M=\pm J$ limit for both systems, several important differences emerge when comparing Ce and Yb.
First, there is a marked difference in the robustness of quasi-isotropy against perturbations of the wavefunction. For $\mathrm{Ce}$, $|K/J|$ increases rapidly as $c$ deviates from $\pm1$ and remains large over a broad intermediate range. By contrast, for $\mathrm{Yb}$, the suppression of $|K/J|$ persists over a substantially wider interval of $c$.
Second, near the ground states naturally realized within the point-charge picture, Ce exhibits a highly anisotropic $\Gamma_7$ ground state, whereas Yb realizes a remarkably isotropic $\Gamma_6$ ground state, consistent with previous studies~\cite{Rau2018,Villanova2023,Jang2024,Pourovskii2025-5d,Pourovskii2025-Ce-Yb}.
Third, Ce is overall more anisotropic than Yb. Even including the limiting pure $M=\pm J$ case, Yb generally exhibits a stronger tendency toward isotropy.

These qualitative distinctions make Yb more favorable than Ce for realizing quasi-isotropic exchange. 
First, the difference is visible in the decomposition of $\ket{J,M}$ into $\ket{m_\ell,m_s}$. 
In Yb, the $\ket{M=\pm 7/2}$ state is a pure $m_\ell=\pm 3$ state with strong in-plane character, and the admixed component from $\ket{M=\mp 1/2}$ also retains in-plane charge density. 
In Ce, by contrast, the $\ket{M=\pm 5/2}$ state already contains an $m_\ell=\pm 2$ component, making the planar orbital character more fragile against admixture of other $M$ components. 
This distinction is also tied to the mirror parity of the relevant orbitals (see \suppref{sm:angular_character}).
Second, the electron--hole asymmetry leads to different crystal-field preferences. 
Because $\mathrm{Ce}^{3+}$ contains one $f$-electron, its charge distribution tends to avoid the negatively charged ligand directions, naturally suppressing the $|M|=5/2$ component. We note that a possible GS for positively charged ligand ions, $\Gamma_{8b}$, shows the flat orbital, and indeed is quasi-isotropic. 
By contrast, the hole character of $\mathrm{Yb}^{3+}$ favors extension within the hopping plane. 
Third, the two ions differ in the virtual charge sectors available in insulating superexchange~\cite{Bordelon2021KCeO2}. 
For Yb, the intermediate states include nonmagnetic configurations in which the holes are transferred to the ligands, so that both rare-earth sites become nonmagnetic; these states provide a major source of isotropic exchange because they do not resolve pseudospin~\cite{Ghioldi2024}. 
By contrast, Ce does not admit a nonmagnetic sector: its virtual processes necessarily involve magnetic intermediate states in which at least one ion is doubly occupied, leading to anisotropic interactions (see Methods and \suppref{sm:perturbation}).
Together, these three features make $\mathrm{Yb}^{3+}$ more favorable than its particle--hole counterpart $\mathrm{Ce}^{3+}$ for realizing quasi-isotropic exchange.

\begin{figure}[h]
\centering
\includegraphics[width=1.0\textwidth]{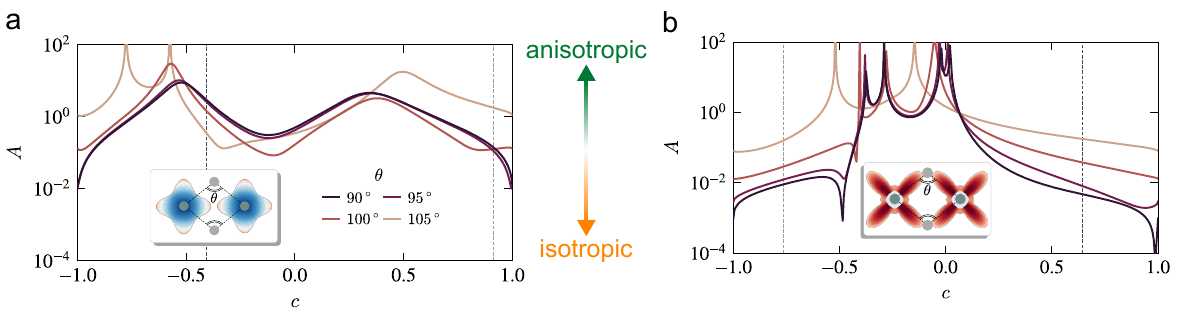}
\caption{
\textbf{Robustness of the isotropic design rule against bond-angle variation.}
Magnetic anisotropy measure in Eq.~\eqref{eq:anisotropy_measure} for several bond angles $\theta$ for Ce (\textbf{a}) and Yb (\textbf{b}) with an edge-sharing octahedral network of ligands. 
The parameter sets used in the perturbation analysis are the same as those adopted in Fig.~\ref{fig:wfn_J}.
}
\label{fig:angle}
\end{figure}

Having demonstrated the rule in the ideal $90^\circ$ geometry, we next examine its robustness against changes in the bond angle $\theta$. 
Figure~\ref{fig:angle} shows the $c$ dependence of the anisotropy for several $\theta$ values in the vicinity of $90^\circ$; because the results for $\theta\le 90^\circ$ and $\theta\ge 90^\circ$ are approximately symmetric within the displayed range, only the latter are shown. 
Once the bond angle deviates from $90^\circ$, the exchange matrix $\mathsf{J}_{12}$ generally acquires symmetric off-diagonal elements. 
To characterize anisotropy in such general cases, we define the anisotropy measure
\begin{align}
    A=\frac{\left|\boldsymbol{\lambda}-\bar{\lambda}\right|}{\left|\bar{\lambda}\right|},
    \label{eq:anisotropy_measure}
\end{align}
where $\boldsymbol{\lambda}=(\lambda_1,\lambda_2,\lambda_3)$ denotes the three eigenvalues of $\mathsf{J}_{12}$ and $\bar{\lambda}$ is their mean. 
By construction, $A=0$ for a purely Heisenberg-like exchange matrix $\mathsf{J}_{12}=\mathrm{diag}(J,J,J)$ and $A$ diverges when $\mathrm{Tr}\mathsf{J}_{12}=0$. 

For both $\mathrm{Ce}$ (\textbf{a}) and $\mathrm{Yb}$ (\textbf{b}), the anisotropy is robustly suppressed as the maximal-$|M|$ amplitude increases, even when the bond angle $\theta$ deviates from $90^\circ$. 
At the same time, the interaction is most isotropic near the ideal $\theta=90^\circ$ geometry, while the anisotropy increases as the bond angle moves away from this value. 
This indicates that the isotropic limit is realized most effectively in the ideal edge-sharing geometry. 
Away from $90^\circ$, the net intra-doublet hopping is geometrically suppressed, so that out-of-doublet processes become relatively more influential in setting the exchange anisotropy (see \suppref{sm:exchange_bond_angle}). 
Nevertheless, the same trend toward more isotropic exchange with increasing $|c|$ remains operative away from the ideal geometry. 
For small and intermediate $c$, by contrast, the anisotropy is already dominated by the out-of-doublet channel and remains large; the detailed profiles of accidental dips and spikes depend sensitively on $\theta$.

\section{Experimental Verification}\label{sec:verification}

We now examine whether the orbital criterion identified above is reflected in experimentally studied Yb-based insulators built from edge-sharing ligand octahedra. 
Because these materials are widely discussed as weakly anisotropic or near-Heisenberg rare-earth magnets, they provide a natural testing ground for the present design principle. 
Ce-based materials remain far less extensively explored. Even among the limited number of compounds studied to date, pronounced Kitaev-type interactions have been reported in systems such as $\mathrm{CsCeSe_2}$~\cite{Xie2024CsCeSe2a,Xie2024CsCeSe2b} and $\mathrm{KCeSe_2}$~\cite{Xie2025KCeSe2}, which is consistent with the Ce--Yb distinction identified in Sec.~\ref{sec:design}.
For each Yb-based material, we identify the dominant rare-earth--ligand--rare-earth superexchange unit, evaluate the corresponding bond angle $\theta$, and rotate the experimental GS Kramers doublet such that $\hat{\mathbf{z}}$ is normal to the dominant hopping plane. 
Writing the rotated doublet as $\ket{\pm}=\sum_M c_M^{\pm} \ket{J,M}$, we define the symmetrized weights
\begin{align}
    w_M = |c_M^\pm|^2 + |c_{-M}^\pm|^2,
    \label{eq:wm}
\end{align}
with $\sum_{M>0} w_M = 1$. 
For the present purpose, the leading experimental descriptor is $w_{7/2}$, while the remaining weights are listed for completeness. 
Table~\ref{tab:yb_material} summarizes the resulting hopping-frame wavefunctions together with representative experimentally constrained exchange interactions. %
We list exchange parameters constrained by inelastic neutron scattering or field-polarized spectroscopy, since these typically provide stronger constraints than thermodynamic properties alone.

A clear empirical consistency is seen across the edge-sharing Yb materials collected in Table~\ref{tab:yb_material}. 
Compounds that are widely regarded as weakly anisotropic---including NaYbO$_2$~\cite{Bordelon2019NaYbO2,Bordelon2020NaYbO2,Ding2019NaYbO2}, KYbSe$_2$~\cite{GrayCrystal2003a,Scheie2024KYbSe2a,Scheie2024KYbSe2b}, CsYbSe$_2$~\cite{Xing2019CsYbSe2,Xing2020CsYbSe2,Pocs2021CsYbSe2,Xie2023CsYbSe2},  YbCl$_3$~\cite{Sala2019YbCl3,Xing2020YbCl3,Sala2023YbCl3, Hao2020YbCl3,Matsumoto2024YbCl3}, YbBr$_3$~\cite{Wessler2020YbBr3,Hernandez2025YbBr3}, and YbAlO$_3$~\cite{Wu2019YbAlO3a,Wu2019YbAlO3b}---all exhibit substantial weight in the maximal angular-momentum sector in the hopping frame, with bond angles remaining in the range $\theta\sim90^\circ$--$100^\circ$. 
These materials provide broad empirical support for the criterion established in Sec.~\ref{sec:design}: large $M=\pm7/2$ weight in the hopping frame tends to generate quasi-isotropic exchange.

At the same time, the descriptor should not be interpreted as a complete one-parameter predictor. 
The $M=\pm 1/2$ component remains appreciable in several compounds, even though the exchange interaction is nearly isotropic. 
This does not contradict the present framework, because the relevant quantity is the orbital shape in the hopping frame rather than a single coefficient in isolation: even when the $M=\pm 1/2$ amplitude is significant, the total doublet can still remain favorably extended toward the ligand directions and thereby sustain a strong intra-doublet contribution. 
Indeed, as shown in Fig.~\ref{fig:wfn_J}\textbf{d}, the system remains isotropic down to small values of $c\gtrsim0.3$ corresponding to $w_{7/2}\gtrsim 0.1$. Since the flat, plane-localized character of the  $J=|M|=7/2$ state remains robust against admixture with other $M$ components, the quasi-isotropic regime persists over a broad region of the manifold.

\begin{table*}[t]
\footnotesize
\centering
\caption{\textbf{Ground-state doublets and exchange interactions in Yb-based insulators.}
For each compound, the bond angle $\theta$ and the orbital weights of the experimental ground-state Kramers doublet rotated to the hopping frame, $w_{M}$ in Eq.~\eqref{eq:wm}, are shown. 
The exchange column lists the representative nearest-neighbour exchange matrix or parameters used in the corresponding experimental analysis. 
The reference column briefly summarizes the experimental technique used to constrain the exchange, together with the corresponding reference.}
\label{tab:yb_material}
\begin{tblr}{
width = 1.0\textwidth,
vspan = even,
colspec = {
Q[c,m,wd=2.0cm]
Q[c,m,wd=0.7cm]
Q[c,m,wd=0.6cm]
Q[c,m,wd=0.6cm]
Q[c,m,wd=0.6cm]
Q[c,m,wd=0.6cm]
Q[c,m,wd=2.55cm]
Q[l,m,wd=2.4cm]
},
row{1,2} = {font=\footnotesize, valign=m},
column{8} = {font=\footnotesize, halign=l},
}
\toprule
\SetCell[r=2]{c,m} Compound &
\SetCell[r=2]{c,m} $\theta$ (deg.) &
\SetCell[c=4]{c,m} $w_{M}$ & & & &
\SetCell[r=2]{c,m} Exchange (meV) &
\SetCell[r=2]{c,m} Ref. \\
\cline{3-6}
& & 7/2 & 5/2 & 3/2 & 1/2 & & & \\
\midrule
\SetCell[r=1]{c,m} NaYbO$_2$~\cite{Ding2019NaYbO2,Bordelon2019NaYbO2}
& \SetCell[r=1]{c,m} 96.54
& \SetCell[r=1]{c,m} 0.401
& \SetCell[r=1]{c,m} 0.041
& \SetCell[r=1]{c,m} 0.103
& \SetCell[r=1]{c,m} 0.455
& $\begin{pmatrix}
0.51 & 0 & 0\\
0 & 0.51 & 0\\
0 & 0 & 0.45
\end{pmatrix}$
& INS(5 T)~\cite{Bordelon2019NaYbO2,Bordelon2020NaYbO2}
\\
\midrule
\SetCell[r=2]{c,m} KYbSe$_2$~\cite{Scheie2024KYbSe2a,GrayCrystal2003a}
& \SetCell[r=2]{c,m} 93.09
& \SetCell[r=2]{c,m} 0.414
& \SetCell[r=2]{c,m} 0.009
& \SetCell[r=2]{c,m} 0.016
& \SetCell[r=2]{c,m} 0.561
& $\begin{pmatrix}
0.199 & 0 & 0\\
0 & 0.202 & 0\\
0 & 0 & 0.196
\end{pmatrix}$
& diffuse scattering, ORF~\cite{Scheie2024KYbSe2a}
\\
\cline{7-9}
& & & & & 
& $J_1 \simeq 0.456$
& INS(4 T)~\cite{Scheie2024KYbSe2b}
\\
\midrule
\SetCell[r=1]{c,m} CsYbSe$_2$~\cite{Xing2019CsYbSe2,Xing2020CsYbSe2,Pocs2021CsYbSe2}
& \SetCell[r=1]{c,m} 93.76
& \SetCell[r=1]{c,m} 0.398
& \SetCell[r=1]{c,m} 0.025
& \SetCell[r=1]{c,m} 0.079
& \SetCell[r=1]{c,m} 0.498
& $J_1\simeq0.395$
& high-fields INS~\cite{Xie2023CsYbSe2}
\\
\midrule
YbCl$_3$~\cite{Sala2019YbCl3}
& 96.75
& 0.386
& 0.113
& 0.222
& 0.279
& $J_1\simeq0.344$
& high-fields INS~\cite{Sala2019YbCl3,Sala2023YbCl3}
\\
\midrule
\SetCell[r=2]{c,m} YbBr$_3$~\cite{Wessler2020YbBr3}
& \SetCell[r=2]{c,m} 93.56
& \SetCell[r=2]{c,m} 0.385
& \SetCell[r=2]{c,m} 0.041
& \SetCell[r=2]{c,m} 0.049
& \SetCell[r=2]{c,m} 0.524
& $J_1\simeq0.69$
& INS~\cite{Wessler2020YbBr3}
\\
\cline{7-9}
& & & & & 
& $J_1\simeq0.326$
& high-field INS~\cite{Hernandez2025YbBr3}
\\
\midrule
\SetCell[r=1]{c,m} YbAlO$_3$~\cite{Buryy2010YbAlO3,Wu2019YbAlO3a}
& \SetCell[r=1]{c,m} 101.45
& \SetCell[r=1]{c,m} 0.416
& \SetCell[r=1]{c,m} 0.310
& \SetCell[r=1]{c,m} 0.040
& \SetCell[r=1]{c,m} 0.234
& $J_1\simeq0.21$
& INS~\cite{Wu2019YbAlO3a,Wu2019YbAlO3b}
\\
\bottomrule
\end{tblr}%
\end{table*}

\begin{figure}[h]
\centering
\includegraphics[width=0.8\textwidth]{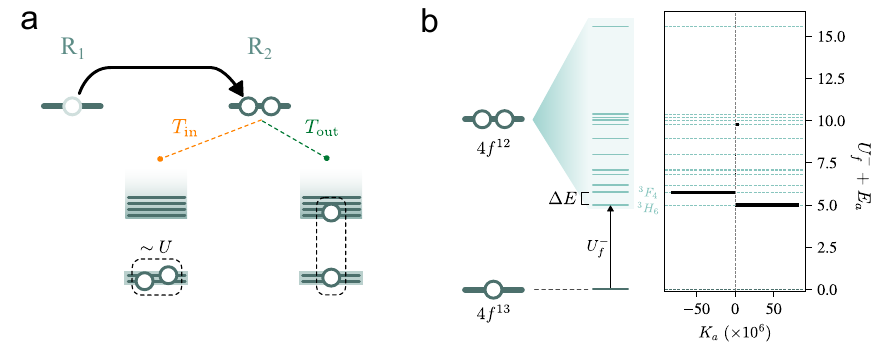}
\caption{
\textbf{Schematic illustration of the intermediate states and cancellation of the anisotropic interaction.}
\textbf{a} Schematic illustration of the hopping process between rare-earth ions that creates the double occupancy at R$_2$. Double occupancy created by the intra-doublet hopping $T_{\rm in}$ and out-of-doublet hopping $T_{\rm out}$ are schematically shown. 
\textbf{b} Energy spectrum for single- and double-hole configurations in the Yb ion: $4f^{13}$ and $4f^{12}$ configurations. The $4f^{12}$ configuration lies $U_{f}^{-}$ above the $4f^{13}$ ground state, and its spectrum is spread in the range of $\sim 10$ eV due to the strong Coulomb interaction. 
In the right panel, we show the anisotropic interaction $K$ decomposed into contributions from individual virtual $4f^{12}$ states labeled by $a$, $K_a$, for the octahedral case with $c=\sqrt{15}/6$ and $\theta=90^\circ$. The parameters are the same as those used in Fig.~\ref{fig:wfn_J}\textbf{d}. 
}
\label{fig:f2_spectrum}
\end{figure}

Finally, we stress that the present criterion is not a necessary condition for isotropic exchange. 
The geometric route discussed here promotes isotropy by suppressing the relative importance of out-of-doublet processes. 
However, quasi-isotropic exchange may also arise through cancellations among out-of-doublet anisotropic contributions from different intermediate states. 
One illustrative example is the artificial limit in which the relevant $4f^{12}$ intermediate states become degenerate, so that their anisotropic contributions cancel one another exactly, irrespective of the detailed hopping structure~\cite{Ghioldi2024}. 
In Fig.~\ref{fig:f2_spectrum}, we show the energy spectrum of the $4f^{12}$ intermediate states, which lie $U_f^-$ above the single-hole $4f^{13}$ state. In the right panel, we show the intermediate-state decomposition of the anisotropic exchange interaction $K$, estimated in Fig.~\ref{fig:wfn_J}\textbf{d}. 
Although the intermediate states are broadly distributed in energy, the anisotropic exchange is dominated by the two lowest-energy intermediate states, separated by an energy $\Delta E$. Since $K$ vanishes in the degenerate limit~\cite{Ghioldi2024}, we can estimate
\begin{align}
    K \sim k\left(\frac{1}{U_f^-} - \frac{1}{U_f^-+\Delta E}\right),
\end{align}
with $k$ conceptually written as $k \sim \tilde{T}_{\rm out}^2$ by using the normalized out-of-doublet hopping $\tilde{T}_{\rm out}$. 
This implies the presence of an additional small parameter $\Delta E/U_f^-$  that reduces  the anisotropic contribution relative to the isotropic one
$
K \sim \frac{k}{(U_f^-)^2}\Delta E \ll J.
$
While this mechanism may be operative in some materials, it is less straightforward to anticipate, as the relative weights of the different $4f^{12}$ intermediate states depend in a nontrivial way on the hopping amplitudes between the $f$-ions and the ligand orbitals. The orbital criterion identified here should therefore be viewed as a particularly transparent and practically useful design principle, rather than the only possible microscopic route to weak exchange anisotropy.

\section{Conclusions}
\label{sec:conclusion}

In this work, we established a microscopic design principle for quasi-isotropic superexchange in Ce- and Yb-based insulators. 
Starting from a multiorbital Hubbard model and employing degenerate perturbation theory, we disentangled the exchange processes into isotropic and anisotropic contributions: the isotropic interaction is governed primarily by the intra-doublet hopping channel connecting the GS Kramers doublets, whereas the anisotropic interaction originates from out-of-doublet processes involving excited multiplets. 
This separation leads naturally to a simple orbital criterion: increasing the weight of the maximal angular-momentum component, $|M|=J$, in the ground-state doublet enhances the intra-doublet hopping and drives the interaction toward the isotropic limit. 
We demonstrated this trend explicitly for ideal and moderately distorted edge-sharing octahedral geometries, and further showed that it is broadly consistent with experimentally reported wavefunctions and exchange interactions in many Yb-based materials.

These results also provide a quantum-chemical framework for understanding the microscopic distinction between Ce and Yb. 
In particular, the stronger tendency toward quasi-isotropic exchange in Yb-based systems can be traced to (1) less sensitive orbital directionality of the $J=7/2$ manifold against perturbations from mixing different $M$ components, (2) the hole-like extension toward ligand directions, and (3) the different virtual charge sectors available in the superexchange. 
The first two points would also apply to metallic systems, while the last point is specific to insulating systems; in metallic systems, such differences in the intermediate states do not generally arise because exchange is mediated both by occupied and unoccupied orbitals~\cite{SimethW2023_CeIn3,Ghioldi2024}. 
The orbital criterion proposed here, together with this microscopic understanding, provides a practical guide for material screening and for the design of pseudospin-$1/2$ rare-earth magnets with quasi-isotropic interactions.

The implications of this work extend well beyond the specific materials analyzed here. More fundamentally, our results identify the dominant microscopic design principle governing quasi-isotropic superexchange in rare-earth quantum magnets, providing a predictive framework for engineering exchange interactions in spin-orbit-entangled materials. While the strong spin--orbit coupling of rare-earth ions has long suggested that their magnetic interactions are intrinsically anisotropic and highly material dependent, we show that quasi-isotropic superexchange follows a remarkably simple orbital selection rule governed by the shape of the crystal-field ground-state wavefunction. In this sense, our work provides the rare-earth counterpart of the orbital design principles embodied in the Goodenough--Kanamori--Anderson framework, transforming the search for rare-earth quantum magnets from a largely empirical endeavor into a rational strategy based on microscopic quantum chemistry.

The predictive power of this framework is already reflected in existing materials. Several compounds listed in Table~\ref{tab:yb_material} belong to the delafossite-derived hexagonal family and lie in close proximity to a quantum spin-liquid regime because of the competition between nearest-neighbor and next-nearest-neighbor antiferromagnetic Heisenberg interactions \cite{Scheie2024KYbSe2a,Scheie2024KYbSe2b,Xie2023CsYbSe2,Bordelon2019NaYbO2,Bordelon2020NaYbO2}. Equally importantly, closely related hexagonal lattices with competing nearest-neighbor ferromagnetic and next-nearest-neighbor antiferromagnetic interactions are expected to stabilize long-wavelength spiral order that evolves into magnetic-field-induced skyrmion crystals in the presence of moderate easy-axis anisotropy~\cite{Butenko2010,Wilson2014,Lin2016,Leonov2016,Hayami2016,Leonov2017}. While such skyrmion crystals have so far been realized primarily in nearly isotropic rare-earth magnets based on Gd$^{3+}$ and Eu$^{2+}$~\cite{Tokura2020,Kurumaji2019,Hirschberger2019,Kakihana2019,Khanh2020,Takagi2022}, our results identify Yb$^{3+}$-based hexagonal magnets as particularly natural candidates in which strong quantum fluctuations, quasi-isotropic exchange, and topological magnetic textures can coexist.

More broadly, our work establishes a microscopic design paradigm for rare-earth quantum materials, demonstrating that the complexity introduced by strong spin-orbit coupling can be distilled into simple orbital selection rules. By placing the engineering of rare-earth superexchange on the same predictive footing as transition-metal superexchange, this framework opens a pathway toward the rational discovery of quantum spin liquids, topological magnetic textures, and other emergent phases in spin-orbit-entangled $4f$ materials.

\section{Methods}

\subsection{Spin-Orbit Coupled Kramers Doublet}

The local low-energy degrees of freedom are Kramers doublets generated by strong spin--orbit coupling (SOC) and subsequent crystal-field splitting. 
The atomic SOC Hamiltonian is
\begin{align}
    \mathcal{H}_{\rm SOC}=\lambda \sum_{m_{\ell},m_{\ell}',m_s,m_s'}\mathbf{l}_{m_{\ell}m_{\ell}'}\cdot\mathbf{s}_{m_sm_s'}\hat{f}^{\dagger}_{i,\alpha}\hat{f}^{\;}_{i,\alpha'},
    \label{eq:H_SOC}
\end{align}
where $\hat{f}^{(\dagger)}_{i,\alpha}$ with $\alpha=(m_{\ell},m_s)$ is the annihilation (creation) operator for an $f$ electron with orbital angular momentum $m_{\ell}$ and spin angular momentum $m_s$ at site $i$, and $\mathbf{l}_{m_{\ell}m_{\ell}'}$ and $\mathbf{s}_{m_sm_s'}$ are the matrix elements of orbital and spin angular momentum operators, respectively. 
For $\mathrm{Ce}^{3+}$ and $\mathrm{Yb}^{3+}$, the SOC splits the 14 single-particle states into the $J=5/2$ and $J=7/2$ multiplets, with the $J=5/2$ ($J=7/2$) multiplet forming the ground manifold for Ce (Yb).

The crystal electric field (CEF) further splits the $(2J+1)$-fold ground manifold into Kramers doublets. 
Within a fixed $J$ manifold, the CEF Hamiltonian is written as
\begin{align}
    \mathcal{H}_{\rm CEF}=\sum_{pq}B_{pq}\hat{O}_{pq},
    \label{eq:H_CEF}
\end{align}
where $\hat{O}_{pq}$ are Stevens operators. 
For the ideal octahedral environment considered here, the ground doublets are $\Gamma_7$ given by 
\begin{align}
    \ket{\Gamma_7;\pm}
    &=\sqrt{\frac{1}{6}}\Ket{J=\frac{5}{2},M=\pm\frac{5}{2}}-\sqrt{\frac{5}{6}}\Ket{J=\frac{5}{2},M=\mp\frac{3}{2}},\\
    \label{eq:Gamma7}
    &=\sqrt{\frac{1}{6}}\left[
        \sqrt{\frac{1}{7}}
        \Ket{m_{\ell}=\pm 2,m_s=\pm\frac{1}{2}}
        -
        \sqrt{\frac{6}{7}}
        \Ket{m_{\ell}=\pm 3,m_s=\mp\frac{1}{2}}
    \right] \nonumber\\
    &\quad
    -\sqrt{\frac{5}{6}}\left[
        \sqrt{\frac{5}{7}}
        \Ket{m_{\ell}=\mp 2,m_s=\pm\frac{1}{2}}
        -
        \sqrt{\frac{2}{7}}
        \Ket{m_{\ell}=\mp 1,m_s=\mp\frac{1}{2}}
    \right],
\end{align}
for Ce, and $\Gamma_6$ given by
\begin{align}
    \ket{\Gamma_6;\pm}
    &=\frac{\sqrt{15}}{6}\Ket{J=\frac{7}{2}, M=\pm\frac{7}{2}}
    +\frac{\sqrt{21}}{6}\Ket{J=\frac{7}{2}, M=\mp\frac{1}{2}},
    \label{eq:Gamma6}\\
    &=
    \frac{\sqrt{15}}{6}\Ket{m_{\ell}=\pm 3,m_s=\pm\frac{1}{2}} \nonumber\\
    &\quad
    +\frac{\sqrt{21}}{6}\left[
        \sqrt{\frac{3}{7}}
        \Ket{m_{\ell}=\mp 1,m_s=\pm\frac{1}{2}}
        +
        \sqrt{\frac{4}{7}}
        \Ket{m_{\ell}=0,m_s=\mp\frac{1}{2}}
    \right].
\end{align}
for Yb.
Here, the $x$, $y$, and $z$ axes are taken to be the directions toward the octahedral ligand ions.

\subsection{Hopping}

Ligand-mediated hopping between the rare-earth ions generates an effective interaction between their ground-state Kramers doublets. Due to the localized nature of $f$ electrons in rare-earth ions, we only consider the hybridization between rare-earth ions and ligands, leading to the superexchange. 
The hybridization term for the geometry given in Fig.~\ref{fig:perturbation}\textbf{a} is given by
\begin{equation}
    \mathcal{V}
    =
    \sum_{i=1,2}\sum_{\lambda=A,B}\sum_{\alpha\beta}
    \left(
    t_{i\lambda}^{\alpha\beta} \hat{f}_{i,\alpha}^{\dagger} \hat{p}^{\;}_{\lambda,\beta}
    + {\rm h.c.}
    \right),
    \label{eq:H_hopping}
\end{equation}
where $\hat{p}_{\lambda,\beta}^{(\dagger)}$ is the annihilation (creation) operator of the ligand $p$ orbital. The hopping between $\alpha=(m_{\ell},m_s)$ and $\beta=(m_{\ell}',m_s')$ is constructed from the Slater--Koster overlaps $t_{pf\sigma}$ and $t_{pf\pi}$~\cite{Takegahara1980}, as
\begin{align}
    t_{i\lambda}^{\alpha\beta}=\delta_{m_s,m_s'}\sum_{m=-1}^{1}\left[\mathsf{D}^{3}(\phi,\theta,0)\right]_{m_{\ell}m}
    t_m
    \left[\mathsf{D}^{1}(\phi,\theta,0)^{\dagger}\right]_{mm_{\ell}'},
\end{align}
where $t_m=t_{pf\sigma}$ for $m=0$ and $t_m=t_{pf\pi}$ for $|m|=1$, $[\mathsf{D}^J(\alpha,\beta,\gamma)]_{MM'}=\braket{J,M|e^{-i\alpha\hat{J}_z}e^{-i\beta\hat{J}_y}e^{-i\gamma\hat{J}_z}|J,M'}$ is the Wigner D-matrix with the angular momentum operators $\hat{J}_{\alpha}$ and its eigenstates $\ket{J,M}$, and $\theta$ and $\phi$ are the polar and azimuthal angles of the bond vector from ligand $\lambda$ to rare-earth site $i$.

\subsection{Coulomb Interaction}

The hopping between rare-earth ions generates the double occupancy of electrons and holes, which costs a Coulomb interaction, resulting in an exchange interaction. 
For the shell with angular momentum $\ell$, the atomic Coulomb interaction is given by
\begin{align}
    \mathcal{H}_{\rm C}=\frac{1}{2}\sum_{\substack{m_1,m_2\\m_3,m_4}}\sum_{m_s,m_s'}\sum_{k=0}^{\ell}
    &\delta_{m_1+m_2,m_3+m_4}(-1)^{m_1-m_4}F^{2k}c^{2k}(m_1,m_4)c^{2k}(m_2,m_3) \notag\\ 
    &\hat{f}^{\dagger}_{i,(m_1,m_s)}
    \hat{f}^{\dagger}_{i,(m_2,m_s')}
    \hat{f}^{\;}_{i,(m_3,m_s')}
    \hat{f}^{\;}_{i,(m_4,m_s)},
    \label{eq:H_Coulomb}
\end{align}
where $F^{2k}$ are the Slater--Condon parameters~\cite{Slater1954,Condon1931,Tinkham1964} and $c^{2k}$ is the Gaunt coefficient~\cite{Racah1942,Tinkham1964}.

\subsection{Microscopic Origin of Superexchange}
\label{method:perturbation}

The microscopic Hamiltonian is written as
\begin{equation}
    \mathcal{H}=\mathcal{H}_0+\mathcal{V},
    \label{eq:H_full}
\end{equation}
where $\mathcal{H}_0=\mathcal{H}_f+\mathcal{H}_p$ contains the local $4f$ and ligand terms, and $\mathcal{V}$ is the $f$--$p$ hybridization in Eq.~\eqref{eq:H_hopping}. 
On each rare-earth ion, the ground charge sector is $f^1$ for $\mathrm{Ce}^{3+}$ and $f^{13}$ for $\mathrm{Yb}^{3+}$, whereas the ligand ions are taken to be in the closed-shell $p^6$ configuration.

For the local $4f$ Hamiltonian, we include the Coulomb interaction and SOC. 
The crystal field is used only to determine the pseudospin-$1/2$ ground-state wavefunction and is neglected in the energy denominators, following Ref.~\cite{Rau2018}, because its energy scale is much smaller than the atomic Coulomb and charge-transfer scales. 
We adopt $(F^2,F^4,F^6)=(6.80, 4.38, 2.61)$\,eV and $\lambda=0.067$\,eV for $\mathrm{Ce}^{3+}$~\cite{Yeung2013}, and $(F^2,F^4,F^6)=(14.184, 9.846, 6.89)$\,eV and $\lambda=0.38$\,eV for $\mathrm{Yb}^{3+}$~\cite{Meftah2013}. 
The energy costs of the rare-earth intermediate states are denoted by $U_f^-$ and $U_f^+$ for the $4f^{n-1}$ and $4f^{n+1}$ sectors, respectively, and the ligand charge-transfer energy is denoted by $\Delta$. 
In the calculations shown in the main text, we use $U_f^- = U_f^+ = 5$ eV and $\Delta = 4$ eV.

The effective nearest-neighbour spin Hamiltonian is obtained by expanding in $\mathcal{V}$ up to fourth order. 
The resulting virtual processes can be grouped according to whether the intermediate rare-earth configurations are magnetic or nonmagnetic~\cite{Ghioldi2024}, which provides a useful way to organize the isotropic and anisotropic contributions discussed in the main text. 
For the Ce case, starting from the ground-state charge configuration of $(\mathrm{RE_1},\mathrm{X_A},\mathrm{X_B},\mathrm{RE_2})$ given by $(f^{1},p^6,p^6,f^1)$, the intermediate states are given by the $f^{(2:1)}$ and $f^{(2:2)}$ contributions, as 
\begin{align}
&f^{(2:1)}&:&~(f^2,p^6,p^6,f^0),~\left(+~1\leftrightarrow2,~A\leftrightarrow B\right),\\
&f^{(2:2)}&:&~(f^2,p^5,p^5,f^2),~(f^2,p^6,p^4,f^2),~\left(+~1\leftrightarrow2,~A\leftrightarrow B\right).
\end{align}
The $f^{(2:1)}$ contribution includes a single rare-earth ion in the magnetic configuration ($f^2$), while the $f^{(2:2)}$ contribution is given by the intermediate states with both rare-earth ions in the magnetic configuration.
In contrast, for the Yb case, the initial state is $(f^{13},p^6,p^6,f^{13})$, and the intermediate states are given by the $f^{(0)}$ and $f^{(2:1)}$ contributions, as 
\begin{align}
&f^{(0)}&:&~(f^{14},p^5,p^5,f^{14}),~(f^{14},p^{6},p^{4},f^{14}),~\left(+~1\leftrightarrow2,~A\leftrightarrow B\right),\\
&f^{(2:1)}&:&~(f^{12},p^6,p^6,f^{14}),~\left(+~1\leftrightarrow2,~A\leftrightarrow B\right),
\end{align}
Notably, in the $f^{(0)}$ intermediate states, both rare-earth ions exhibit the closed-shell configuration, namely a nonmagnetic configuration, and thus this process is the major source of the isotropic interaction. 
The full perturbative expressions, as well as further discussion on the source of isotropic and anisotropic interactions, are given in ~\suppref{sm:perturbation} and numerically verified by exact diagonalization in \suppref{sm:perturbation_verification}.

\subsection{Hopping Frame}
In practice, the crystal-field parameters $B_{pq}$ are determined by fitting inelastic neutron scattering spectra using the crystal-field Hamiltonian in Eq.~\eqref{eq:H_CEF}. To minimize the number of independent fitting parameters, the quantization axis is conventionally chosen along the highest-symmetry direction of the local ligand environment.
Furthermore, the CEF $x$ and $y$ axes cannot be uniquely fixed from the experiments. 

To test the criterion introduced in Sec.~\ref{sec:verification}, we first infer the experimental CEF frame and then rotate it to the hopping frame introduced above, in which the $z$ axis is normal to the superexchange plane.
First, we compute the CEF parameters $B_{pq}$ by using the point charge model, and compare the sign structures to experimental $B_{pq}$ to infer the CEF $x$ and $y$ axes. 
Then, from the experimentally reported crystal structure, we determine the positions of ligand ions bridging two rare-earth ions, and compute the direction perpendicular to the plane spanned by the rare-earth and ligand ions, to make a new $z$ axis. The new $x$ axis is taken in the direction of the other rare-earth ions, and the $y$ axis accordingly. 

For the experimentally obtained GS wavefunction $\ket{\pm}_{\rm exp}=\sum_mc_{{\rm exp},m}^{\pm}\ket{J,m}_{\rm exp}$, we rotate the experimental CEF frame $(\hat{\mathbf{x}}_{\rm exp},\hat{\mathbf{y}}_{\rm exp},\hat{\mathbf{z}}_{\rm exp})$ to the hopping frame $(\hat{\mathbf{x}},\hat{\mathbf{y}},\hat{\mathbf{z}})$. The wavefunction in the new frame $\ket{\pm}=\sum_{m}c_m^{\pm}\ket{J,m}$ is calculated by means of the Wigner D-matrix as
\begin{align}
    c_{m}^{\pm}=\sum_{m'} \left(D^J_{m'm}(\alpha,\beta,\gamma)\right)^*c_{{\rm exp},m'}^{\pm},
\end{align}
where $\alpha$, $\beta$, and $\gamma$ are Euler angles for this rotation. 
The explicit wavefunction in the original frame and detailed comparisons between experimental and point-charge CEF for each material are given in the Supplementary Information.

\backmatter

\bmhead{Supplementary information}

Additional supporting data are found in Supplementary Materials.

\bmhead{Acknowledgements}

The authors thank S.-H. Jang, Y. Motome, H. Nakai, P. Park, J. Rau, Y. Wang, and L. Zhang for valuable discussions, and M. Bordelon, A. Podlesnyak, A. Scheie, L. Wu, S. Wilson, and Q. Zhang for providing experimental information. 
K.S. is supported by JSPS KAKENHI Grant Numbers JP24K22870 and JP26H02011. Work by C.D.B. and K.S. was partially supported by the US Department of Energy, Office of Science, Basic Energy Sciences, Materials Sciences and Engineering Division under Award No. DE-SC0018660. F.R. was supported by the U.S. DOE Office of Basic Energy Sciences Materials Sciences and Engineering Division project ``Quantum Fluctuations in Narrow Band Systems''.

\putbib[ref-insulators]
\end{bibunit}

\beginsupplementary
\begin{bibunit}[sn-aps]

\section{Ionic Hamiltonian and multiplet structure}
\label{sm:multiplet_structure}

In this section, we summarize the single-ion ingredients entering the perturbation theory. We first introduce the rare-earth Hamiltonian and then discuss the resulting multiplet structures for Ce- and Yb-based compounds.

\subsection{Rare-earth Hamiltonian}
\label{sm:atomic_hamiltonian}

Within a fixed $4f$ shell with orbital angular momentum $\ell=3$, we write the single-ion Hamiltonian as
\begin{align}
    \mathcal{H}_{f}
    =\epsilon_f \hat{n}_f
    +\mathcal{H}_{\rm C}
    +\mathcal{H}_{\rm SOC}
    +\mathcal{H}_{\rm CEF},
    \label{eq:Hf_ion_electron}
\end{align}
with 
\begin{align}
    \hat{n}_f=\sum_{m s}\hat f_{m s}^{\dagger}\hat f_{m s}^{\;}.
\end{align}
Here $m=-3,\ldots,3$ and $s=\pm1/2$. The parameter $\epsilon_f$ denotes the spherically symmetric one-electron energy of the $4f$ shell. 
The orbital energy $\epsilon_f$ is not directly observable in the low-energy spin model and is absorbed, together with the monopole Coulomb energy, into the charge-excitation energies $U_f^{\pm}$ introduced below.

The leading energy scale in the rare-earth Hamiltonian is the Coulomb interaction, which splits the $4f^n$ configuration into multiplets labeled by the total orbital angular momentum $L$ and total spin angular momentum $S$.
The Coulomb interaction can be written in the symmetrized form
\begin{align}
    \mathcal{H}_{\rm C}
    =\frac{1}{2}
    \sum_{\alpha\beta\gamma\delta}
    U_{\alpha\beta\gamma\delta}
    \hat f_{\alpha}^{\dagger}
    \hat f_{\beta}^{\dagger}
    \hat f_{\delta}^{\;}
    \hat f_{\gamma}^{\;},
    \label{eq:HC_general}
\end{align}
where $\alpha=(m,s)$. The matrix element is obtained by expanding the Coulomb potential as
\begin{align}
    \frac{1}{|\mathbf r_1-\mathbf r_2|}
    =\sum_{k=0}^{\infty}\frac{4\pi}{2k+1}
    \frac{r_<^k}{r_>^{k+1}}
    \sum_{q=-k}^{k}Y_{kq}^{*}(\hat{\mathbf r}_1)Y_{kq}(\hat{\mathbf r}_2),
    \label{eq:coulomb_multipole}
\end{align}
where $r_<=\min(r_1,r_2)$, $r_>=\max(r_1,r_2)$, and $Y_{kq}(\hat{\mathbf r})$ are spherical harmonics in the Condon--Shortley phase convention.
This separates the matrix element into a radial part and an angular part. The radial part is parameterized by the Slater--Condon integral
\begin{align}
    F^{k}
    =e^2\int_{0}^{\infty}dr_1\int_{0}^{\infty}dr_2
    |u_{4f}(r_1)|^2 |u_{4f}(r_2)|^2
    \frac{r_<^k}{r_>^{k+1}},
    \label{eq:slater_integral}
\end{align}
where $u_{4f}(r)=rR_{4f}(r)$ is the radial wavefunction of the $4f$ orbital, and $e$ is the electron charge. The angular part is expressed by Gaunt coefficients
\begin{align}
    c^{k}(m,m')
    =(-1)^{m}(2\ell+1)
    \begin{pmatrix}
        \ell & k & \ell \\ 0 & 0 & 0
    \end{pmatrix}
    \begin{pmatrix}
        \ell & k & \ell \\ -m & m-m' & m'
    \end{pmatrix},
    \label{eq:gaunt_def}
\end{align}
where the parentheses denote the Wigner $3j$-symbol.
Within the same $4f$ shell, parity and the triangle rule allow only $k=0,2,4,6$. Thus the Coulomb interaction is parametrized by the four Slater integrals $F^0,F^2,F^4,F^6$. The monopole term $F^0$ mainly controls the energy cost of changing the valence, while $F^2,F^4,F^6$ split a fixed $4f^n$ configuration into LS multiplets, as described below. In the limit of $F^2=F^4=F^6=0$, the Coulomb interaction does not split the multiplets, so that the energy of a $4f^n$ configuration is given by $n_f\epsilon_f+\frac{1}{2}n_f(n_f-1)F^0$.

The next relevant energy scale is the SOC, which splits each LS multiplet into $J$ multiplets. The SOC Hamiltonian is given by
\begin{align}
    \mathcal{H}_{\rm SOC}
    =\lambda\sum_{m,m'}\sum_{s,s'}
    \mathbf l_{mm'}\cdot\mathbf s_{ss'}
    \hat f_{m s}^{\dagger}\hat f_{m's'}^{\;}.
    \label{eq:HSOC_onebody}
\end{align}
For a single $4f$ electron, this splits the ${}^2F$ term into $j=5/2$ and $j=7/2$ multiplets with
\begin{align}
    E_{j}^{(1)}=\frac{\lambda}{2}
    \left[j(j+1)-\ell(\ell+1)-\frac{3}{4}\right].
    \label{eq:soc_single_electron}
\end{align}

For two $4f$ particles, Hund's rules first select the LS term with the largest $S$ and then that with the largest $L$. The two-particle configurations relevant here are the $4f^2$ intermediate states of Ce and, after the particle--hole transformation described below, the two-hole $4f^{12}$ intermediate states of Yb.
In both cases, the Hund's-rule LS term is ${}^3H$, with $L=5$ and $S=1$.
First-order SOC splits this term into $J=4,5,6$ multiplets. Within the ${}^3H$ term, the SOC splitting is given by
\begin{align}
    E_{\rm SOC}({}^3H_J)
    =\frac{\zeta_H}{2}
    \left[J(J+1)-L(L+1)-S(S+1)\right]
    \label{eq:soc_3H}
\end{align}
For a less-than-half-filled $4f^2$ shell, $\zeta_H>0$, so that the lowest multiplet is ${}^3H_4$. In the two-hole representation of the $4f^{12}$ configuration, the sign of the effective spin--orbit coupling is reversed, $\zeta_H<0$, making ${}^3H_6$ the lowest multiplet.

For later use, we record the ionic energies of the lowest multiplets.
With the Coulomb interaction in Eq.~\eqref{eq:HC_general} and the SOC in Eq.~\eqref{eq:soc_3H}, the lowest-energy single-electron and single-hole multiplets are given by
\begin{align}
    E(4f^1,{}^2F_{5/2})
    &=
    \epsilon_f-2\lambda,
    \label{eq:E_f1_2F5/2}
    \\
    E(4f^{13},{}^2F_{7/2})
    &=
    \epsilon_h - \frac{3}{2}\lambda,
    \label{eq:E_f13_2F7/2}
\end{align}
where we have defined 
\begin{align}
    \epsilon_h=-\epsilon_f-(4\ell+1)F^0
    \label{eq:epsilon_h}
\end{align}
as the single-hole energy, which includes the monopole Coulomb energy. 
To first order in the SOC, the lowest two-particle energies are given by
\begin{align}
    E(4f^2,{}^3H_4)
    &\simeq
    2\epsilon_f+F^0+\varepsilon_{\rm C}({}^3H)-3\lambda,
    \label{eq:E_f2_3H4}
    \\
    E(4f^{12},{}^3H_6)
    &\simeq
    2\epsilon_h+F^0+\varepsilon_{\rm C}({}^3H)-\frac{5}{2}\lambda,
    \label{eq:E_h2_3H6}
\end{align}
with the non-monopole Coulomb energy of the ${}^3H$ term 
\begin{align}
    \varepsilon_{\rm C}({}^3H)
    =-\frac{1}{9}F^2
    -\frac{17}{363}F^4
    -\frac{25}{14157}F^6.
    \label{eq:epsilonC_3H}
\end{align}
In the numerical calculations, we diagonalize the full Coulomb and SOC Hamiltonian rather than using the perturbative expansion; the resulting multiplet energies therefore need not coincide with the approximate expressions above.
For Yb, we employ the particle--hole transformation that maps the $f^{13}$ configuration onto a single-hole $h^1$ configuration. The transformation and the correspondence between parameters in the electron and hole representations are discussed in \suppref{sm:particle_hole}.

\subsection{Crystal electric field}
\label{sm:cef_review}

\begin{figure}[h]
\centering
\includegraphics[width=1.0\textwidth]{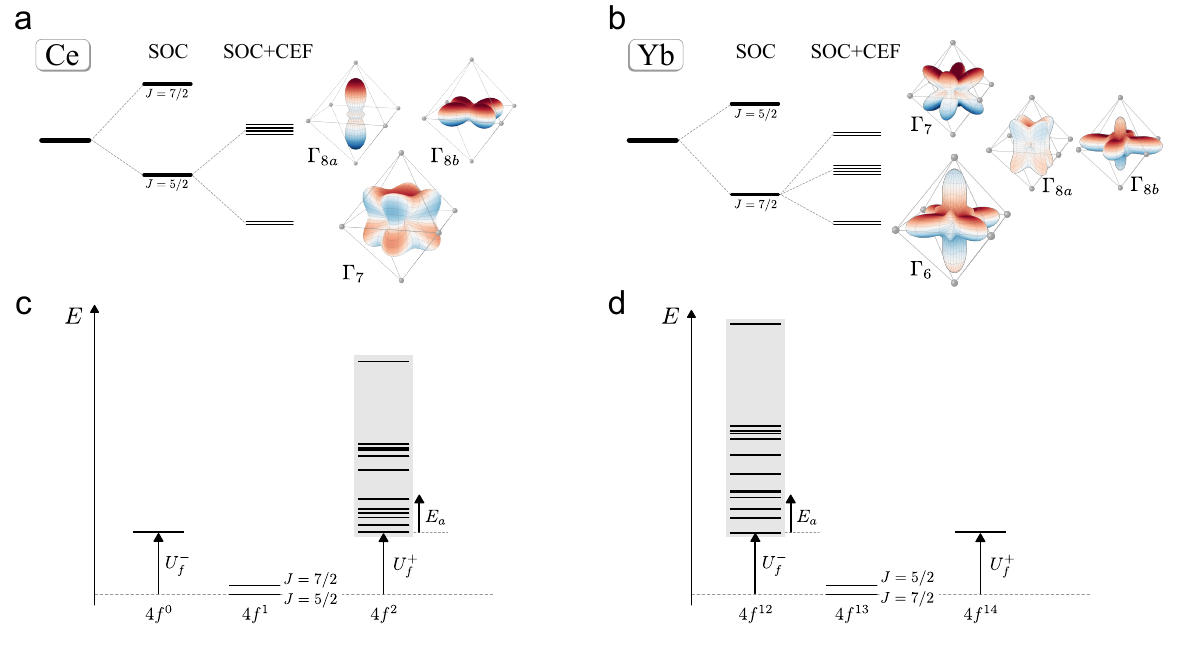}
\caption{
\textbf{Energy spectra of Ce and Yb considered in the perturbation analysis.}
Energy splitting in $\mathrm{Ce}^{3+}$ (\textbf{a}) and $\mathrm{Yb}^{3+}$ (\textbf{b}) associated with the SOC and the octahedral CEF. For each doublet, we show the wavefunction along with the octahedral ligand cage. The shape represents the electron (Ce) and hole (Yb) charge distribution, and the color represents the magnetic charge density~\cite{Kusunose2008}.
Energy spectra of the $4f^{n-1}$ and $4f^{n+1}$ configurations, corresponding to the states obtained by removing or adding a single electron to the $4f^n$ ground state, for Ce (\textbf{c}) and Yb (\textbf{d}).  In the two-particle sector, the Hund's-rule LS term is ${}^3H$; SOC selects ${}^3H_4$ as the lowest $4f^2$ state for Ce and ${}^3H_6$ as the lowest two-hole $4f^{12}$ state for Yb.
}
\label{figS:perturbation}
\end{figure}

The CEF is the nonspherical part of the electrostatic potential generated by the ligands. In a point-charge description,
\begin{align}
    \phi_{\rm CEF}(\mathbf r)
    =\sum_j\frac{Q_j}{|\mathbf r-\mathbf R_j|},
    \label{eq:cef_potential}
\end{align}
where $Q_j$ and $\mathbf R_j$ are the charge and position of the $j$-th ligand. After projecting the many-body Hamiltonian for the CEF, 
\begin{align}
    \mathcal{H}_{\rm CEF}=-e\sum_{i=1}^{N_f}\phi_{\rm CEF}(\mathbf r_i),
    \label{eq:HCEF_electron}
\end{align}
into a fixed $J$ manifold, the Wigner--Eckart theorem allows one to replace the coordinate tensors by Stevens operators $\hat O_{pq}$ built from $\mathbf J$,
\begin{align}
    \mathcal{H}_{\rm CEF}=\sum_{p,q}B_{pq}\hat O_{pq}.
    \label{eq:HCEF_stevens}
\end{align}
In an ideal octahedral environment, the symmetry-allowed Hamiltonian reduces to
\begin{align}
    \mathcal{H}_{\rm CEF}=B_4O_4+B_6O_6,
    \label{eq:HCEF_oct}
\end{align}
with
\begin{align}
    O_4&=O_{40}+5O_{44},\\
    O_6&=O_{60}-21O_{64}.
\end{align}
For Ce$^{3+}$ in the $J=5/2$ manifold, the rank-six operator does not split the manifold, and the point-charge octahedral environment yields a $\Gamma_7$ ground doublet. For Yb$^{3+}$, the same negatively charged ligand cage is more naturally described in the one-hole representation. Because a hole has positive charge, its density extends toward the negatively charged ligands and yields the octahedral $\Gamma_6$ ground doublet. The resulting level schemes and wavefunctions are shown in Figs.~\ref{figS:perturbation}\textbf{a} and \ref{figS:perturbation}\textbf{b}. The visualization of the wavefunction follows Ref.~\cite{Kusunose2008}: The shape represents the directional distribution of the corresponding charge density, and the color represents the magnetic (monopole) charge density, defined as $\rho_{\rm m}(\mathbf{r})=-\boldsymbol{\nabla}\cdot\mathbf{M}(\mathbf{r})$ with the magnetization $\mathbf{M}(\mathbf{r})$.

\subsection{Energy costs of virtual charge sectors}
\label{sm:configuration_energies}

The perturbation theory requires the energies of virtual charge sectors relative to the trivalent ground configuration. Let $E_0(4f^n)$ denote the lowest energy of the $4f^n$ configuration. We define $U_f^-$ as the energy cost of removing one electron from the $4f^N$ ground configuration and $U_f^+$ as the cost of adding one electron:
\begin{align}
    U_f^{-}&=E_0(4f^{N-1})-E_0(4f^N),
    \label{eq:Ufminus_def_electron}
    \\
    U_f^{+}&=E_0(4f^{N+1})-E_0(4f^N).
    \label{eq:Ufplus_def_electron}
\end{align}

For Ce$^{3+}$, $N=1$ and the ground multiplet of $4f^1$ is given by ${}^2F_{5/2}$. 
Using $E_0(4f^0)=0$, Eq.~\eqref{eq:E_f1_2F5/2}, and Eq.~\eqref{eq:E_f2_3H4}, the charge excitation energies are given by
\begin{align}
    U_{f}^{-}
    &= -\epsilon_f+2\lambda,
    \label{eq:Ufm_Ce}
    \\
    U_{f}^{+}
    &\simeq \epsilon_f+F^0
    -\frac{1}{9}F^2
    -\frac{17}{363}F^4
    -\frac{25}{14157}F^6
    -\lambda.
    \label{eq:Ufp_Ce}
\end{align}
The excited $4f^2$ multiplets are written as $U_{f}^{+}+E_a$, with $E_a=0$ for the lowest ${}^3H_4$ state. The energy levels of the $4f^{n}$ configurations with $n=0,1,2$ are shown in Fig.~\ref{figS:perturbation}\textbf{c}.

For Yb$^{3+}$, the ionic ground state is $4f^{13}$ ($N=13$), which is a one-hole configuration. By using the particle--hole transformation described in \suppref{sm:particle_hole}, and Eqs.~\eqref{eq:E_f13_2F7/2} and \eqref{eq:E_h2_3H6}, the charge excitation energies are given by
\begin{align}
    U_{f}^{+}
    &= -\epsilon_h+\frac{3}{2}\lambda,
    \label{eq:Ufp_Yb}
    \\
    U_{f}^{-}
    &\simeq \epsilon_h
    +F^0
    -\frac{1}{9}F^2
    -\frac{17}{363}F^4
    -\frac{25}{14157}F^6
    -\lambda.
    \label{eq:Ufm_Yb}
\end{align}
The corresponding energy levels of the $4f^{n}$ configurations with $n=12,13,14$ are shown in Fig.~\ref{figS:perturbation}\textbf{d}.

\subsection{Ligand Hamiltonian}
\label{sm:ligand_hamiltonian}

We next introduce the ligand Hamiltonian. Each ligand $\lambda=A,B$ has six $p$ spin-orbitals, labeled by $\beta=(m,s)$ with $m=-1,0,1$ and $s=\pm1/2$. Neglecting ligand SOC and Hund's coupling, we use the minimal form
\begin{align}
    \mathcal{H}_{p}
    &=\epsilon_p\hat{n}_{p}
    +\frac{U_p}{2}\sum_{\beta,\beta'} \hat{p}_{\beta}^{\dagger} \hat{p}_{\beta'}^{\dagger} \hat{p}_{\beta'}^{\;} \hat{p}_{\beta}^{\;},\\
    &=\epsilon_p\hat{n}_{p}
    +\frac{U_p}{2}\hat{n}_{p}(\hat{n}_{p}-1),
    \label{eq:H_ligand_electron}
\end{align}
with
\begin{align}
    \hat{n}_{p}=\sum_{\beta}\hat{p}_{\beta}^{\dagger}
    \hat{p}_{\beta}^{\;}.
\end{align}
The ligand ground state is the closed-shell $p^6$ configuration, and the superexchange processes involve virtual states with one or two ligand holes.
From Eq.~\eqref{eq:H_ligand_electron}, the energy cost of one ligand hole is
\begin{align}
    \Delta
    = E_0(p^5)-E_0(p^6)
    =-\epsilon_p-5U_p.
    \label{eq:Delta_from_ep}
\end{align}
Similarly, the energy cost of two holes on the same ligand is
\begin{align}
    E_0(p^4)-E_0(p^6)=2\Delta+U_p.
    \label{eq:two_ligand_holes}
\end{align}
Thus, within the relevant charge sectors $n_p=4,5,6$, the ligand Hamiltonian is fully specified by the two parameters $\Delta$ and $U_p$:
\begin{align}
    \mathcal{H}_{p}
    = \Delta\sum_{\beta}\ket{p^5_{\beta}}\bra{p^5_{\beta}}
    + (2\Delta+U_p)\sum_{\mu}\ket{p^4_{\mu}}\bra{p^4_{\mu}}.
\end{align}
Here, $\ket{p^5_{\beta}}$ denotes a single-hole state labeled by $\beta$, and $\ket{p^4_{\mu}}$ denotes a two-hole state labeled by $\mu$. The parameter $\Delta$ is the energy cost of creating a single ligand hole, whereas $U_p$ is the Coulomb interaction between two ligand holes.

\section{Hopping amplitudes}

\subsection{Slater--Koster integrals for interatomic matrix elements}
We denote an atomic wavefunction centered at the origin by $\psi_{\ell m}(\mathbf{r})$, where $\ell$ and $m$ are the orbital angular-momentum quantum numbers, and decompose it into radial and angular parts as
\begin{align}
\psi_{\ell m}(\mathbf{r})=F_\ell(r)Y_{\ell m}(\hat{\mathbf{r}}).
\end{align}
Within the two-center approximation, we evaluate the matrix element between the atomic wavefunctions $\psi_{\ell'm'}$ located around the position $\mathbf{X}$ and $\psi_{\ell m}$ located around the origin, $t_{\ell'm',\ell m}(\mathbf{X})=\braket{\psi_{\ell'm'}(\mathbf{r}-\mathbf{X})|\mathcal{V}|\psi_{\ell m}(\mathbf{r})}$. 
When $\mathbf{X}$ is parallel to the angular-momentum quantization axis $z$, the hopping element is nonzero only if the $z$ component of orbital angular momentum is conserved, $m'=m$. The number of independent hopping parameters is then reduced to $\min(\ell,\ell')+1$~\cite{Harrison1980}.
We introduce the Slater--Koster parameters $(\ell' \ell m)$ as the two-center integral for $\mathbf{X}\parallel \hat{\mathbf{z}}$, i.e.,
\begin{align}
(\ell' \ell m)=\braket{\psi_{\ell' m'}(\mathbf{r}-X\hat{z})|\mathcal{H}|\psi_{\ell m}(\mathbf{r})}.
\end{align}
For hopping from a $p$ orbital ($\ell=1$) to an $f$ orbital ($\ell'=3$), the independent Slater--Koster parameters are $t_{pf\sigma}=(310)$ and $t_{pf\pi}=(311)$.

We next consider a general bond vector $\mathbf{X}=X(\sin\beta\cos\alpha,\sin\beta\sin\alpha,\cos\beta)$, where $\alpha$ and $\beta$ are its azimuthal and polar angles, respectively.
We introduce a rotated coordinate system $(x',y',z')$ in which $\mathbf{X}$ is parallel to the $z'$ axis.
To obtain $t_{\ell'm',\ell m}(\mathbf{X})$ for an arbitrary $\mathbf{X}$, we express the angular part $Y_{\ell m}(\hat{\mathbf{r}})=\braket{\hat{\mathbf{r}}|\ell m}$ in terms of the rotated spherical harmonics $Y'_{\ell m}(\hat{\mathbf{r}})=\braket{\hat{\mathbf{r}}|\hat{R}(\alpha,\beta,0)|\ell m}$, where $\hat{R}(\alpha,\beta,0)=e^{-i\alpha\hat{J}_z}e^{-i\beta\hat{J}_y}$. Here, $\hat{J}_{\alpha}$ acts on the orbital part of the wavefunction.
Expanding $\hat{R}$ in the spherical-harmonic basis defines the Wigner $D$ matrix $D_{mm'}^{\ell}(\alpha,\beta,0)$ through
\begin{align}
    \hat{R}(\alpha,\beta,0)\ket{\ell m}
    =\sum_{m'=-\ell}^{\ell}\ket{\ell m'}D_{m'm}^{\ell}(\alpha,\beta,0).
\end{align}
Here,
\begin{align}
D_{m'm}^{\ell}(\alpha,\beta,\gamma)&=
\braket{\ell m' | e^{-i\alpha\hat{J_z}}e^{-i\beta\hat{J_y}}e^{-i\gamma\hat{J_z}} | \ell m}\\
&=\exp(-i\alpha m')d_{m'm}^{\ell}(\beta)\exp(-i\gamma m),
\end{align}
and $d_{m'm}^{\ell}(\beta)$ is the Wigner small d-matrix, which is a real function of $\beta$. By defining the matrix $\mathsf{D}^{\ell}(\alpha,\beta,\gamma)$ with $[\mathsf{D}^{\ell}(\alpha,\beta,\gamma)]_{mm'}=D_{mm'}^{\ell}(\alpha,\beta,\gamma)$, the spherical harmonics in the original and rotated frames are related by
\begin{align}
\begin{pmatrix}
\cdots, & \hat{R}(\alpha, \beta, 0)\ket{\ell m}, & \cdots
\end{pmatrix}
=
\begin{pmatrix}
\cdots, & \ket{\ell m}, & \cdots
\end{pmatrix}
\mathsf{D}^{\ell}(\alpha,\beta,0)
,\\
\begin{pmatrix}
\cdots, & \ket{\ell m}, & \cdots
\end{pmatrix}
=
\begin{pmatrix}
\cdots, & \hat{R}(\alpha, \beta, 0)\ket{\ell m}, & \cdots
\end{pmatrix}
\left[\mathsf{D}^{\ell}(\alpha,\beta,0)\right]^{\dag}.
\end{align}
The overlap integral for an arbitrary $\mathbf{X}$ is therefore
\begin{align}
t_{\ell' m',\ell m}(\mathbf{X})
&=
\braket{\psi_{\ell'm'}(\mathbf{r}-\mathbf{X})|\mathcal{H}|\psi_{\ell m}(\mathbf{r})}\\
&=
\sum_{m_1=-\ell'}^{\ell'}\sum_{m_2=-\ell}^{\ell}
D_{m'm_1}^{\ell'}(\alpha,\beta,0)\left(D_{mm_2}^{\ell}(\alpha,\beta,0)\right)^*\braket{\psi_{\ell' m_1}(\mathbf{r}'-\mathbf{X}')|\mathcal{H}|\psi_{\ell m_2}(\mathbf{r}')}\\
&=
\sum_{m_1=-\min(\ell,\ell')}^{\min(\ell,\ell')}
D_{m'm_1}^{\ell'}(\alpha,\beta,0)\left(D_{mm_1}^{\ell}(\alpha,\beta,0)\right)^*
(\ell'\ell m_1),\\
&=
\left[\mathsf{D}^{\ell'}(\alpha,\beta,0)
\mathsf{T}^{\text{SK}}_{\ell'\ell}
\mathsf{D}^{\ell}(\alpha,\beta,0)^{\dag}\right]_{m'm}.
\label{eq:SK_hopping_matrix_general}
\end{align}
Here, we define the $(2\ell'+1)\times(2\ell+1)$ matrix composed of the Slater--Koster parameters 
\begin{align}
    \left[\mathsf{T}^{\text{SK}}_{\ell'\ell}\right]_{mm'}=(\ell'\ell m)\delta_{m,m'}.
\end{align}

\begin{figure}[h!]
    \centering
    \includegraphics[width=0.5\linewidth]{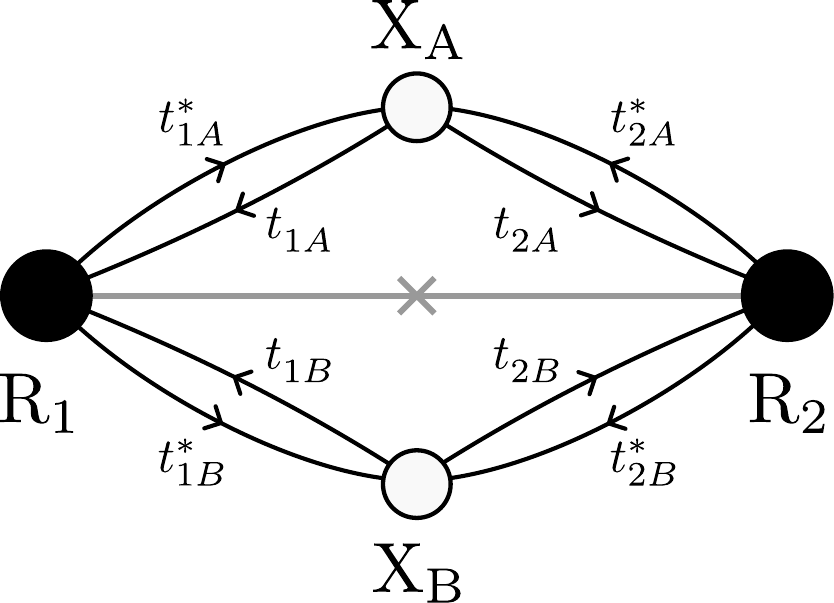}
    \caption{Sketch of a centrosymmetric bond connecting two rare-earth sites ($R_i$) via two ligand ions ($X_\lambda$). The thin solid lines illustrate the hopping processes between the ions.}
    \label{fig:bond-sketch}
\end{figure}

For the $p$--$f$ hopping relevant to this work, the hopping Hamiltonian is
\begin{align}
\mathcal{V}
&=\sum_{i=1,2}\sum_{\lambda=A,B}
\left[t_{i\lambda}^{\alpha\beta}\hat f_{i\alpha}^{\dagger}\hat p_{\lambda\beta}^{\;}+{\rm h.c.}\right],
\label{eq:hopping_pf}
\end{align}
with
\begin{align}
\mathsf{t}_{i\lambda}=\mathsf{D}^{3}(\alpha_{i\lambda},\beta_{i\lambda},0)
\mathsf{T}^{\text{SK}}_{3,1}
\mathsf{D}^{1}(\alpha_{i\lambda},\beta_{i\lambda},0)^{\dag}.
\end{align}
Here, $\alpha_{i\lambda}$ and $\beta_{i\lambda}$ are the azimuthal and polar angles, respectively, of the vector from ligand $\lambda$ to rare-earth site $i$.
We consider the bond geometry illustrated in Fig.~\ref{fig:bond-sketch}. The $z$ axis is perpendicular to the plane containing the two rare-earth ions and a ligand, the bond has an inversion center at the midpoint between the rare-earth ions, and a mirror plane is perpendicular to the bond.
In this geometry, we have $\alpha_{1A}=\pi + \alpha_{2B}=\pi +\alpha$, $\alpha_{2A}=\pi + \alpha_{1B}=2\pi - \alpha$, and $\beta_{i\lambda}=\pi/2$ for all $i$ and $\lambda$. 
The $\mathrm{Yb}$--$\mathrm{O}$--$\mathrm{Yb}$ bond angle is given by $\theta = \pi - 2\alpha$. 
Hereafter, we use the same Slater--Koster parameters $t_{pf\sigma}$ and $t_{pf\pi}$ for all bonds. The hopping matrices are therefore parametrized by $t_{pf\sigma}$, $t_{pf\pi}$, and the bond angle $\theta$.

\subsection{Effective hopping mediated by ligands}

Within the Slater--Koster scheme, we compute the hopping matrix elements in the Hamiltonian $\mathcal{V}$ for the given bond geometry parametrized by the angle $\alpha$. 
From the $f$--$p$ hopping matrices, we define the effective hopping from rare-earth site 2 to site 1 through ligand $\lambda$ as
\begin{align}
\mathsf{T}_{\lambda}=\mathsf{t}_{1\lambda} \mathsf{t}_{2\lambda}^{\dag}.
\label{eq:T_lambda}
\end{align}
Due to the inversion symmetry of the system, the hopping mediated by the ligand $B$ is given by
\begin{align}
\mathsf{T}_B = \mathsf{T}_A^{\dag},
\end{align}
and the total effective hopping is therefore
\begin{align}
\mathsf{T}=\mathsf{T}_A+\mathsf{T}_B = \mathsf{T}_A+\mathsf{T}_A^{\dag}.
\label{eq:T_total}
\end{align}
The diagonal components of the effective hopping for the $J=5/2$ multiplet are given by
\begin{align}
T^{J=5/2}\left(\pm\frac12, \pm\frac12\right)
&=
\frac{t_{pf\sigma}^{2}}{28}
\left(
6\cos^2\theta+2\sqrt{6}\rho\sin^2\theta+\rho^2(9+\cos^2\theta)
\right),
\\
T^{J=5/2}\left(\pm\frac32, \pm\frac32\right)
&=
\frac{t_{pf\sigma}^{2}}{112}
\left(
12\cos^2\theta+4\sqrt{6}\rho\sin^2\theta+\rho^2(-50+102\cos^2\theta)
\right),
\\
T^{J=5/2}\left(\pm\frac52, \pm\frac52\right)
&=
\frac{5t_{pf\sigma}^{2}}{112}
\left(
6(\cos2\theta+\cos4\theta)+6\sqrt{6}\rho
(\cos2\theta-\cos4\theta) \notag\right.\\
&\qquad\left.+\rho^2(11\cos2\theta+9\cos4\theta)
\right),
\end{align}
and those for the $J=7/2$ multiplet are given by
\begin{align}
T^{J=7/2}\left(\pm\frac12, \pm\frac12\right)
&=
\frac{3 t_{pf\sigma}^{2}}{112} 
\left(
6\cos^2\theta + 2\sqrt{6}\rho\sin^2\theta + \rho^2(16+\cos^2\theta)
\right),
\\
T^{J=7/2}\left(\pm\frac32, \pm\frac32\right)
&=
\frac{5t_{pf\sigma}^{2}}{224}
\left(
12\cos^2\theta + 4\sqrt{6}\rho\sin^2\theta + \rho^2(-8+18\cos^2\theta)
\right),
\\
T^{J=7/2}\left(\pm\frac52, \pm\frac52\right)
&=
\frac{5t_{pf\sigma}^{2}}{224}
\left(
2(\cos2\theta+\cos4\theta) + 2\sqrt{6}\rho
(\cos2\theta-\cos4\theta) \notag\right.\\
&\qquad\left.
+3\rho^2(9\cos2\theta+\cos4\theta)
\right),
\\
T^{J=7/2}\left(\pm\frac72, \pm\frac72\right)
&=
\frac{5t_{pf\sigma}^{2}}{32}
\left(
2(\cos2\theta+\cos4\theta) + 2\sqrt{6}\rho
(\cos2\theta-\cos4\theta) \notag\right.\\
&\qquad\left.
+ 3\rho^2(\cos2\theta+\cos4\theta)
\right). 
\end{align}

\section{Particle--hole transformation for Yb}
\label{sm:particle_hole}

Here, we outline the particle--hole transformation used to describe the $4f^{13}$ configuration of Yb$^{3+}$ in the one-hole representation. We formulate the transformation for the rare-earth Hamiltonian and then apply the same procedure to the ligand Hamiltonian.
The rare-earth Hamiltonian in Eq.~\eqref{eq:Hf_ion_electron} is written in the electron representation. For Yb$^{3+}$, however, the physical configuration $4f^{13}$ is more simply described as one missing electron from the closed $4f^{14}$ shell. We define the $4f$ hole operator by
\begin{align}
    \hat h_{\alpha}^{\dagger}=\hat f_{\alpha}^{\;},
    \qquad
    \hat h_{\alpha}^{\;}=\hat f_{\alpha}^{\dagger}.
    \label{eq:ph_def}
\end{align}
The hole number operator is written as $\hat{n}_h=\sum_{\alpha}\hat h_{\alpha}^{\dagger}\hat h_{\alpha}^{\;}=\Omega_f - \hat{n}_f$, where $\Omega_f=2(2\ell+1)$ is the total number of orbitals including spin. 
For a general one-body electron operator
\begin{align}
    \mathcal{O}=\sum_{\alpha\beta}A_{\alpha\beta}
    \hat f_{\alpha}^{\dagger}\hat f_{\beta}^{\;},
\end{align}
normal ordering with respect to the filled shell gives
\begin{align}
    \mathcal{O}
    ={\rm Tr}\,\mathsf{A}-
    \sum_{\alpha\beta}A_{\beta\alpha}
    \hat h_{\alpha}^{\dagger}\hat h_{\beta}^{\;}.
    \label{eq:onebody_operator_ph}
\end{align}
Since the trace is a constant, the one-body part of the rare-earth Hamiltonian in the electron representation, $\epsilon_f\hat{n}_f+\mathcal{H}_{\rm SOC}$, transforms into the hole representation as
\begin{align}
    \epsilon_f\hat{n}_f+\mathcal{H}_{\rm SOC}
    = \Omega_f \epsilon_f - \epsilon_f \hat{n}_h
    -\lambda\sum_{m,m'}\sum_{s,s'}
    \mathbf l_{mm'}\cdot\mathbf s_{ss'}
    \hat h_{m s}^{\dagger}\hat h_{m's'}^{\;},
     \label{eq:onebody_ph}
\end{align}
where we have used $\mathrm{Tr}\,(\mathbf l\cdot\mathbf s)=0$.

The two-body Coulomb interaction transforms similarly. 
First, we consider the monopole term $\hat{n}_f(\hat{n}_f-1)F^0/2$. By inserting $\hat{n}_f=\Omega_f-\hat{n}_h$ into the monopole term, we find
\begin{align}
    \frac{1}{2}F^0\hat{n}_f(\hat{n}_f-1)
    =\frac{F^0}{2} \Omega_f (\Omega_f - 1)
    -F^0(\Omega_f - 1) \hat{n}_h
    +\frac{1}{2}F^0\hat{n}_h(\hat{n}_h-1),
\end{align}
where the first term is a constant representing the Coulomb energy of the filled shell, the second term is a spherically symmetric one-hole energy that will be absorbed into $\epsilon_h$, and the third term is the monopole Coulomb energy of the holes.
The non-monopole terms retain the same Slater-integral structure as Eq.~\eqref{eq:HC_general}, because each contribution proportional to $F^{2k}$ can be written in terms of a traceless tensor operator of rank $2k$.
Therefore, the Coulomb interaction in the hole representation is given by
\begin{align}
\mathcal{H}_{\rm C}
=
\frac{F^0}{2} \Omega_f (\Omega_f - 1)
-F^0(\Omega_f - 1) \hat{n}_h
+\frac{1}{2}\sum_{\alpha\beta\gamma\delta}
U_{\alpha\beta\gamma\delta}
\hat h_{\alpha}^{\dagger}
\hat h_{\beta}^{\dagger}
\hat h_{\delta}^{\;}
\hat h_{\gamma}^{\;},
\end{align}
where the matrix element $U_{\alpha\beta\gamma\delta}$ is the same as in the electron representation. 

Together with the one-body part, the rare-earth Hamiltonian in the hole representation is given by
\begin{align}
    \mathcal{H}_{f}
    =&\left(\Omega_f \epsilon_f + \frac{F^0}{2} \Omega_f (\Omega_f - 1)\right)
    +\epsilon_{h} \hat{n}_h
    -\lambda\sum_{m,m'}\sum_{s,s'}
    \mathbf l_{mm'}\cdot\mathbf s_{ss'}
    \hat h_{m s}^{\dagger}\hat h_{m's'}^{\;}
    \nonumber\\
    &+\frac{1}{2}\sum_{\alpha\beta\gamma\delta}
    U_{\alpha\beta\gamma\delta}    \hat h_{\alpha}^{\dagger}
    \hat h_{\beta}^{\dagger}
    \hat h_{\delta}^{\;}
    \hat h_{\gamma}^{\;}.
    \label{eq:Hf_ion_hole}
\end{align}
The first term is the constant energy of the filled shell, whereas $\epsilon_h\hat n_h$ is the spherically symmetric one-hole contribution. As defined in Eq.~\eqref{eq:epsilon_h}, $\epsilon_h$ includes the monopole Coulomb contribution; the energy of the $4f^{13}$ ground multiplet is therefore $\epsilon_h-3\lambda/2$, as shown in Eq.~\eqref{eq:E_f13_2F7/2}.

Applying the same transformation to the ligand gives
\begin{align}
    \mathcal{H}_{p}
    =&\left(\Omega_p \epsilon_p + \frac{U_p}{2} \Omega_p (\Omega_p - 1)\right)
    +\Delta \hat{n}_q
    +\frac{U_p}{2}\hat{n}_q(\hat{n}_q-1),
\end{align}
where $\Omega_p=6$ is the total number of $p$ spin-orbitals, $\hat q_{\beta}^{\dagger}=\hat p_{\beta}^{\;}$ is the hole operator for the ligand orbital $\beta$, $\hat{n}_q$ is the corresponding number operator, and $\Delta$ is the energy cost of creating a single ligand hole defined in Eq.~\eqref{eq:Delta_from_ep}.

Finally, the electron hopping between a rare-earth ion and a ligand in Eq.~\eqref{eq:hopping_pf} transforms as
\begin{align}
    \mathcal{V}
    =&\sum_{i=1,2}\sum_{\lambda=A,B}
    \left[
        t_{i\lambda}^{\alpha\beta}\hat f_{i\alpha}^{\dagger}\hat p_{\lambda\beta}^{\;}
        + \left(t_{i\lambda}^{\alpha\beta}\right)^*\hat p_{\lambda\beta}^{\dagger}\hat f_{i\alpha}^{\;}
    \right]\nonumber\\
    =&\sum_{i=1,2}\sum_{\lambda=A,B}
    \left[
        -\left(t_{i\lambda}^{\alpha\beta}\right)^*\hat h_{i\alpha}^{\dagger}\hat q_{\lambda\beta}^{\;}
        - t_{i\lambda}^{\alpha\beta}\hat q_{\lambda\beta}^{\dagger}\hat h_{i\alpha}^{\;}
    \right].  
    \label{eq:hopping_ph}
\end{align}
Thus, the hopping matrix $\mathsf{t}_{i\lambda}$ in the electron representation becomes $\mathsf{t}_{i\lambda}^{h}=-\mathsf{t}_{i\lambda}^*$ in the hole representation.

\section{Microscopic theory of superexchange}
\label{sm:perturbation}

Here, we outline the fourth-order perturbation theory used to derive the nearest-neighbor effective spin Hamiltonian from its superexchange origin for the bond geometry illustrated in Fig.~\ref{fig:bond-sketch}. 

The microscopic Hamiltonian is written as
\begin{equation}
    \mathcal{H} = \mathcal{H}_0 + \mathcal{V},
    \label{eq:Hfull}
\end{equation}
with $\mathcal{V}$ representing the hybridization between the rare-earth $4f$ and ligand $p$ orbitals, and the unperturbed part $\mathcal{H}_0$ defined as the sum of the atomic Hamiltonians:
\begin{align}
    \mathcal{H}_0 = \sum_{i=1,2} \mathcal H_{f,i} + \sum_{\lambda=A,B} \mathcal H_{p, \lambda},
    \label{eq:H0}
\end{align}
where the first term, $\mathcal H_{f,i} = U_f^{-(+)} |i 0 \rangle \langle i 0 | + \mathcal{H}_{{\rm SI}, f^2} $, accounts for the energy of the empty $f$ orbitals for Ce (Yb). The energy reference corresponds to the $f^1$ state $|\sigma\rangle$, where $\sigma$ runs over the 14 $f$ states. Energy splittings due to spin-orbit coupling (SOC) and the crystal electric field (CEF) are explicitly considered for these $f^1$ states; note that the system does not occupy any excited $f^1$ states outside of the ground state doublet. The second term, $\mathcal{H}_{{\rm SI}, f^2}$, accounts for the Coulomb repulsion and SOC between the $f$ electrons (or holes) when the $f$ orbital is doubly occupied, while the CEF is neglected in this configuration. This Hamiltonian can be written as
\begin{equation}
    \mathcal{H}_{{\rm SI}, f^2} = \sum_a (U_f^{-(+)} + E_a) |\eta_a\rangle \langle \eta_a | , 
\end{equation}
where the eigenenergies are $U_f^{+(-)} + E_a$, with $U_{f}^{+(-)}$ being the lowest energy of the two-electron (two-hole) states relative to the single-electron (single-hole) ground state. For two-electron ($\text{Ce}^{3+}$) or two-hole ($\text{Yb}^{3+}$) states, there are 91 eigenstates $|\eta_a\rangle$ that can be grouped into 13 $J$ multiplets, with values of $J$ ranging from 0 to 6 (with more than one multiplet appearing for certain values of $J$). Since the $2J + 1$ eigen-energies for each multiplet must be degenerate due to rotational invariance, $E_a$ takes only 13 distinct values.

Regarding the ligand orbitals, the first term of $\mathcal H_{p,\lambda} = \sum_{s} \Delta |\lambda s \rangle \langle\lambda s | + \mathcal{H}_{{\rm SI}, p^2}$ accounts for the energy of one hole in the $p$ orbital. The second term, $\mathcal{H}_{{\rm SI}, p^2}$, partially accounts for the Coulomb repulsion while ignoring any energy differences among the $p^2$ states. We also neglect the ligand SOC. This Hamiltonian is expressed as
\begin{equation}
    \mathcal{H}_{{\rm SI}, p^2} = \sum_\mu U_p  | p_{\lambda \mu}^2 \rangle \langle  p_{\lambda \mu}^2 |,
\end{equation}
where $U_p$ is the energy of the $| p_{\lambda \mu}^2 \rangle$ states. There are 15 possible states with two electrons (or holes) in the $p$ orbitals.

Finally, the hybridization term is given by
\begin{equation}
    \mathcal{V} = \sum_{i \lambda} \sum_{\alpha \beta} [ t_{i\lambda}^{\alpha\beta} f_{i\alpha}^{\dag} p_{\lambda\beta}^{} + (t_{i\lambda}^{\alpha\beta})^* p_{\lambda\beta}^{\dag} f_{i\alpha}^{}   ]
\end{equation}
where the hopping integrals between the $f$ and $p$ orbitals are parameterized by the $14\times6$ matrix $t_{i\lambda}^{}$.

The effective spin Hamiltonian is obtained at fourth order in degenerate perturbation theory as~\cite{LindgrenI1974}
\begin{equation}\label{eq:Heff}
    \mathcal{H}_{\rm eff} = P \mathcal{H}_0 P + P \mathcal{V} G(\varepsilon_0) \mathcal{V} G(\varepsilon_0) \mathcal{V} G(\varepsilon_0) \mathcal{V} P 
\end{equation}
where the operator $P$ projects onto the ground state manifold, and $\varepsilon_0$ is the unperturbed ground-state energy. The resolvent operator is given by
\begin{equation}
    G(\varepsilon) = P^\perp \frac{1}{\varepsilon - \mathcal{H}_0} P^\perp,
\end{equation}
where the complementary projector is defined as $ P^\perp = 1 -  P$.

\begin{figure}[h]
    \centering
    \includegraphics[width=0.95\linewidth]{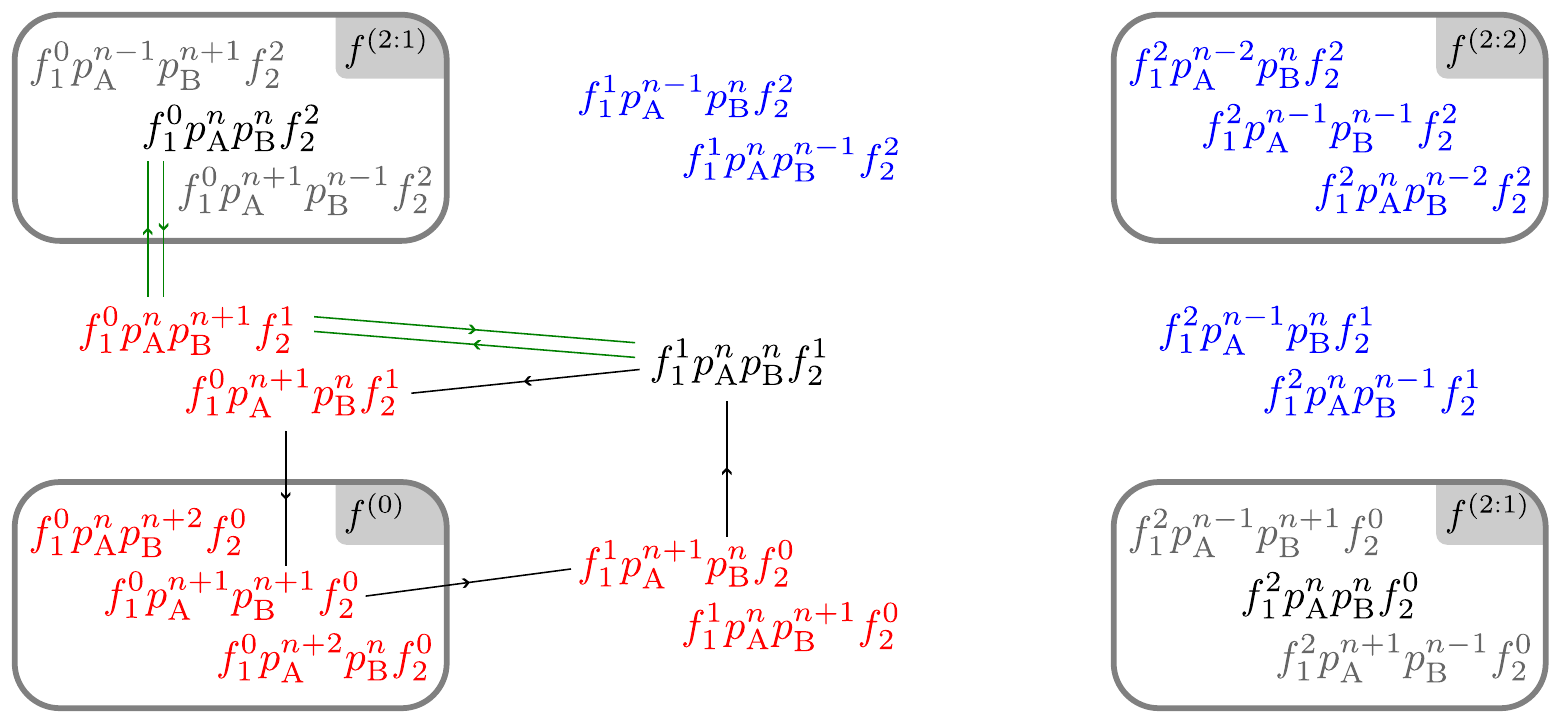}
    \caption{Complete configuration space of the intermediate virtual states involved in the fourth-order perturbation theory. The ground-state configuration is located at the center of the diagram. Single $fp$-hopping events transition the system to adjacent states within the same row or column. Fourth-order virtual processes comprise all four-step closed paths returning to the ground state. For the unrestricted general case where ligand occupancies of $n\pm2$ are allowed, the system possesses 24 virtual processes in the $f^0$ channel, 48 in the $f^{2:1}$ channel, and 24 in the $f^{2:2}$ channel. For Yb-based systems, the hole occupancy ($n\ge0$) restricts the accessible configurations to those highlighted in black and red. Conversely, for Ce-based systems, the physical limit $n \le 6$ restricts the active configurations to those highlighted in black and blue.}
    \label{fig:gral-diags}
\end{figure}

The superexchange interaction between the localized $4f$ ions is obtained via this fourth-order expansion. Figure~\ref{fig:gral-diags} maps out the complete space of intermediate virtual state configurations involved in the perturbation theory for a system with $n$ electrons (or holes) in the ligand orbitals. While the initial rare-earth configuration is defined here as having one electron (or hole), the structure of the virtual state space remains identical when considering an initial configuration with $m$ electrons in the $f$ shells. 

All fourth-order processes are traced in Fig.~\ref{fig:gral-diags} by starting from the initial ground-state configuration located at the center. Within this diagram, transitions are restricted to horizontal or vertical steps representing individual $fp$-hopping events. Consequently, the first and third virtual states correspond to configurations reached via a single vertical or horizontal displacement, whereas the second intermediate state is located at the corners of the paths. We classify these virtual processes into the $f^{0}$, $f^{2:1}$, and $f^{2:2}$ channels, as indicated in the respective corner boxes. The $f^{0}$ label signifies that both $f$ orbitals are empty in the intermediate state, $f^{2:1}$ indicates that one $f$ orbital is empty while the other is doubly occupied, and $f^{2:2}$ means both $f$ shells are doubly occupied. This classification simplifies the subsequent analysis of the symmetry properties and weights of these processes in the final magnetic exchange interaction.

To evaluate the exchange coefficients, we must sum over all unique four-step closed loops that start and end at the central configuration ($f_{1}^1 p_{A}^n p_{B}^n f_{2}^1$). In the unrestricted general case (where ligand occupancies ranging from $n-2$ to $n+2$ are energetically accessible) there are 24, 48, and 24 virtual processes for the $f^{0}$, $f^{2:1}$, and $f^{2:2}$ channels, respectively, as summarized in Table~\ref{tab:f-states}.

For the specific case of Yb-based insulators, the system is naturally framed within the hole picture, meaning there is one hole in the initial $f$ orbitals and a completely filled ligand shell ($n=0$ holes). This restriction ($n=0$) strictly forbids intermediate states requiring a reduction in ligand hole occupancy ($n-1$ or $n-2$), completely eliminating the $f^{2:2}$ channel and restricting the $f^{2:1}$ channel. Meanwhile, all 24 virtual paths in the $f^{0}$ channel remain active (see the configurations highlighted in red in Fig.~\ref{fig:gral-diags}). In the $f^{2:1}$ channel, configurations requiring $n-1$ intermediate holes are forbidden, leaving only 8 out of the original 48 virtual processes. Figure~\ref{fig:Yb-diags}\textbf{a} depicts the configuration space involved in these allowed $f^{0}$ channels, panels \textbf{b} to \textbf{g} show the six fundamental types of virtual processes extracted from these four-step loops, and Fig.~\ref{fig:Yb-diags}\textbf{h} shows the allowed states within the $f^{2:1}$ channel. The remaining symmetry-equivalent processes are generated by swapping the rare-earth ions and the ligand sites.

Conversely, for Ce-based insulators where the ligand shell is fully occupied by electrons ($n=6$), Fig.~\ref{fig:Ce-diags}\textbf{a} displays the allowed configurations within the $f^{2:2}$ channel. This channel hosts six distinct types of virtual processes, which are isolated in panels \textbf{b} to \textbf{g}. Figure~\ref{fig:Ce-diags}\textbf{h} highlights the allowed configurations within the corresponding $f^{2:1}$ channel, with the resulting virtual processes detailed in panels \textbf{i} and \textbf{j}. The remaining symmetry-equivalent configurations are similarly obtained by exchanging the rare-earth and ligand indices.

\begin{table}[h!]
\centering
\begin{tabular}{|l|ccc|}
\hline
 & \textbf{Gral} & \textbf{Yb$^{3+}$} & \textbf{Ce$^{3+}$} \\ \hline
$f^0$     & 24 & 24 & 0  \\
$f^{2:1}$ & 48 & 8  & 8  \\
$f^{2:2}$ & 24 & 0  & 24 \\ \hline
\end{tabular}
\caption{Allowed virtual processes per channel for the general case ($n\pm2$ allowed), Yb$^{3+}$ (restricted to $n+1/n+2$), and Ce$^{3+}$ (restricted to $n-1/n-2$) configurations.}
\label{tab:f-states}
\end{table}

\subsection{Yb-based insulator}
In this section we implement the fourth-order perturbation theory for the Yb$^{3+}$ case. For this, we use the \emph{hole} picture with one hole in the $f$-orbitals, and no holes in the $p$-orbitals, $n=0$. As we discussed before, there are 24 virtual processes in the $f^(0)$ channel, and only 8 for the $f^{(2:1)}$.

\begin{figure}[h!]
    \centering
    \includegraphics[width=0.95\linewidth]{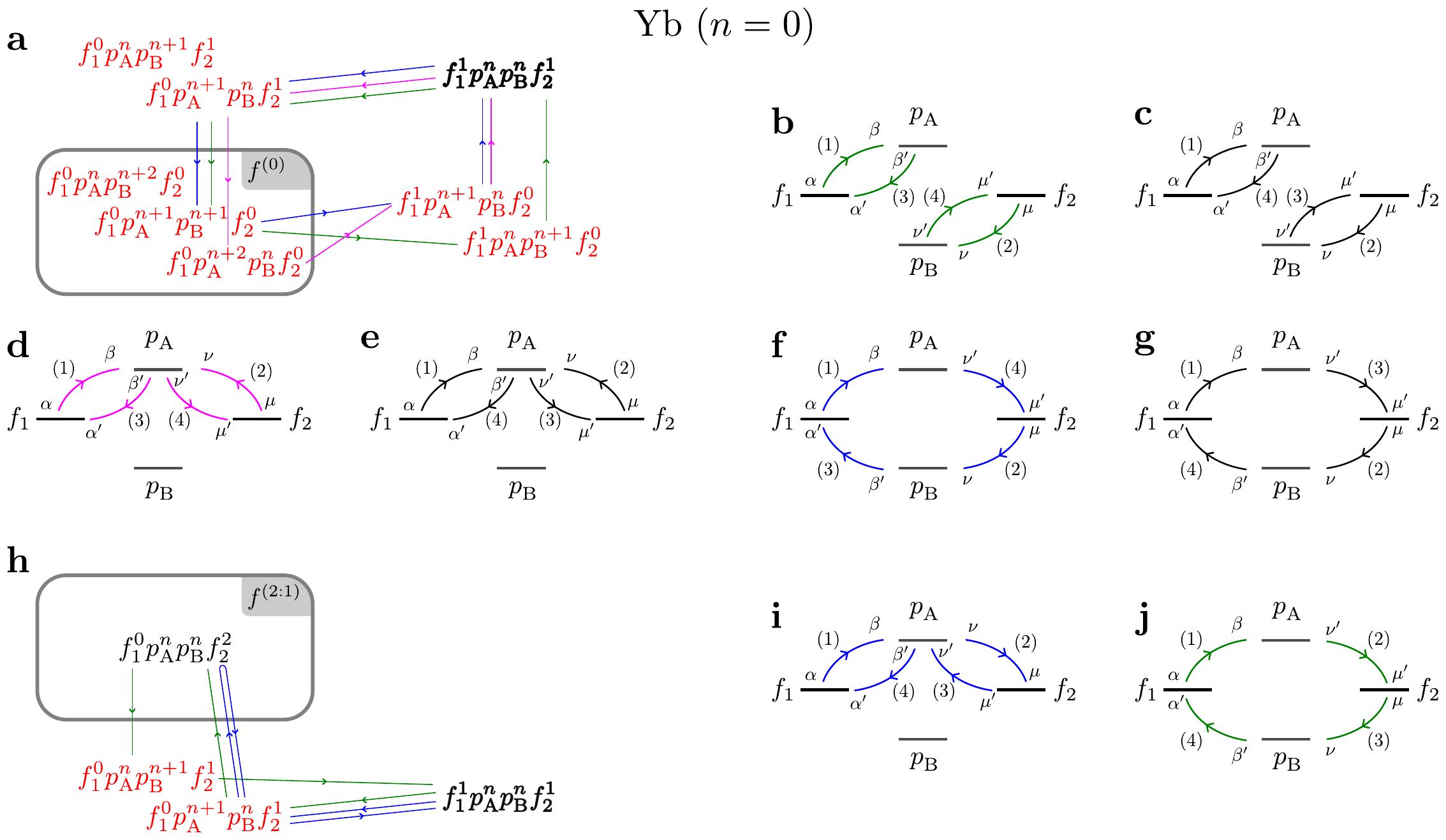}
    \caption{Allowed virtual state configurations involved in the perturbation theory for Yb-based insulators, mapped in the hole picture with empty ligand orbitals ($n=0$). These configurations constitute a subset of the general cases shown in Fig.~\ref{fig:gral-diags}. Panel \textbf{a} depicts the virtual states involved in the $f^{(0)}$ channel, while panels \textbf{b}-\textbf{g} show the six distinct diagram types extracted from it. The virtual states within the $f^{(2:1)}$ channel are presented in panel \textbf{h}, with the corresponding virtual processes detailed in panels \textbf{i} and \textbf{j}. Exchanging the sites of the ligands and/or the rare-earth ions yields the complete diagrammatic series.}
    \label{fig:Yb-diags}
\end{figure}

In order to compute the term $P \mathcal{V} G(\varepsilon_0) \mathcal{V} G(\varepsilon_0) \mathcal{V} G(\varepsilon_0) \mathcal{V} P$ for each virtual process included in Fig.~\ref{fig:Yb-diags}, we list in Table~\ref{tab:excited_states-Yb} the energy difference $(\varepsilon_0 - \mathcal{H}_0)$ between the ground state configuration ($f_{1}^1 p_{A}^0  p_{B}^0 f_{2}^1$) and the excited states. Note that the configurations that are not listed in the table can be obtained by swapping the rare-earth and/or ligand sites, which leaves the energy invariant. 
\begin{table}[h!]
\centering
\caption{Excited State Configurations and Energy Differences.}
\label{tab:excited_states-Yb}
\vskip 2mm
\renewcommand{\arraystretch}{1.4} 
\begin{tabular}{ll}
\hline
\textbf{Excited State } \qquad & $(\varepsilon_0^{} - \mathcal{H}_0 )$ \\ 
\hline
$f_{1}^0 p_{A}^1 p_{B}^0 f_{2}^1$ & $-(\Delta + U_{f}^+ )$ \\
$f_{1}^0 p_{A}^2 p_{B}^0 f_{2}^0$ & $-[ 2(\Delta + U_{f}^+ )+ U_p  ]$ \\
$f_{1}^0 p_{A}^1 p_{B}^1 f_{2}^0$ & $-2(\Delta + U_{f}^+) $ \\
$f_{1}^0 p_{A}^0 p_{B}^0 f_{2}^2$ & $-(U_{f}^+ + U_{f}^- + E_{a_2})$ \\ 
\hline
\end{tabular}
\end{table}

\subsubsection{$f^{(0)}$ channel}

We begin with the contribution from the virtual process in Fig.~\ref{fig:Yb-diags}\textbf{b} (see green arrows loop in \textbf{a}) to the low-energy Hamiltonian using Eq.~\eqref{eq:Heff}. We denote this contribution as $\mathcal{H}_{b}^{f^{(0)}}$, which reads as
\begin{align}
    &\mathcal{H}_{b}^{f^{(0)}} %
    = -  \sum_{\substack{\alpha \beta \mu \nu \\ \alpha' \beta' \mu' \nu'}}  \frac{ t_{1A}^{\alpha \beta}  t_{2B}^{\mu \nu}  (t_{1A}^{\alpha' \beta'})^*  (t_{2B}^{\mu' \nu'})^*  }{ 2(\Delta + U_f^+)^3  } P f_{2\mu'}^\dag p_{B\nu'}^{} \ f_{1\alpha'}^\dag p_{A\beta'}^{} \ p_{B\nu}^\dag f_{2\mu}^{} \ p_{A\beta}^\dag f_{1\alpha}^{} P  \nonumber \\
    & =  \sum_{\substack{\alpha \beta \mu \nu \\ \alpha' \beta' \mu' \nu'}}  \frac{ t_{1A}^{\alpha \beta}  t_{2B}^{\mu \nu}  (t_{1A}^{\alpha' \beta'})^*  (t_{2B}^{\mu' \nu'})^*  }{ 2(\Delta + U_f^+)^3  }  (P_p p_{B\nu'}^{}  p_{A\beta'}^{}  p_{B\nu}^\dag  p_{A\beta}^\dag P_p )  (P_f f_{1\alpha'}^\dag f_{1\alpha}^{} P_f)  (P_f f_{2\mu'}^\dag f_{2\mu}^{} P_f)  \nonumber \\
    & = - \sum_{\substack{\sigma_1 \sigma_1' \\ \sigma_2 \sigma_2'}} \sum_{\alpha \alpha'  \mu \mu'} \frac{ [\mathring{T}_{1A}^{}]^{\alpha \alpha'}  [\mathring{T}_{2B}^{}]^{\mu \mu'}  }{ 2(\Delta + U_f^+)^3  }  \delta_{\alpha' \sigma_1} \delta_{\alpha \sigma_1'}   \  \delta_{\mu' \sigma_2} \delta_{\mu \sigma_2'}    |\sigma_1 \rangle \langle \sigma_1|  |\sigma_2' \rangle \langle \sigma_2'| \nonumber \\
    &= %
    - \sum_{\substack{\sigma_1 \sigma_1' \\ \sigma_2 \sigma_2'}} \frac{ [\mathring{T}_{1A}^{}]^{\sigma_1' \sigma_1} [\mathring{T}_{2B}^{}]^{\sigma_2' \sigma_2}  }{ 2(\Delta + U_f^+)^3  } |\sigma_1 \rangle \langle \sigma_1'|  |\sigma_2 \rangle \langle \sigma_2'| \ .
\end{align}
Here we have decomposed the GS projector into a projector for the $p$- and $f$-orbitals, $P=P_c P_f$. For the ligands it yields $ P_p p_{B\nu'}^{}  p_{A\beta'}^{}  p_{B\nu}^\dag  p_{A\beta}^\dag P_p = -\delta_{\nu\nu'} \delta_{\beta\beta'} $, and for the rare-earth sites
\begin{equation}
    P_f f_{i\alpha}^\dag f_{i\alpha'}^{} P_f = \sum_{\sigma \sigma'} |\sigma \rangle \langle \sigma | f_{i\alpha'}^{} f_{i\alpha}^\dag |\sigma' \rangle \langle \sigma'| = \sum_{\sigma \sigma'} \delta_{\alpha \sigma} \delta_{\alpha' \sigma'}  |\sigma \rangle \langle \sigma'| \ .
\end{equation}
We have defined the \textbf{effective local hopping matrix} $ \mathring{T}_{i\lambda}^{} = t_{i\lambda}^{}  t_{i\lambda}^{\dag} $, where the overhead circle means that the particle return to the original site after the two $fp$-hoppings.

For the virtual process depicted in Fig.~\ref{fig:Yb-diags}\textbf{c} (see magenta arrows loop in \textbf{a}), the effective Hamiltonian reads as
\begin{align}
    \mathcal{H}_{c}^{f^{(0)}} &= -  \sum_{\substack{\alpha \beta \mu \nu \\ \alpha' \beta' \mu' \nu'}}  \frac{ t_{1A}^{\alpha \beta}  t_{2B}^{\mu \nu}  (t_{2B}^{\mu' \nu'})^* (t_{1A}^{\alpha' \beta'})^*  }{ 2(\Delta + U_f^+)^3  } P f_{1\alpha'}^\dag p_{A\beta'}^{} \   f_{2\mu'}^\dag p_{B\nu'}^{} \ p_{B\nu}^\dag f_{2\mu}^{} \ p_{A\beta}^\dag f_{1\alpha}^{} P  \nonumber \\
    & = %
    - \sum_{\sigma_1 \sigma_1'  \sigma_2 \sigma_2'} \frac{ [\mathring{T}_{1A}^{}]^{\sigma_1' \sigma_1} [\mathring{T}_{2B}^{}]^{\sigma_2' \sigma_2}  }{ 2(\Delta + U_f^+)^3  } |\sigma_1 \rangle \langle \sigma_1'| \ |\sigma_2 \rangle \langle \sigma_2'|
\end{align}
where $P_p  p_{A\beta'}^{} p_{B\nu'}^{}  p_{B\nu}^\dag  p_{A\beta}^\dag P_p = -\delta_{\nu \beta} \delta_{\nu' \beta'}$.

For the virtual process shown in Fig.~\ref{fig:Yb-diags}\textbf{d} the effective Hamiltonian reads as
\begin{align}
    &\mathcal{H}_{d}^{f^{(0)}} %
    = -  \sum_{\substack{\alpha \beta \mu \nu \\ \alpha' \beta' \mu' \nu'}}  \frac{ t_{1A}^{\alpha \beta}  t_{2A}^{\mu \nu} (t_{1A}^{\alpha' \beta'})^*  (t_{2A}^{\mu' \nu'})^*  }{ (\Delta + U_f^+)^2 [2(\Delta+U_f^+) + U_p]  } P  f_{2\mu'}^\dag p_{A\nu'}^{} \ f_{1\alpha'}^\dag p_{A\beta'}^{} \   p_{A\nu}^\dag f_{2\mu}^{} \ p_{A\beta}^\dag f_{1\alpha}^{} P  \nonumber \\
    & =   \sum_{\substack{\alpha \beta \mu \nu \\ \alpha' \beta' \mu' \nu'}}  \frac{  t_{1A}^{\alpha \beta}  t_{2A}^{\mu \nu} (t_{1A}^{\alpha' \beta'})^*  (t_{2A}^{\mu' \nu'})^*   }{ (\Delta + U_f^+)^2 [2(\Delta+U_f^+) + U_p]  }  (P_p  p_{A\nu'}^{}  p_{A\beta'}^{} p_{A\nu}^\dag  p_{A\beta}^\dag P_p )  (P_f f_{1\alpha'}^\dag f_{1\alpha}^{} P_f)  (P_f f_{2\mu'}^\dag f_{2\mu}^{} P_f)  \nonumber \\
    & = \sum_{\sigma_1 \sigma_1'  \sigma_2 \sigma_2'} \sum_{\alpha \alpha'  \mu \mu'} \frac{ [{T}_{A}^{}]^{\alpha \mu'}  [{T}_{A}^{\dag}]^{\mu \alpha'}  - [\mathring{T}_{1A}^{}]^{\alpha \alpha'} [\mathring{T}_{2A}^{}]^{\mu \mu'}   }{ (\Delta + U_f^+)^2 [2(\Delta+U_f^+) + U_p]   }  (  \delta_{\alpha' \sigma_1} \delta_{\alpha \sigma_1'} )   (  \delta_{\mu' \sigma_2} \delta_{\mu \sigma_2'} )   |\sigma_1 \rangle \langle \sigma_1|  |\sigma_2' \rangle \langle \sigma_2'| \nonumber \\
   &= \sum_{\sigma_1 \sigma_1'  \sigma_2 \sigma_2'} \frac{ [{T}_{A}^{}]^{\sigma_1' \sigma_2}  [{T}_{A}^{\dag}]^{\sigma_2' \sigma_1} - [\mathring{T}_{1A}^{}]^{\sigma_1' \sigma_1} [\mathring{T}_{2A}^{}]^{\sigma_2' \sigma_2}  }{ (\Delta + U_f^+)^2 [2(\Delta+U_f^+) + U_p] } |\sigma_1 \rangle \langle \sigma_1'|  |\sigma_2 \rangle \langle \sigma_2'| 
\end{align}
where we used $P_p  p_{A\nu'}^{}  p_{A\beta'}^{} p_{A\nu}^\dag  p_{A\beta}^\dag P_p = \delta_{\nu \beta'} \delta_{\beta \nu'} - \delta_{\beta \beta'} \delta_{\nu \nu'} $. The \textbf{single-ligand effective hopping matrix} between the two $f$-orbitals through the ligand $\lambda$ are defined as $ T_\lambda = t_{1\lambda}^{} t_{2\lambda}^{\dag}$.

For the virtual process in Fig.~\ref{fig:Yb-diags}\textbf{e} the effective Hamiltonian reads as
\begin{align}
    \mathcal{H}_{e}^{f^{(0)}} &= -  \sum_{\substack{\alpha \beta \mu \nu \\ \alpha' \beta' \mu' \nu'}}  \frac{ t_{1A}^{\alpha \beta}  t_{2A}^{\mu \nu}  (t_{2A}^{\mu' \nu'})^*  (t_{1A}^{\alpha' \beta'})^*   }{ (\Delta + U_f^+)^2 [2(\Delta+U_f^+) + U_p]  } P  f_{1\alpha'}^\dag p_{A\beta'}^{} \ f_{2\mu'}^\dag p_{A\nu'}^{} \   p_{A\nu}^\dag f_{2\mu}^{} \ p_{A\beta}^\dag f_{1\alpha}^{} P  \nonumber \\
   &= \sum_{\sigma_1 \sigma_1'  \sigma_2 \sigma_2'} \frac{ [{T}_{A}^{}]^{\sigma_1' \sigma_2}  [{T}_{A}^{\dag}]^{\sigma_2' \sigma_1} - [\mathring{T}_{1A}^{}]^{\sigma_1' \sigma_1} [\mathring{T}_{2A}^{}]^{\sigma_2' \sigma_2}  }{ (\Delta + U_f^+)^2 [2(\Delta+U_f^+) + U_p] } |\sigma_1 \rangle \langle \sigma_1'| \ |\sigma_2 \rangle \langle \sigma_2'| 
\end{align}
where the projection in the ligand orbitals is $P_p  p_{A\beta'}^{} p_{A\nu'}^{}  p_{A\nu}^\dag  p_{A\beta}^\dag P_p = \delta_{\nu \nu'} \delta_{\beta \beta'} -  \delta_{\nu \beta'} \delta_{\beta \nu'} $.

For the virtual process in Fig.~\ref{fig:Yb-diags}\textbf{f} (see blue arrows loop in \textbf{a}), the effective Hamiltonian reads as
\begin{align}
    &\mathcal{H}_{f}^{f^{(0)}} %
    = -  \sum_{\substack{\alpha \beta \mu \nu \\ \alpha' \beta' \mu' \nu'}}  \frac{ t_{1A}^{\alpha \beta}  t_{2B}^{\mu \nu}  (t_{1B}^{\alpha' \beta'})^*  (t_{2A}^{\mu' \nu'})^*   }{ 2(\Delta + U_f^+)^3  } P  f_{2\mu'}^\dag p_{A\nu'}^{} \ f_{1\alpha'}^\dag p_{B\beta'}^{} \   p_{B\nu}^\dag f_{2\mu}^{} \ p_{A\beta}^\dag f_{1\alpha}^{} P  \nonumber \\
    & = \sum_{\substack{\alpha \beta \mu \nu \\ \alpha' \beta' \mu' \nu'}}   \frac{ t_{1A}^{\alpha \beta}  t_{2B}^{\mu \nu}  (t_{1B}^{\alpha' \beta'})^*  (t_{2A}^{\mu' \nu'})^*   }{ 2(\Delta + U_f^+)^3  }  (P_p  p_{A\nu'}^{} p_{B\beta'}^{}  p_{B\nu}^\dag  p_{A\beta}^\dag P_p )  (P_f f_{1\alpha'}^\dag f_{1\alpha}^{} P_f)  (P_f f_{2\mu'}^\dag f_{2\mu}^{} P_f)  \nonumber \\
   &= \sum_{\sigma_1 \sigma_1'  \sigma_2 \sigma_2'} \frac{ [{T}_{A}^{}]^{\sigma_1' \sigma_2}  [{T}_{B}^{}]^{\sigma_2' \sigma_1 } }{ 2(\Delta + U_f^+)^3 } |\sigma_1 \rangle \langle \sigma_1'|  |\sigma_2 \rangle \langle \sigma_2'| 
\end{align}

The last $f^{(0)}$ virtual process is depicted in Fig.~\ref{fig:Yb-diags}\textbf{g}. The effective Hamiltonian from the process reads as
\begin{align}
    \mathcal{H}_{g}^{f^{(0)}} &= -  \sum_{\substack{\alpha \beta \mu \nu \\ \alpha' \beta' \mu' \nu'}}  \frac{ t_{1A}^{\alpha \beta}  t_{2B}^{\mu \nu}   (t_{2A}^{\mu' \nu'})^*  (t_{1B}^{\alpha' \beta'})^* }{ 2(\Delta + U_f^+)^3  } P f_{1\alpha'}^\dag p_{B\beta'}^{} \  f_{2\mu'}^\dag p_{A\nu'}^{}  \   p_{B\nu}^\dag f_{2\mu}^{} \ p_{A\beta}^\dag f_{1\alpha}^{} P  \nonumber \\
   &= \sum_{\sigma_1 \sigma_1'  \sigma_2 \sigma_2'} \frac{ [{T}_{A}^{}]^{\sigma_1' \sigma_2}  [{T}_{B}^{\dag}]^{\sigma_2' \sigma_1 } }{ 2(\Delta + U_f^+)^3 } |\sigma_1 \rangle \langle \sigma_1'|  |\sigma_2 \rangle \langle \sigma_2'| 
\end{align}
\subsubsection{$f^{(2:1)}$ channel}

We now turn to the virtual processes that contribute to the $f^{(2:1)}$ channel depicted in Fig.~\ref{fig:Yb-diags}\textbf{h}. 
In these processes we introduce the function $\gamma_{\sigma \sigma', \alpha \alpha'}^a$ \cite{Ghioldi2024} that accounts for the projection of $f^2$ states that are a direct product of two single-particle ($f^1$) states into the eigenstates $|\eta_a \rangle$ of the single-ion Hamiltonian,
\begin{equation}
    \gamma_{\sigma \sigma', \alpha \alpha'}^a \equiv \langle 0| f_{\sigma}^{} f_{\alpha'}^{} |\eta_a \rangle \langle \eta_a | f_{\alpha}^{\dag} f_{\sigma'}^{\dag}  | 0 \rangle = \langle \sigma | f_{\alpha'}^{} |\eta_a \rangle \langle \eta_a | f_{\alpha}^{\dag} | \sigma' \rangle
\end{equation}
where $\sigma$ and $\sigma'$ are, respectively, the final and initial $f^1$ states, while $\alpha$ ($\alpha'$) is the state of the second particle created (annihilated).
The projection of the $f$-orbitals after double occupy the rare-earth sites is
\begin{align}
    P_f f_{\alpha'}^{} | \eta_a \rangle \langle \eta_a | f_{\alpha}^\dag P_f & = \sum_{\sigma \sigma'} |\sigma \rangle \langle \sigma |  f_{\alpha'}^{} | \eta_a \rangle \langle \eta_a | f_{\alpha}^\dag |\sigma' \rangle \langle \sigma' | =  \sum_{\sigma \sigma'}  \gamma_{\sigma \sigma', \alpha \alpha'}^a  |\sigma \rangle \langle \sigma' | \ .%
\end{align}

For the virtual process shown in Fig.~\ref{fig:Yb-diags}\textbf{i} (see blue arrows loop in \textbf{h}) the effective Hamiltonian reads as
\begin{align}
    &\mathcal{H}_{i}^{f^{(2:1)}} %
    = - \sum_{\substack{\alpha \beta \mu \nu \\ \alpha' \beta' \mu' \nu'}} \sum_a \frac{ t_{1A}^{\alpha \beta}  (t_{2A}^{\mu \nu})^*  t_{2A}^{\mu' \nu'}  (t_{1A}^{\alpha' \beta'})^*  }{ (\Delta + U_f^+)^2 (U_f^+ + U_f^- + E_a)  } \nonumber \\
    & \qquad \times P f_{1\alpha'}^\dag p_{A\beta'}^{} p_{A\nu'}^\dag f_{2\mu'}^{} |\eta_{2a}\rangle \langle \eta_{2a}| f_{2\mu}^\dag p_{A\nu}^{} p_{A\beta}^\dag f_{1\alpha}^{} P  \nonumber \\
    & = - \sum_{\substack{\alpha \beta \mu \nu \\ \alpha' \beta' \mu' \nu'}}   \sum_a \frac{ t_{1A}^{\alpha \beta}  (t_{2A}^{\mu \nu})^*  t_{2A}^{\mu' \nu'}  (t_{1A}^{\alpha' \beta'})^*  }{ (\Delta + U_f^+)^2 (U_f^+ + U_f^- + E_a)  } \nonumber \\
    &  \qquad \times  (P_p p_{A\beta'}^{} p_{A\nu'}^\dag  p_{A\nu}^{} p_{A\beta}^\dag P_p )  (P_f f_{1\alpha'}^\dag f_{1\alpha}^{} P_f)  (P_f f_{2\mu'}^{} |\eta_{2a}\rangle \langle \eta_{2a}|f_{2\mu}^{\dag} P_f)  \nonumber \\
    & = - \sum_{\sigma_1^{} \sigma_1' \sigma_2^{} \sigma_2' } \sum_{\alpha  \mu   \alpha'  \mu'} \sum_a \frac{ [T_A^{}]^{\alpha \mu}  [T_A^{\dag}]^{\mu'\alpha'}  }{ (\Delta + U_f^+)^2 (U_f^+ + U_f^- + E_a)  }  \delta_{\alpha' \sigma_1} \delta_{\alpha \sigma_1'} |\sigma_1\rangle \langle \sigma_1'|  \gamma_{\sigma_2 \sigma_2', \mu \mu'}^a  |\sigma_2 \rangle \langle \sigma_2' | \nonumber \\
    & = - \sum_{\sigma_1^{} \sigma_1' \sigma_2^{} \sigma_2' } \sum_{\mu  \mu' a}  \frac{ [T_A^{}]^{\sigma_1' \mu}  [T_A^\dag]^{\mu' \sigma_1} \gamma_{\sigma_2^{} \sigma_2',\mu\mu'}^a }{ (\Delta + U_f^+)^2 (U_f^+ + U_f^- + E_a)  }  |\sigma_1^{}\rangle \langle\sigma_1'|  \ |\sigma_2^{}\rangle \langle\sigma_2'|
\end{align}
where $ P_p p_{A\beta'}^{} p_{A\nu'}^\dag  p_{A\nu}^{} p_{A\beta}^\dag P_p = \delta_{\nu\beta} \delta_{\nu'\beta'} $.

For the virtual process in Fig.~\ref{fig:Yb-diags}\textbf{j} (see green arrows loop in \textbf{h}) the effective Hamiltonian reads as
\begin{align}
    \mathcal{H}_{j}^{f^{(2:1)}} &= - \sum_{\substack{\alpha \beta \mu \nu \\ \alpha' \beta' \mu' \nu'}} \sum_a \frac{ t_{1A}^{\alpha \beta}  (t_{2A}^{\mu' \nu'})^*   t_{2B}^{\mu \nu} (t_{1B}^{\alpha' \beta'})^*   }{ (\Delta + U_f^+)^2 (U_f^+ + U_f^- + E_a)  }  \nonumber \\
    & \qquad \times P f_{1\alpha'}^\dag p_{B\beta'}^{} p_{B\nu}^\dag f_{2\mu}^{}  |\eta_{2a}\rangle \langle \eta_{2a}| f_{2\mu'}^\dag p_{A\nu'}^{} p_{A\beta}^\dag f_{1\alpha}^{} P  \nonumber \\
    & = - \sum_{\sigma_1^{} \sigma_1' \sigma_2^{} \sigma_2' } \sum_{\mu  \mu' a}  \frac{ [T_A^{}]^{\sigma_1' \mu'}  [T_B^\dag]^{\mu \sigma_1} \gamma_{\sigma_2^{} \sigma_2',\mu\mu'}^a }{ (\Delta + U_f^+)^2 (U_f^+ + U_f^- + E_a)  }  |\sigma_1^{}\rangle \langle\sigma_1'|  \ |\sigma_2^{}\rangle \langle\sigma_2'|
\end{align}
with $ P_p p_{B\beta'}^{} p_{B\nu}^\dag  p_{A\nu'}^{} p_{A\beta}^\dag P_p  = \delta_{\nu'\beta} \delta_{\nu\beta'} $.

As we mentioned before, by swapping A and B ligands and/or swapping the rare-earth sites, we obtain the remaining virtual processes.
In order to make the notation simpler, we write the effective Hamiltonian as
\begin{equation}
    \mathcal{H} = \sum_{\sigma_1^{} \sigma_1' \sigma_2^{} \sigma_2' } \mathcal{K}_{\sigma_1^{} \sigma_1' \sigma_2^{} \sigma_2'}^{} \  |\sigma_1^{}\rangle \langle\sigma_1'|  \ |\sigma_2^{}\rangle \langle\sigma_2'|
\end{equation}
with $\mathcal{K}_{\sigma_1^{} \sigma_1' \sigma_2^{} \sigma_2'}^{}$ being the rank-4 exchange tensor.

The exchange tensor after including the other virtual processes for the $f^{(0)}$ channel obtained from Figs.~\ref{fig:Yb-diags}\textbf{b} and \textbf{c} combined read as
\begin{align}
    \mathcal{K}_{\sigma_1^{} \sigma_1' \sigma_2^{} \sigma_2'}^{f^{(0), b+c-tot}} = - \sum_{\lambda} 2 \frac{ [\mathring{T}_{1\lambda}^{}]^{\sigma_1' \sigma_1} [\mathring{T}_{2\bar \lambda}^{}]^{\sigma_2' \sigma_2}  }{ (\Delta + U_f^+)^3  } 
\end{align}
All the virtual processes related to Figs.~\ref{fig:Yb-diags}\textbf{d} and \textbf{e} can be combined into
\begin{align}
    \mathcal{K}_{\sigma_1^{} \sigma_1' \sigma_2^{} \sigma_2'}^{f^{(0), d+e-tot}} =  \sum_\lambda 4 \frac{ [{T}_{\lambda}^{}]^{\sigma_1' \sigma_2}  [{T}_{\lambda}^{\dag}]^{\sigma_2' \sigma_1} - [\mathring{T}_{1\lambda}^{}]^{\sigma_1' \sigma_1} [\mathring{T}_{2\lambda}^{}]^{\sigma_2' \sigma_2}  }{ (\Delta + U_f^+)^2 [2(\Delta+U_f^+) + U_p] } 
\end{align}
and Figs.~\ref{fig:Yb-diags}\textbf{f} and \textbf{g} can be combined into
\begin{align}
    \mathcal{K}_{\sigma_1^{} \sigma_1' \sigma_2^{} \sigma_2'}^{f^{(0), f+g-tot}}  =  \sum_\lambda 2 \frac{ [{T}_{\lambda}^{}]^{\sigma_1' \sigma_2}  [{T}_{\bar \lambda}^{\dag}]^{\sigma_2' \sigma_1 } }{ (\Delta + U_f^+)^3 } 
\end{align}
Note that $T_{\lambda}^{} \rightarrow T_{\lambda}^{\dag}$ when swapping rare-earth sites.

The total $f^{(0)}$ channel yields,
\begin{align}
    \mathcal{K}_{\sigma_1^{} \sigma_1' \sigma_2^{} \sigma_2'}^{f^{(0)}} 
    = \frac{4}{(\Delta + U_f^+)^2} \sum_\lambda  \bigg[ & \frac{ [{T}_{\lambda}^{}]^{\sigma_1' \sigma_2}  [{T}_{\lambda}^{\dag}]^{\sigma_2' \sigma_1} - [\mathring{T}_{1\lambda}^{}]^{\sigma_1' \sigma_1} [\mathring{T}_{2\lambda}^{}]^{\sigma_2' \sigma_2}  }{ 2(\Delta+U_f^+) + U_p }  \nonumber \\
    + & \frac{ [{T}_{\lambda}^{}]^{\sigma_1' \sigma_2}  [{T}_{\bar \lambda}^{\dag}]^{\sigma_2' \sigma_1} - [\mathring{T}_{1\lambda}^{}]^{\sigma_1' \sigma_1} [\mathring{T}_{2\bar \lambda}^{}]^{\sigma_2' \sigma_2}  }{2(\Delta + U_f^+) } \bigg] 
\end{align}

As for the $f^{(2:1)}$ channel, the sites and ligands swapping in the process depicted in Fig.~\ref{fig:Yb-diags}\textbf{i} yields,
\begin{align}
    \mathcal{K}_{\sigma_1^{} \sigma_1' \sigma_2^{} \sigma_2'}^{f^{(2:1), i-tot}} = -  \sum_\lambda \sum_{\mu  \mu' a}  \frac{ [T_\lambda^{}]^{\sigma_1' \mu}  [T_\lambda^\dag]^{\mu' \sigma_1} \gamma_{\sigma_2^{} \sigma_2',\mu\mu'}^a   +    [T_\lambda^{}]^{\mu' \sigma_2} [T_\lambda^{\dag}]^{\sigma_2' \mu}  \gamma_{\sigma_1^{} \sigma_1',\mu\mu'}^a }{ (\Delta + U_f^+)^2 (U_f^+ + U_f^- + E_a)  } 
\end{align}
For the process in Fig.~\ref{fig:Yb-diags}\textbf{j},
\begin{align}
    \mathcal{K}_{\sigma_1^{} \sigma_1' \sigma_2^{} \sigma_2'}^{f^{(2:1), j-tot}} = -  \sum_\lambda \sum_{\mu  \mu' a}  \frac{ [T_\lambda^{}]^{\sigma_1' \mu'}  [T_{\bar \lambda}^\dag]^{\mu \sigma_1} \gamma_{\sigma_2^{} \sigma_2',\mu\mu'}^a  +  [T_{\lambda}^{}]^{\mu \sigma_2} [T_{\bar \lambda}^{\dag}]^{\sigma_2' \mu'}  \gamma_{\sigma_1^{} \sigma_1',\mu\mu'}^a }{ (\Delta + U_f^+)^2 (U_f^+ + U_f^- + E_a)  }  
\end{align}

The total $f^{(2:1)}$ channel yields,
\begin{align}
    &\mathcal{K}_{\sigma_1^{} \sigma_1' \sigma_2^{} \sigma_2'}^{(2:1)} = - \sum_\lambda \sum_{\mu  \mu' a}  \bigg[ ( [T_\lambda^{}]^{\sigma_1' \mu}  [T_\lambda^\dag]^{\mu' \sigma_1} + [T_\lambda^{}]^{\sigma_1' \mu'}  [T_{\bar \lambda}^\dag]^{\mu \sigma_1}) \gamma_{\sigma_2^{} \sigma_2',\mu\mu'}^a \nonumber \\
    & \quad +  ( [T_\lambda^{}]^{\mu' \sigma_2} [T_\lambda^{\dag}]^{\sigma_2' \mu}  +  [T_{\lambda}^{}]^{\mu \sigma_2} [T_{\bar \lambda}^{\dag}]^{\sigma_2' \mu'} ) \gamma_{\sigma_1^{} \sigma_1',\mu\mu'}^a \bigg] \frac{1}{ (\Delta + U_f^+)^2 (U_f^+ + U_f^- + E_a)  }  
\end{align}

We can simplify the expression using the \textbf{total effective hopping matrix} $T \equiv T_A + T_B $ and the \textbf{total effective local hopping matrix} $\mathring{T}_{i} \equiv \mathring{T}_{iA} + \mathring{T}_{iB} $.

The exchange interaction for the $f^{(0)}$ channel turns out
\begin{align}%
    \mathcal{K}_{\sigma_1^{} \sigma_1' \sigma_2^{} \sigma_2'}^{f^{(0)}} = \frac{2}{(\Delta + U_f^+)^3} & \bigg[ [{T}_{}^{}]^{\sigma_1' \sigma_2}  [{T}_{}^{\dag}]^{\sigma_2' \sigma_1} - [\mathring{T}_{1}^{}]^{\sigma_1' \sigma_1} [\mathring{T}_{2}^{}]^{\sigma_2' \sigma_2}  \nonumber \\
    &- \kappa \sum_\lambda \left( [{T}_{\lambda}^{}]^{\sigma_1' \sigma_2}  [{T}_{\lambda}^{\dag}]^{\sigma_2' \sigma_1} - [\mathring{T}_{1\lambda}^{}]^{\sigma_1' \sigma_1} [\mathring{T}_{2\lambda}^{}]^{\sigma_2' \sigma_2}  \right) \bigg] %
    \label{eq:f0-Yb}
\end{align}
with
\begin{equation}
    \kappa = \frac{U_p}{2(\Delta + U_f^+) + U_p} \ .
\end{equation}

As for the $f^{(2:1)}$ channel interaction tensor,
\begin{equation}%
     \mathcal{K}_{\sigma_1^{} \sigma_1' \sigma_2^{} \sigma_2' }^{(f^{2:1})}  
     = - \sum_{\mu  \mu' a} \frac{ [T]^{\sigma_1' \mu}  [T^\dag]^{\mu' \sigma_1^{}}  \gamma_{\sigma_2^{} \sigma_2',\mu\mu'}^a   +  [T]^{\mu'\sigma_2^{}}  [T^\dag]^{\sigma_2'\mu}  \gamma_{\sigma_1^{} \sigma_1',\mu\mu'}^a }{ (\Delta + U_f^+)^2 (U_f^+ + U_f^- + E_a) }  %
    \label{eq:f21-Yb}
\end{equation}

\subsection{Ce-based insulator}
\begin{figure}[h!]
    \centering
    \includegraphics[width=0.95\linewidth]{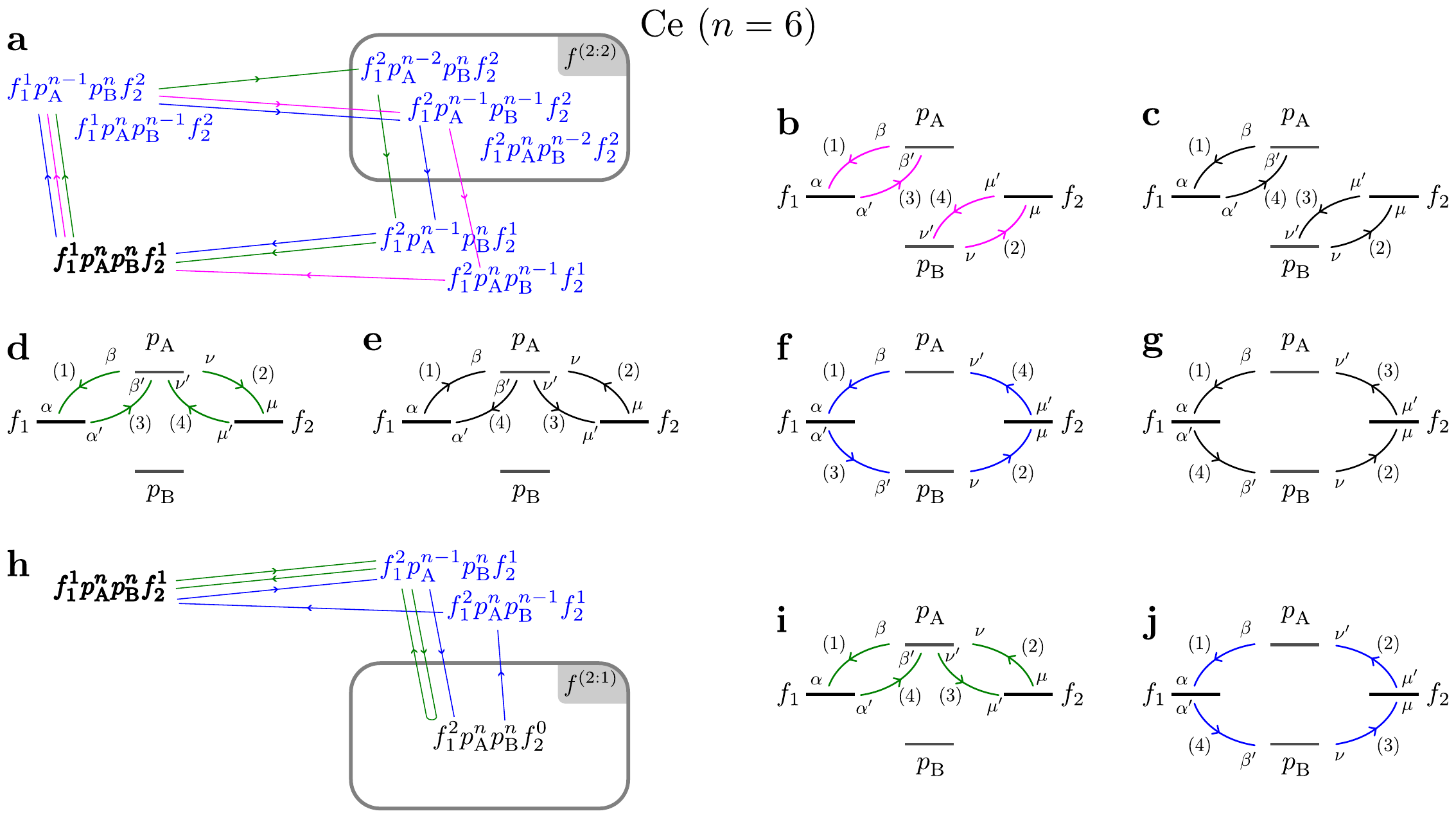}
    \caption{Virtual state configurations for the perturbation theory of Ce-based insulators ($n=6$, filled ligand shell), representing a subset of the general configurations in Fig.~\ref{fig:gral-diags}. (\textbf{a}) Virtual states in the $f^{(2:2)}$ channel, with (\textbf{b}--\textbf{g}) showing the six diagram types extracted from it. (\textbf{h}) Virtual states in the $f^{(2:1)}$ channel, with (\textbf{i}, \textbf{j}) detailing the resulting virtual processes. The complete diagrammatic series is generated by swapping the rare-earth and/or ligand sites.}
    \label{fig:Ce-diags}
\end{figure}

When considering Ce-based insulators, we adopt the electron picture where the ligand orbitals are completely filled ($n=6$). As shown in Fig.~\ref{fig:gral-diags}, the allowed intermediate states in the perturbation theory correspond to the upper-right quadrant, highlighted by the state configurations in black and blue. From a perturbative perspective, the fundamental difference between Ce- and Yb-based insulators stems from the lack of particle-hole symmetry in the ligand states. To derive the effective spin Hamiltonian, all relevant virtual processes are summarized in Fig.~\ref{fig:Ce-diags}. Notably, within this framework, the $f^{(0)}$ processes are entirely absent and are replaced instead by $f^{(2:2)}$ channels.

A similar procedure to the Yb case is applied to evaluate the contribution of each virtual process to the total exchange tensor. This calculation requires the energy differences between the ground-state configuration ($f_{1}^1 p_{A}^6 p_{B}^6 f_{2}^1$) and the intermediate excited states listed in Table~\ref{tab:excited_states-Ce}.
\begin{table}[h!]
\centering
\caption{Excited State Configurations and Energy Differences}
\label{tab:excited_states-Ce}
\vskip 2mm
\renewcommand{\arraystretch}{1.4} 
\begin{tabular}{ll}
\hline
\textbf{Excited State} \qquad & $(\varepsilon_0^{} - \mathcal{H}_0 )$ \\ 
\hline
$f_{1}^1 p_{A}^5 p_{B}^6 f_{2}^2$ & $-(\Delta + U_{f}^+ + E_{a_2})$ \\
$f_{1}^2 p_{A}^4 p_{B}^6 f_{2}^2$ & $-[ 2(\Delta + U_{f}^+ )+ U_p  + E_{a_1} + E_{a_2} ]$ \\
$f_{1}^2 p_{A}^5 p_{B}^5 f_{2}^2$ & $-[ 2(\Delta + U_{f}^+) + E_{a_1} + E_{a_2}]$ \\
$f_{1}^2 p_{A}^6 p_{B}^6 f_{2}^0$ & $-(U_{f}^- + U_{f}^+ + E_{a_1})$ \\ 
\hline
\end{tabular}
\end{table}
\subsubsection{$f^{(2:2)}$ channel}

For the virtual process in Fig.~\ref{fig:Ce-diags}\textbf{b} (see magenta arrows loop in \textbf{a}) the exchange tensor reads as 
\begin{align}
    \mathcal{K}_{\sigma_1\sigma_1'\sigma_2\sigma_2'}^{f^{(2:2),b}} &= - 
    \sum_{\substack{\alpha\beta\mu\nu\\\alpha'\beta'\mu'\nu'}}\sum_{a_1a_2}
    \frac{
        (t_{1A}^{\alpha\beta})(t_{2B}^{\mu\nu})(t_{1A}^{\alpha'\beta'})^*(t_{2B}^{\mu'\nu'})^* \ \delta_{\beta\beta'} \delta_{\nu\nu'} 
     \gamma^{a_1}_{\sigma_1\sigma_1',\alpha\alpha'}
     \gamma^{a_2}_{\sigma_2\sigma_2',\mu\mu'}
    }{
        (\Delta+U_f^++E_{a_1})(\Delta+U_f^++E_{a_2})[2(\Delta + U_f^{+})+E_{a_1}+E_{a_2}]
    }   \nonumber \\
    &=-\sum_{\alpha\alpha' \mu \mu'}\sum_{a_1a_2}
    \frac{
        [ \mathring{T}_{1A}^{} ]^{\alpha\alpha'}
        [ \mathring{T}_{2B}^{\dag} ]^{\mu\mu'}
        \gamma^{a_1}_{\sigma_1\sigma_1',\alpha\alpha'}
        \gamma^{a_2}_{\sigma_2\sigma_2',\mu\mu'}
    }{
        (\Delta+U_f^++E_{a_1})(\Delta+U_f^++E_{a_2})[2(\Delta + U_f^{+})+E_{a_1}+E_{a_2}]
    }.
\end{align}

For the virtual process in Fig.~\ref{fig:Ce-diags}\textbf{c}, the exchange tensor reads as
\begin{align}
   \mathcal{K}_{\sigma_1\sigma_1'\sigma_2\sigma_2'}^{f^{(2:2),c}}
    &=-\sum_{\substack{\alpha\beta\mu\nu\\\alpha'\beta'\mu'\nu'}}\sum_{a_1a_2}
    \frac{
        (t_{1A}^{\alpha\beta})(t_{2B}^{\mu\nu})(t_{2B}^{\mu'\nu'})^*(t_{1A}^{\alpha'\beta'})^*  \ \delta_{\beta\beta'} \delta_{\nu\nu'} 
     \gamma^{a_1}_{\sigma_1\sigma_1',\alpha\alpha'}
     \gamma^{a_2}_{\sigma_2\sigma_2',\mu\mu'}
    }{
        (\Delta+U_f^++E_{a_1})^2[2(\Delta + U_f^{+})+E_{a_1}+E_{a_2}]
    } \nonumber \\
    &=-\sum_{\alpha \alpha' \mu\mu'}\sum_{a_1a_2}
    \frac{
        [ \mathring{T}_{1A}^{} ]^{\alpha\alpha'}
        [ \mathring{T}_{2B}^{\dag}]^{\mu\mu'}
        \gamma^{a_1}_{\sigma_1\sigma_1',\alpha\alpha'}
        \gamma^{a_2}_{\sigma_2\sigma_2',\mu\mu'}
    }{
        (\Delta+U_f^++E_{a_1})^2[2(\Delta + U_f^{+})+E_{a_1}+E_{a_2}]
    }.
\end{align}

For the virtual process in Fig.~\ref{fig:Ce-diags}\textbf{d} (see green arrows loop in \textbf{a}), the exchange tensor reads as 
\begin{align}
    &\mathcal{K}_{\sigma_1\sigma_1'\sigma_2\sigma_2'}^{f^{(2:2),d}}
    =\sum_{\substack{\alpha\beta\mu\nu\\\alpha'\beta'\mu'\nu'}}\sum_{a_1a_2}
    \frac{
        (t_{1A}^{\alpha\beta})(t_{2A}^{\mu\nu})(t_{1A}^{\alpha'\beta'})^*(t_{2A}^{\mu'\nu'})^* \ (\delta_{\beta\nu'}\delta_{\nu\beta'}-\delta_{\beta\beta'}\delta_{\nu\nu'})
    \gamma^{a_1}_{\sigma_1\sigma_1',\alpha\alpha'}
    \gamma^{a_2}_{\sigma_2\sigma_2',\mu\mu'}
    }{
        (\Delta+U_f^++E_{a_1})(\Delta+U_f^++E_{a_2})[2(\Delta + U_f^{+})+U_p+E_{a_1}+E_{a_2}]
    } \nonumber \\
    & = \sum_{\alpha\mu \alpha'\mu'} \sum_{a_1a_2}
    \frac{
      [T_A^{}]^{\alpha\mu'} [T_A^{\dag}]^{\alpha'\mu} 
      - [\mathring{T}_{1A}^{}]^{\alpha\mu} [\mathring{T}_{2A}^{\dag}]^{\alpha'\mu'}
     \gamma^{a_1}_{\sigma_1\sigma_1',\alpha\mu}
     \gamma^{a_2}_{\sigma_2\sigma_2',\alpha'\mu'} 
    }{
        (\Delta+U_f^++E_{a_1})(\Delta+U_f^++E_{a_2})[2(\Delta + U_f^{+})+U_p+E_{a_1}+E_{a_2}]
    } .
\end{align}

For the virtual process in Fig.~\ref{fig:Ce-diags}\textbf{e}, the exchange tensor reads as 
\begin{align}
    &\mathcal{K}_{\sigma_1\sigma_1'\sigma_2\sigma_2'}^{f^{(2:2),e}}
    =-\sum_{\substack{\alpha\beta\mu\nu\\\alpha'\beta'\mu'\nu'}}\sum_{a_1a_2}
    \frac{
        (t_{1A}^{\alpha\beta})(t_{2A}^{\mu\nu})(t_{2A}^{\mu'\nu'})^*(t_{1A}^{\alpha'\beta'})^* \ (\delta_{\beta\beta'}\delta_{\nu\nu'}-\delta_{\beta\nu'}\delta_{\beta'\nu})
    }{
        (\Delta+U_f^++E_{a_1})^2[2(\Delta + U_f^{+})+U_p+E_{a_1}+E_{a_2}]
    }
    \gamma^{a_1}_{\sigma_1\sigma_1',\alpha\alpha'}
    \gamma^{a_2}_{\sigma_2\sigma_2',\mu\mu} \nonumber \\
    &=\sum_{\alpha\mu  \alpha'\mu'} \sum_{a_1a_2}
    \frac{ \left(
     [T_A]^{\alpha\mu} [T_A^{\dag}]^{\alpha\mu}-
     [\mathring{T}_{1A}^{}]^{\alpha\mu'} [\mathring{T}_{2A}^{\dag}]^{\alpha'\mu}    \right)
    \gamma^{a_1}_{\sigma_1\sigma_1',\alpha\mu'}
    \gamma^{a_2}_{\sigma_2\sigma_2',\alpha'\mu}
    }{
        (\Delta+U_f^++E_{a_1})^2[2(\Delta + U_f^{+})+U_p+E_{a_1}+E_{a_2}]
    } .
\end{align}

For the virtual process in Fig.~\ref{fig:Ce-diags}\textbf{f} (see blue arrows loop in \textbf{a}), the exchange tensor reads as 
\begin{align}
    \mathcal{K}_{\sigma_1\sigma_1'\sigma_2\sigma_2'}^{f^{(2:2),f}}
    &= \sum_{\substack{\alpha\beta\nu\mu\\\alpha'\beta'\mu'\nu'}}\sum_{a_1a_2}
    \frac{
        (t_{1A}^{\alpha\beta})(t_{2B}^{\mu\nu})(t_{1B}^{\alpha'\beta'})^*(t_{2A}^{\mu'\nu'})^* \  \delta_{\beta\nu'}\delta_{\nu\beta'}
    \gamma^{a_1}_{\sigma_1\sigma_1',\alpha\alpha'}
    \gamma^{a_2}_{\sigma_2\sigma_2',\mu\mu'}
    }{
        (\Delta+U_f^++E_{a_1})(\Delta+U_f^++E_{a_2})[2(\Delta + U_f^{+})+E_{a_1}+E_{a_2}]
    } \nonumber \\
    &= \sum_{\alpha\mu \alpha'\mu'}\sum_{a_1a_2}
    \frac{
        [T_{A}^{}]^{\alpha\mu'}
        [T_{B}^{\dag}]^{\mu\alpha'} \ \gamma^{a_1}_{\sigma_1\sigma_1',\alpha\alpha'}
    \gamma^{a_2}_{\sigma_2\sigma_2',\mu\mu'}
    }{
        (\Delta+U_f^++E_{a_1})(\Delta+U_f^++E_{a_2})[2(\Delta + U_f^{+})+E_{a_1}+E_{a_2}]
    } .
\end{align}

For the virtual process in Fig.~\ref{fig:Ce-diags}\textbf{g}, the exchange tensor reads as 
\begin{align}
    \mathcal{K}_{\sigma_1\sigma_1'\sigma_2\sigma_2'}^{f^{(2:2),g}}
    &= \sum_{\substack{\alpha\beta\nu\mu\\\alpha'\beta'\mu'\nu'}}\sum_{a_1a_2}
    \frac{
        (t_{1A}^{\alpha\beta})(t_{2B}^{\mu\nu})(t_{2A}^{\mu'\nu'})^*(t_{1B}^{\alpha'\beta'})^* \  \delta_{\beta\nu}\delta_{\beta'\nu'}
    }{
        (\Delta+U_f^++E_{a_1})^2[2(\Delta + U_f^{+})+E_{a_1}+E_{a_2}]
    }
    \gamma^{a_1}_{\sigma_1\sigma_1',\alpha\alpha'}
    \gamma^{a_2}_{\sigma_2\sigma_2',\mu\mu'}  \nonumber \\
    &= \sum_{\alpha\mu \alpha'\mu'}\sum_{a_1a_2}
    \frac{
       [T_{A}^{}]^{\alpha\mu'}
       [T_{B}^{\dag}]^{\mu\alpha'} \
      \gamma^{a_1}_{\sigma_1\sigma_1',\alpha\alpha'}
      \gamma^{a_2}_{\sigma_2\sigma_2',\mu\mu'}
    }{
        (\Delta+U_f^++E_{a_1})^2[2(\Delta + U_f^{+})+E_{a_1}+E_{a_2}]
    } .
\end{align}

By mutually exchanging the ligand positions ($A \leftrightarrow B$) and/or the rare-earth sites, we generate the complete set of virtual processes within the $f^{(2:2)}$ channel. Summing the individual contributions from all such symmetry-related paths yields the total exchange tensor for the $f^{(2:2)}$ channel, which reads as:
\begin{align}
&\mathcal{K}_{\sigma_1\sigma_1'\sigma_2\sigma_2'}^{(f^{2:2})}
    =
    \sum_{\alpha\mu \alpha'\mu'}\sum_{a_1a_2} g_{a_1,a_2}
    \Big[
        [T]^{\alpha\mu}
        [T^{\dag}]^{\alpha'\mu'}
        -
        [\mathring{T}_1^{}]^{\alpha\mu'}
        [\mathring{T}_2^{\dag} ]^{\alpha'\mu}
         \nonumber \\
    & -\tau_{a_1,a_2}\sum_{\lambda=A,B}\left(
            [T_{\lambda}^{}]^{\alpha\mu}
            [T_{\lambda}^{\dag}]^{\alpha'\mu'}
            -
            [\mathring{T}_{1\lambda}]^{\alpha\mu'}
            [\mathring{T}_{2\lambda}]^{\alpha'\mu}
        \right)
    \Big]
    \gamma^{a_1}_{\sigma_1\sigma_1',\alpha\mu'}
    \gamma^{a_2}_{\sigma_2\sigma_2',\alpha'\mu} \ ,
 \label{eq:f22-Ce}
\end{align}
with the function of energies
\begin{equation}
    g_{a_1,a_2} = \bigg( \frac{1}{ \Delta+U_f^++E_{a_1} } + \frac{1}{ \Delta+U_f^++E_{a_2} } \bigg)^2
    \frac{1}{2(\Delta+U_f^+)+E_{a_1}+E_{a_2}}
\end{equation}
and
\begin{align}
    \tau_{a_1,a_2} = \frac{U_p}{2(\Delta+U_f^+)+U_p+E_{a_1}+E_{a_2}}. 
\end{align}

\subsubsection{$f^{(2:1)}$ channel}

We next evaluate the exchange contributions arising from the $f^{(2:1)}$ channel, where the allowed intermediate virtual states are schematically depicted in Fig.~\ref{fig:Ce-diags}\textbf{h}. 

For the specific virtual process tracking the green-arrow loop shown in Fig.~\ref{fig:Ce-diags}\textbf{i}, the corresponding exchange tensor component takes the form
\begin{align}
    \mathcal{K}_{\sigma_1\sigma_1'\sigma_2\sigma_2'}^{f^{(2:1),i}} =
    -\sum_{\alpha,\mu',a_1}
    \frac{ [ T_A^{} ]^{\alpha\sigma_2'}  [ T_A^{\dag} ]^{\sigma_2\mu'}  }{ (\Delta + U_f^++E_{a_1})^2 (U_f^+ + U_f^- + E_{a_1}) }
    \gamma_{\sigma_1\sigma_1',\alpha\mu'}^{a_1},
    \label{eq:K_f21_P2_1}
\end{align}
whereas the virtual process given by the blue-arrow loop associated with Fig.~\ref{fig:Ce-diags}\textbf{j} yields
\begin{align}
    \mathcal{K}_{\sigma_1\sigma_1'\sigma_2\sigma_2'}^{f^{(2:1),j}} =
    -\sum_{\alpha,\mu',a_1}
    \frac{ [ T_A^{} ]^{\alpha\sigma_2'}  [ T_B^{\dag} ]^{\sigma_2\mu'}  }{ (\Delta + U_f^++E_{a_1})^2 (U_f^+ + U_f^- + E_{a_1}) }
    \gamma_{\sigma_1\sigma_1',\alpha\mu'}^{a_1}.
    \label{eq:K_f21_P1_1}
\end{align}

By accounting for all symmetry-related operations (specifically exchanging the ligand indices ($A \leftrightarrow B$) and the rare-earth sites) and summing over the complete set of paths, we obtain the total exchange tensor for the $f^{(2:1)}$ channel:
\begin{align}
\mathcal{K}_{\sigma_1\sigma_1'\sigma_2\sigma_2'}^{(f^{2:1})}
    =-\sum_{\alpha,\mu',a}
    \frac{ 
    [T]^{\sigma_1\mu'}  [ T^{\dag} ]^{\alpha\sigma_1'} \gamma_{\sigma_2\sigma_2',\alpha\mu'}^{a}
    +
    [ T ]^{\alpha\sigma_2'}  [ T^{\dag} ]^{\sigma_2\mu'} \gamma_{\sigma_1\sigma_1',\alpha\mu'}^{a}
    }{ (\Delta + U_f^++E_{a})^2 (U_f^+ + U_f^- + E_{a}) }.
 \label{eq:f21-Ce} 
\end{align}

\subsection{Spin-conserved hopping in centrosymmetric Kramers systems}

We consider the effective hopping restricted to the ground-state Kramers doublet, represented by the $2\times2$ matrix $T^{\rm in}$.

As a consequence of time-reversal (TR) symmetry within both the rare-earth Kramers doublets and the ligand $p$-orbitals, the $fp$-hopping matrix can be generally expressed as
\begin{equation}
    t_{i\lambda}^{\rm in} = c_{i\lambda} \ e^{i \bm \sigma \cdot \bm n_{i\lambda} \phi_{i\lambda}}
\end{equation}
where $c_{i\lambda}$ is a real scalar determining the hopping amplitude, $\bm \sigma$ is the vector of Pauli matrices, $\bm n_{i\lambda}$ is a unit vector specifying the spin-rotation axis, and $\phi_{i\lambda}$ is a phase angle governing the spin-flip. The single-ligand effective hopping matrix through a ligand $\lambda$ is then given by
\begin{equation}
    T_{\lambda}^{\rm in} = t_{1\lambda}^{\rm in}  [t_{2\lambda}^{\rm in}]^\dag = c_{1\lambda} c_{2\lambda} \ e^{i \bm \sigma \cdot (\bm n_{1\lambda} \phi_{1\lambda} + \bm n_{2\lambda} \phi_{2\lambda} )} = c_{\lambda} \ e^{i \bm \sigma \cdot \bm n_{\lambda} \phi_{\lambda}}
\end{equation}
In a centrosymmetric bond geometry, the inversion symmetry relation $t_{1\lambda} = t_{2 \bar \lambda}$ implies that $T_{\lambda} = T_{\bar \lambda}^\dag$, where $\bar \lambda$ denotes the inversion-partner ligand site. Hence, the total effective hopping matrix yields
\begin{equation}
    T^{\rm in} = T_{\lambda}^{\rm in} + T_{\bar \lambda}^{\rm in} =  T_{\lambda}^{\rm in} + [T_{\lambda}^{\rm in}]^\dag = c_{\lambda} \ e^{i \bm \sigma \cdot \bm n_{\lambda} \phi_{\lambda}} + c_{\lambda} \ e^{-i \bm \sigma \cdot \bm n_{\lambda} \phi_{\lambda}} = 2 c_{\lambda} \cos(\bm \sigma \cdot \bm n_{\lambda} \phi_{\lambda}) \mathbb{I}_{2\times2}
\end{equation}
It follows that $T^{\rm in}=a\mathbb{I}_{2\times 2}$, demonstrating that the total effective intersite hopping strictly conserves the spin.

Alternatively, if the system exhibits a mirror reflection symmetry across a plane perpendicular to the bond axis and positioned at its center, the $fp$ hopping matrices satisfy $t_{1\lambda} = t_{2\lambda}$. Under this spatial constraint, the single-ligand effective hopping simplifies to
\begin{equation}
    T_{\lambda}^{\rm in} = t_{1\lambda}^{\rm in} [t_{2\lambda}^{\rm in}]^\dag = t_{1\lambda}^{\rm in} [t_{1\lambda}^{\rm in}]^\dag = |c_{1\lambda} \ e^{i \bm \sigma \cdot \bm n_{1\lambda} \phi_{1\lambda}}|^2 = c_{1\lambda}^2 \mathbb{I}_{2 \times 2}
\end{equation}

Similarly, for the restricted, effective local hopping of the form $\mathring{T}_{i\lambda}^{\rm in} = t_{i\lambda}^{\rm in} [t_{i\lambda}^{\rm in}]^\dag$, evaluating the product within the Kramers subspace gives
\begin{equation}
    \mathring{T}_{i\lambda}^{\rm in} = t_{i\lambda}^{\rm in} [t_{i\lambda}^{\rm in}]^\dag = [c_{i\lambda} \ e^{i \bm \sigma \cdot \bm n_{i\lambda} \phi_{i\lambda}}] [c_{i\lambda} \ e^{-i \bm \sigma \cdot \bm n_{i\lambda} \phi_{i\lambda}}] = c_{i\lambda}^2 \mathbb{I}_{2\times2}
\end{equation}

With the spin-conserving nature of the effective hopping matrices demonstrated for the ground-state Kramers doublet, we proceed to project the $f^{(0)}$ and $f^{(2:1)}$ channels for the Yb case. First, we observe that the $f^{(0)}$ channel exchange tensor in Eq.~\eqref{eq:f0-Yb} naturally constitutes an \emph{intra-doublet} contribution:
\begin{align}
    \mathcal{K}_{\sigma_1^{} \sigma_1' \sigma_2^{} \sigma_2'}^{f^{(0)}} &= \frac{2}{(\Delta + U_f^+)^3} \left[ [{T}_{}^{}]^{\uparrow \uparrow}  [{T}_{}^{\dag}]^{\uparrow \uparrow} \delta_{\sigma_1' \sigma_2} \delta_{\sigma_2' \sigma_1} - [\mathring{T}_{1}^{}]^{\uparrow \uparrow} [\mathring{T}_{2}^{}]^{\uparrow \uparrow} \delta_{\sigma_1' \sigma_1} \delta_{\sigma_2' \sigma_2} \right. \nonumber \\
    & - \kappa \sum_\lambda \left. \left( [{T}_{\lambda}^{}]^{\sigma_1' \sigma_2}  [{T}_{\lambda}^{\dag}]^{\sigma_2' \sigma_1} - [\mathring{T}_{1\lambda}^{}]^{\uparrow \uparrow} [\mathring{T}_{2\lambda}^{}]^{\uparrow \uparrow} \delta_{\sigma_1' \sigma_1} \delta_{\sigma_2' \sigma_2} \right) \right] .
\end{align}
Within this expression, the term proportional to $\delta_{\sigma_1' \sigma_1} \delta_{\sigma_2' \sigma_2}$ represents a non-magnetic density-density interaction and does not contribute to the magnetic coupling. Conversely, the term containing $\delta_{\sigma_1' \sigma_2} \delta_{\sigma_2' \sigma_1}$ yields a purely isotropic magnetic exchange. The remaining term involving the single-ligand effective hopping $[T_{\lambda}]^{\sigma \sigma'}$ can introduce magnetic anisotropy, but its magnitude is explicitly modulated by the parameter $\kappa$. Consequently, in the absence of inter-orbital ligand Coulomb repulsion ($U_p = 0$), $\kappa$ vanishes and the magnetic interaction becomes entirely isotropic. Furthermore, in the presence of the aforementioned mirror symmetry plane at the bond center, the individual pathway $[T_{\lambda}]^{\sigma \sigma'}$ itself becomes spin-preserving. As a result, the $f^{(0)}$ channel remains purely isotropic for any value of $U_p$ under this high spatial symmetry.

Next, we partition the $f^{(2:1)}$ channel from Eq.~\eqref{eq:f21-Yb} into its \emph{intra-doublet} and \emph{out-of-doublet} contributions. The \emph{intra-doublet} channel captures the processes where the intermediate transitions connect states exclusively within the ground-state Kramers doublets, reading as
\begin{equation}
     \mathcal{K}_{\sigma_1^{} \sigma_1' \sigma_2^{} \sigma_2' }^{(f_{\rm in}^{2:1})}   
     = -  \sum_{a} \frac{ 2 [T]^{\uparrow \uparrow}  [T^\dag]^{\uparrow \uparrow}  \gamma_{\sigma_2^{} \sigma_2',\sigma_1'\sigma_1^{}}^a  }{(\Delta + U_f^+)^2 (U_f^+ + U_f^- + E_a) }   
\end{equation}
Here, the symmetry of the exchange interaction is dictated by the structural tensor $\gamma_{\sigma_2^{} \sigma_2',\sigma_1'\sigma_1^{}}^a$, which maintains an isotropic form.

The remaining processes within the $f^{(2:1)}$ channel constitute the \emph{out-of-doublet} subchannel, where the hopping matrices connect the ground-state Kramers doublets to their respective orthogonal states. This out-of-doublet tensor takes the form
\begin{equation}\label{eq:f21out}
     \mathcal{K}_{\sigma_1^{} \sigma_1' \sigma_2^{} \sigma_2' }^{(f_{\rm out}^{2:1})}   
     = -  \sum_{\mu  \mu' a}' \frac{  [T]^{\sigma_1' \mu}  [T^\dag]^{\mu' \sigma_1^{}}  \gamma_{\sigma_2^{} \sigma_2',\mu\mu'}^a   +  [T]^{\mu'\sigma_2^{}}  [T^\dag]^{\sigma_2'\mu}  \gamma_{\sigma_1^{} \sigma_1',\mu\mu'}^a  }{(\Delta + U_f^+)^2 (U_f^+ + U_f^- + E_a) }   
\end{equation}
where the prime on the summation denotes that the indices $\mu$ and $\mu'$ run strictly over the 12 excited atomic states lying outside the ground-state Kramers doublet manifold.

Next, we calculate the exchange tensor for the \emph{intra-doublet} channel in the Ce case. Within this regime, the tensor elements reduce to $\gamma^{a}_{\sigma \sigma',\alpha\alpha'} = \delta_{\sigma\sigma'} \delta_{\alpha\alpha'} - \delta_{\sigma\alpha} \delta_{\sigma'\alpha'}$ for indices restricted to the Kramers doublet states. 
Using the identities $\sum_{\alpha \alpha'} \gamma^{a_1}_{\sigma_1\sigma_1',\alpha\alpha'} \gamma^{a_2}_{\sigma_2\sigma_2',\alpha'\alpha} = \delta_{\sigma_1 \sigma_2'} \delta_{\sigma_1' \sigma_2} $ and $\sum_{\alpha \alpha'} \gamma^{a_1}_{\sigma_1\sigma_1',\alpha\alpha}\gamma^{a_2}_{\sigma_2\sigma_2',\alpha'\alpha'} = \delta_{\sigma_1 \sigma_1'} \delta_{\sigma_2 \sigma_2'}$, the final expression for the \emph{intra-doublet} channel tensor in Eq.~\eqref{eq:f22-Ce} is given by
\begin{align}
\mathcal{K}_{\sigma_1\sigma_1'\sigma_2\sigma_2'}^{(f_{\rm in}^{2:2})}
    &=
    \sum_{a_1a_2} g_{a_1,a_2}
    \bigg[
        [T]^{\uparrow\uparrow}
        [T^{\dag}]^{\uparrow \uparrow} \delta_{\sigma_1 \sigma_2'} \delta_{\sigma_1' \sigma_2} \nonumber \\
    &  -\tau_{a_1,a_2} \sum_{\alpha\alpha'\mu\mu'} \sum_{\lambda=A,B}
            [T_{\lambda}]^{\alpha\mu'}
            [T_{\lambda}^{\dag}]^{\mu\alpha'}
            \gamma^{a_1}_{\sigma_1\sigma_1',\alpha\alpha'}
             \gamma^{a_2}_{\sigma_2\sigma_2',\mu\mu'}    \nonumber \\
    & - \big( \ [\mathring{T}_1]^{\uparrow\uparrow}
         [\mathring{T}_2^{\dag} ]^{\uparrow\uparrow}
     - \tau_{a_1,a_2}\sum_{\lambda=A,B}
            [\mathring{T}_{1\lambda}]^{\uparrow\uparrow}
            [\mathring{T}_{2\lambda}]^{\uparrow\uparrow} \
    \big) \
    \delta_{\sigma_1 \sigma_1'} \delta_{\sigma_2 \sigma_2'} \bigg] .
\end{align}
Here, the $\delta_{\sigma_1' \sigma_1} \delta_{\sigma_2' \sigma_2}$ term represents a non-magnetic density-density interaction, while the $\delta_{\sigma_1' \sigma_2} \delta_{\sigma_2' \sigma_1}$ term yields an isotropic magnetic coupling. The final term involving $[T_{\lambda}]^{\sigma \sigma'}$ can introduce anisotropy, modulated by the parameter $\tau_{a_1,a_2}$, which vanishes when $U_p = 0$. Notably, if the system possesses a mirror symmetry plane perpendicular to the bond and positioned at its center, $T_{\lambda}$ also preserves spin. Under these symmetric conditions, the intra-doublet channel becomes entirely isotropic regardless of the strength of the ligand Coulomb repulsion $U_p$.

Regarding the $f^{(2:1)}$ channel given in the Eq.~\eqref{eq:f21-Ce} for the Ce-based system, its projection onto the \emph{intra-doublet} subspace is achieved by exploiting the symmetry property $\gamma_{\sigma_2\sigma_2',\sigma_1'\sigma_1}^{a} = \gamma_{\sigma_1 \sigma_1', \sigma_2'\sigma_2}^{a}$. Under this relation, the \emph{intra-doublet} channel tensor reduces to
\begin{align}
    \mathcal{K}_{\sigma_1\sigma_1'\sigma_2\sigma_2'}^{(f_{\rm in}^{2:1})}
    &=-  \sum_{a}
    \frac{ 
    2 [ T ]^{\uparrow \uparrow}  [ T^{\dag} ]^{\uparrow \uparrow} \, \gamma_{\sigma_1\sigma_1',\sigma_2'\sigma_2}^{a}
    }{ (\Delta + U_f^++E_{a})^2 (U_f^+ + U_f^- + E_{a}) }.
\end{align}

Remarkably, the algebraic structure of this \emph{intra-doublet} interaction tensor is identical to that derived for the Yb case, producing purely isotropic magnetic interactions. Consequently, any emergent magnetic anisotropy originates strictly from the \emph{out-of-doublet} channel, provided that the intermediate $f^2$ states are non-degenerate.

\subsection{Effective Hamiltonian expressed with spin operators}

In this section, we analyze the rigorous algebraic structure of the intermediate interaction matrix governed by time-reversal symmetry, spatial-inversion symmetry about the bond center, and Hermiticity. The effective spin Hamiltonian can be generally formulated in terms of the U(2) generators as
\begin{equation}
    \mathcal{H}_{\rm eff} = \frac{1}{2} \sum_{ij} \sum_{\sigma_1^{} \sigma_1' \sigma_2^{} \sigma_2'} \mathcal{K}_{\sigma_1^{} \sigma_1' \sigma_2^{} \sigma_2' }^{}(i,j) \ f_{i \sigma_1^{}}^\dag f_{i \sigma_1'}^{}  f_{j \sigma_2^{}}^\dag f_{j \sigma_2'}^{} = \bm F^\dag \mathbb{K}(i,j) \bm F
\end{equation}
where the rank-4 tensor $ \mathcal{K}_{\sigma_1^{} \sigma_1' \sigma_2^{} \sigma_2' }^{}(i,j) $ can be mapped onto a $4\times4$ matrix $\mathbb{K}(i,j)$ by utilizing the basis vector $\bm F^\dag=( f_{i\uparrow}^\dag f_{j\uparrow}^{\dag} , f_{i\uparrow}^\dag f_{j\downarrow}^{\dag} , f_{i\downarrow}^\dag f_{j\uparrow}^{\dag} , f_{i\downarrow}^\dag f_{j\downarrow}^{\dag} ) $.

The Hermiticity requirement of the effective Hamiltonian enforces the following constraint:
\begin{equation}
    \mathcal{H}_{\rm eff} = \mathcal{H}_{\rm eff}^\dag \implies \mathcal{K}_{\sigma_1^{} \sigma_1' \sigma_2^{} \sigma_2' }^{}(i,j) = [\mathcal{K}_{\sigma_1' \sigma_1^{} \sigma_2' \sigma_2^{} }^{}(i,j)]^* 
\end{equation}
Furthermore, invariant properties under time-reversal symmetry dictate that
\begin{equation}
    \mathcal{K}_{\sigma_1^{} \sigma_1' \sigma_2^{} \sigma_2' }^{}(i,j) = {\rm sgn}(\sigma_1^{}) {\rm sgn}(\sigma_1') {\rm sgn}(\sigma_2^{}) {\rm sgn}(\sigma_2') [\mathcal{K}_{\bar\sigma_1^{} \bar\sigma_1' \bar\sigma_2^{} \bar\sigma_2' }^{}(i,j)]^*
\end{equation}
where the sign function satisfies ${\rm sgn}(\sigma)=1 \, (-1)$ for spin $\sigma=\uparrow (\downarrow)$, and the bar notation designates the corresponding time-reversed state (i.e., if $\sigma=\uparrow$, then $\bar \sigma=\downarrow$).

By combining the constraints of Hermiticity and time-reversal symmetry, the interaction matrix expressed within this basis assumes the highly constrained form
\begin{align}
    \mathbb{K}(i,j) = \left(
    \begin{array}{cccc}
     a & x  & y & z  \\
     x^* & b & u & -y  \\
     y^* & u^* & b & -x  \\
     z^* & -y^* & -x^* & a 
    \end{array} \right)
\end{align}
where the diagonal parameters $a$ and $b$ are strictly real scalars.

To transform this description into a conventional localized magnetism framework, we map the effective Hamiltonian from the U(2) generator language to physical spin-1/2 operators via the relation
\begin{equation}
    \mathcal{H} = \frac{1}{2} \sum_{ij} \sum_{\alpha\beta} I_{ij}^{\alpha\beta} S_i^\alpha S_j^\beta =  \frac{1}{2} \sum_{ij} \sum_{\sigma_1^{} \sigma_1' \sigma_2^{} \sigma_2'} \frac{1}{4} \sum_{\alpha\beta} I_{ij}^{\alpha\beta} \sigma_{\sigma_1^{} \sigma_1' }^\alpha  \sigma_{\sigma_2^{} \sigma_2' }^\beta \ f_{i\sigma_1^{}}^\dag f_{i\sigma_1'}^{} f_{j\sigma_2^{}}^\dag f_{j\sigma_2'}^{} 
\end{equation}
where we invoke the standard Abrikosov pseudo-fermion representation $ S^\alpha = \frac{1}{2} {\bm f}^{\dag} \sigma^\alpha \bm f $ with the spinor operator defined as ${\bm f}^\dag = ( f_{\uparrow}^\dag , f_{\downarrow}^\dag )$. Identifying the structural tensor across the representations yields
\begin{equation}
    \mathcal{K}_{\sigma_1^{} \sigma_1' \sigma_2^{} \sigma_2' }^{} (i,j) = \frac{1}{4} \sum_{\alpha\beta} I_{ij}^{\alpha\beta} \sigma_{\sigma_1^{} \sigma_1' }^\alpha  \sigma_{\sigma_2^{} \sigma_2' }^\beta 
\end{equation}
which translates into a matrix Kronecker product structure:
\begin{equation}
    \mathbb{K}(i,j) = \frac{1}{4} \sum_{\alpha\beta} I_{ij}^{\alpha\beta} \sigma^\alpha \otimes \sigma^\beta \ .
\end{equation}
The individual magnetic exchange tensor components can be extracted cleanly by taking the matrix trace:
\begin{equation}
    I_{ij}^{\alpha\beta} = {\rm Tr}\left[ \mathbb{K}(i,j) \cdot \sigma^\alpha \otimes \sigma^\beta \right]
\end{equation}
where we make use of the standard identity ${\rm Tr}[\sigma^\alpha \sigma^\beta] = 2 \delta_{\alpha \beta}$.

Because the time-reversal operator maps physical spin components as $T S^\alpha T^{-1} \rightarrow -S^\alpha$, any cross-coupling between a physical spin component and the identity component must vanish identically, ensuring that
\begin{align}
    I^{\alpha 0} =  I^{0 \alpha} = 0
\end{align}
for spatial directions $\alpha=x,y,z$.

Executing the trace operations explicitly connects the unique elements of the fermionic interaction tensor $\mathcal{K}$ to the components of the magnetic exchange tensor $I$:
\begin{equation}
     \mathcal{H}_{\rm eff} = \frac{1}{2} \sum_{ij} \sum_{\alpha \beta} I_{ij}^{\alpha \beta} S_i^\alpha S_j^\beta
\end{equation}
where $\alpha,\beta \in \{x,y,z\}$, yielding the explicit parameterized expressions:
\begin{align}
I_{ij}^{xx} &= \left[\mathcal{K}_{\uparrow\downarrow\downarrow\uparrow}+\mathcal{K}_{\downarrow\uparrow\uparrow\downarrow}+\mathcal{K}_{\uparrow\downarrow\uparrow\downarrow}+\mathcal{K}_{\downarrow\uparrow\downarrow\uparrow}\right] = u + u^* + (z + z^*), \\
I_{ij}^{yy} &= \left[\mathcal{K}_{\uparrow\downarrow\downarrow\uparrow}+\mathcal{K}_{\downarrow\uparrow\uparrow\downarrow}-\mathcal{K}_{\uparrow\downarrow\uparrow\downarrow}-\mathcal{K}_{\downarrow\uparrow\downarrow\uparrow}\right] = u + u^* - (z + z^*), \\
I_{ij}^{zz} &= \left[\mathcal{K}_{\uparrow\uparrow\uparrow\uparrow}+\mathcal{K}_{\downarrow\downarrow\downarrow\downarrow}-\mathcal{K}_{\uparrow\uparrow\downarrow\downarrow}-\mathcal{K}_{\downarrow\downarrow\uparrow\uparrow}\right] = 2(a-b) , \\
I_{ij}^{xy} &=  i \left[\mathcal{K}_{\downarrow\uparrow\uparrow\downarrow}-\mathcal{K}_{\uparrow\downarrow\downarrow\uparrow}+\mathcal{K}_{\uparrow\downarrow\uparrow\downarrow}-\mathcal{K}_{\downarrow\uparrow\downarrow\uparrow}\right] = i[(z-z^*) - (u-u^*) ] , \\
I_{ij}^{yx} &=  i \left[ \mathcal{K}_{\uparrow\downarrow\downarrow\uparrow}-\mathcal{K}_{\downarrow\uparrow\uparrow\downarrow} + \mathcal{K}_{\uparrow\downarrow\uparrow\downarrow} -\mathcal{K}_{\downarrow\uparrow\downarrow\uparrow}\right] = i[(z-z^*) + (u-u^*) ] , \\
I_{ij}^{yz} &= i \left[\mathcal{K}_{\uparrow\downarrow\uparrow\uparrow}-\mathcal{K}_{\uparrow\downarrow\downarrow\downarrow}-\mathcal{K}_{\downarrow\uparrow\uparrow\uparrow} +\mathcal{K}_{\downarrow\uparrow\downarrow\downarrow}\right] = i2(y-y^*) , \\
I_{ij}^{zy} &= i \left[\mathcal{K}_{\uparrow\uparrow\uparrow\downarrow} - \mathcal{K}_{\downarrow\downarrow\uparrow\downarrow}-\mathcal{K}_{\uparrow\uparrow\downarrow\uparrow} + \mathcal{K}_{\downarrow\downarrow\downarrow\uparrow}\right] = i2(x-x^*) , \\
I_{ij}^{zx} &= \left[\mathcal{K}_{\uparrow\uparrow\uparrow\downarrow}-\mathcal{K}_{\downarrow\downarrow\downarrow\uparrow}+\mathcal{K}_{\uparrow\uparrow\downarrow\uparrow}-\mathcal{K}_{\downarrow\downarrow\uparrow\downarrow}\right] = 2(x+x^*) , \\
I_{ij}^{xz} &= \left[\mathcal{K}_{\uparrow\downarrow\uparrow\uparrow}-\mathcal{K}_{\uparrow\downarrow\downarrow\downarrow} +\mathcal{K}_{\downarrow\uparrow\uparrow\uparrow} -\mathcal{K}_{\downarrow\uparrow\downarrow\downarrow}\right] = 2(y+y^*), \\
I_{ij}^{00} &= \left[\mathcal{K}_{\uparrow\uparrow\uparrow\uparrow}+\mathcal{K}_{\downarrow\downarrow\downarrow\downarrow}+\mathcal{K}_{\uparrow\uparrow\downarrow\downarrow}+\mathcal{K}_{\downarrow\downarrow\uparrow\uparrow}\right] = 2(a+b)
\end{align}

Finally, we introduce the spatial inversion constraints specific to centrosymmetric bonds, which impose the following cross-site condition:
\begin{equation}
    \mathcal{K}_{\sigma_1^{} \sigma_1' \sigma_2^{} \sigma_2'^{}}(i,j) = \mathcal{K}_{\sigma_2^{} \sigma_2' \sigma_1^{} \sigma_1'^{}}(j,i) = \mathcal{K}_{\sigma_2^{} \sigma_2' \sigma_1^{} \sigma_1'^{}}(i,j)
\end{equation}
It directly follows from this permutation symmetry that the matrix elements collapse to $x=y$ and that $u$ is strictly a real parameter. Consequently, the symmetric parts of the off-diagonal magnetic exchange elements equate perfectly ($I_{ij}^{yz}=I_{ij}^{zy}$, $I_{ij}^{xz}=I_{ij}^{zx}$, and $I_{ij}^{xy}=I_{ij}^{yx}$), proving that the antisymmetric Dzyaloshinskii-Moriya interaction vector vanishes completely ($D^x = D^y = D^z = 0$).

\subsection{\emph{Out-of-doublet} channel contributions in the degenerate regime}

It is instructive to first analyze the $f^{(2:1)}$ \emph{out-of-doublet} channel contribution in the limit where all $f^2$ state configurations are completely degenerate ($E_a=E$ $\forall a$). In this idealized regime, the closure relation and summation over the complete set of $f^2$ states yields~\cite{Ghioldi2024}
\begin{equation}\label{eq:gam-deg}
   \sum_{a} \gamma_{\sigma \sigma',\mu\mu'}^a = \delta_{\sigma\sigma'} \delta_{\mu\mu'} - \delta_{\sigma \mu} \delta_{\sigma'\mu'} \ .
\end{equation}
Prior to computing the resulting magnetic exchange components in this limit, we establish a useful identity for the effective hopping matrices. Invoking time-reversal symmetry alone, one can demonstrate the following relation:
\begin{equation}\label{eq:TR-deg}
    \sum_{\mu} [T]^{\mu\sigma} [T^{\dag}]^{\sigma'\mu} 
    = {\rm sgn}(\sigma) {\rm sgn}(\sigma') \sum_{\mu} [T^{\dag}]^{\bar\sigma\bar\mu} [T]^{\bar\mu\bar\sigma'} 
    = {\rm sgn}(\sigma) {\rm sgn}(\sigma') \sum_{\mu} [T]^{\bar\mu\bar\sigma'} [T^{\dag}]^{\bar\sigma\bar\mu} \ .
\end{equation}
To simplify the subsequent algebraic notation, we define the energy-dependent function $g = [(\Delta + U_f^+)^2 (U_f^+ + U_f^- + E)]^{-1}$.

We begin by evaluating the magnetic component $I_{\rm out}^{zz} = 2 (\mathcal{K}_{\uparrow\uparrow\uparrow\uparrow}^{(f^{2:1}_{\rm out})} - \mathcal{K}_{\uparrow\uparrow\downarrow\downarrow}^{(f^{2:1}_{\rm out})})$. Substituting the tensor contraction from Eq.~\eqref{eq:gam-deg} into the general expression Eq.~\eqref{eq:f21out} leads to
\begin{align}
    \mathcal{K}_{\uparrow\uparrow\uparrow\uparrow}^{(f^{2:1}_{\rm out})} - \mathcal{K}_{\uparrow\uparrow\downarrow\downarrow}^{(f^{2:1}_{\rm out})} 
    =& -  \sum_{\mu  \mu'}' ( [T]^{\uparrow \mu}  [T^\dag]^{\mu' \uparrow}   \delta_{\mu\mu'}   +  [T]^{\mu'\uparrow}  [T^\dag]^{\uparrow\mu}  \delta_{\mu\mu'}  ) g  \nonumber \\
    &+   \sum_{\mu  \mu'}'  ( [T]^{\uparrow \mu}  [T^\dag]^{\mu' \uparrow}  \delta_{\mu\mu'}   +  [T]^{\mu'\downarrow}  [T^\dag]^{\downarrow\mu}  \delta_{\mu\mu'}  ) g \nonumber \\
    = &  \sum_{\mu }' ( - [T]^{\mu\uparrow}  [T^\dag]^{\uparrow\mu}  + [T]^{\mu\downarrow}  [T^\dag]^{\downarrow\mu}  ) g = 0
\end{align}
where the final summation vanishes identically due to the time-reversal constraint established in Eq.~\eqref{eq:TR-deg}. This proof demonstrates that the \emph{out-of-doublet} channel yields no contribution to the $I^{zz}$ exchange coupling in the degenerate limit. Note, however, that no such cancellation occurs for the non-magnetic density-density term $I_{\rm out}^{00}=2 (\mathcal{K}_{\uparrow\uparrow\uparrow\uparrow}^{(f^{2:1}_{\rm out})} + \mathcal{K}_{\uparrow\uparrow\downarrow\downarrow}^{(f^{2:1}_{\rm out})})$.

Next, we consider the tensor element $\mathcal{K}_{\uparrow\uparrow\uparrow\downarrow}^{(f^{2:1}_{\rm out})}$, which evaluates as
\begin{align}
    \mathcal{K}_{\uparrow\uparrow\uparrow\downarrow}^{(f^{2:1}_{\rm out})} 
    = -  \sum_{\mu  \nu }' [  [T]^{\uparrow \nu}  [T^\dag]^{\mu \uparrow}  (-\delta_{\uparrow\nu} \delta_{\downarrow\mu})     +  [T]^{\mu\uparrow}  [T^\dag]^{\downarrow\nu}  \delta_{\mu\nu} ] g 
    = -  \sum_{\mu }' [T]^{\mu\uparrow}  [T^\dag]^{\downarrow\mu}   g = 0
\end{align}
Here, the final sum is driven to zero directly by Eq.~\eqref{eq:TR-deg}. Following an analogous procedure, it is straightforward to show that $\mathcal{K}_{\uparrow\downarrow\uparrow\uparrow}^{(f^{2:1}_{\rm out})}=0$.
The remaining components vanish structurally without requiring any specific time-reversal properties of the hopping integrals:
\begin{equation}
    \mathcal{K}_{\uparrow\downarrow\downarrow\uparrow}^{(f^{2:1}_{\rm out})} \sim \delta_{\uparrow\mu} \delta_{\downarrow\mu'} = 0 \text{ and }  \mathcal{K}_{\uparrow\downarrow \uparrow\downarrow}^{(f^{2:1}_{\rm out})} \sim \delta_{\uparrow\nu} \delta_{\downarrow\mu} = 0
\end{equation}
These expressions are zero because $\gamma_{\uparrow \downarrow,\mu\mu'}^a = - \delta_{\uparrow \mu} \delta_{\downarrow\mu'}$, where the indices $\mu$ and $\mu'$ are restricted to the \emph{out-of-doublet} states orthogonal to the ground-state doublet manifold. This exact property also dictates the cancellation of the $f^{(2:2)}$ \emph{out-of-doublet} channel contributions to the tensor elements $\mathcal{K}_{\uparrow\downarrow\downarrow\uparrow}^{(f^{2:2}_{\rm out})}$, $\mathcal{K}_{\uparrow\uparrow\uparrow\downarrow}^{(f^{2:2}_{\rm out})}$, $\mathcal{K}_{\uparrow\downarrow\uparrow\uparrow}^{(f^{2:2}_{\rm out})}$, and $\mathcal{K}_{\uparrow\downarrow\uparrow\downarrow}^{(f^{2:2}_{\rm out})}$. Furthermore, the net contribution to $I^{zz}$ vanishes due to the identity $\mathcal{K}_{\uparrow\uparrow\uparrow\uparrow}^{(f^{2:2}_{\rm out})} = \mathcal{K}_{\uparrow\uparrow\downarrow\downarrow}^{(f^{2:2}_{\rm out})}$, which follows directly from $\gamma_{\uparrow \uparrow,\mu\mu'}^a = \delta_{\mu\mu'}$~\cite{Ghioldi2024}.
The remaining components of the exchange tensor follow from Hermiticity and time-reversal symmetry constraints on the elements studied above.

Remarkably, these systematic cancellations occur without invoking any spatial or geometric symmetries of the intersite hopping paths. This analysis proves that the emergent strength of all anisotropic magnetic components depends on the finite energy splittings $E_a$ of the intermediate $f^2$ multiplet states.

\section{Numerical verification of the perturbative expansion}
\label{sm:perturbation_verification}

\subsection{Hamiltonian and Hilbert space for exact diagonalization}

To verify the perturbative expansion discussed in \suppref{sm:perturbation}, we compare it with exact diagonalization of a cluster containing two rare-earth ions and two ligand sites, as illustrated in Fig.~\ref{fig:bond-sketch}. The rare-earth Hamiltonian $\mathcal{H}_f$ includes the SOC term $\mathcal{H}_{\rm SOC}$ in Eq.~\eqref{eq:H_SOC} and the Coulomb interaction $\mathcal{H}_{\rm C}$ in Eq.~\eqref{eq:H_Coulomb} of the main text. Instead of an explicit CEF Hamiltonian, we introduce an additional potential that stabilizes the target ground-state Kramers doublet $\ket{\sigma}$ ($\sigma=\uparrow,\downarrow$):
\begin{equation}
\mathcal{H}_{\epsilon} = \epsilon \sum_{\mu \neq \uparrow, \downarrow} f_{\mu}^\dag f_{\mu}^{}.
\label{eq:H_epsilon}
\end{equation}
We set $\epsilon=0.01$ eV. This term raises all states outside the target doublet by $\epsilon$, thereby controlling their separation from the desired low-energy subspace in the exact diagonalization.
The ligand Hamiltonian $\mathcal{H}_p$ includes the on-site energy and the Coulomb interaction on the ligand, as discussed in \suppref{sm:ligand_hamiltonian}. 
The hopping Hamiltonian $\mathcal{V}$ is given by Eq.~\eqref{eq:hopping_pf}. 

For the Ce system, the total number of electrons is set to be 14, and the ground-state electronic configuration is $(n_{f,1},n_{p,A},n_{p,B},n_{f,2})=(1,6,6,1)$, where $n_{f,i}$ and $n_{p,\alpha}$ denote the number of electrons on the $i$-th rare-earth site and the $\alpha$-th ligand site, respectively. The dimension of the total Hilbert space is $\sim 10^{10}$, which is unrealistic for exact diagonalization. Therefore, we restrict the Hilbert space to the subspace relevant for the superexchange process, namely 
\begin{align*}
(n_{f,1},n_{p,A},n_{p,B},n_{f,2}) = & 
(1,6,6,1),~\\
&
(1,5,6,2),~
(1,6,5,2),~
(2,5,6,1),~
(2,6,5,1),~\\
&
(0,6,6,2),~
(2,6,6,0),~
(2,5,5,2),~
(2,4,6,2),~
(2,6,4,2),
\end{align*}
where the first line corresponds to the ground-state configuration, the second line corresponds to the intermediate states generated by single and triple hopping processes, and the third line corresponds to the intermediate states generated by two hopping processes. 
All other electronic configurations either contribute only at higher order or are irrelevant to the exchange interaction. This restriction reduces the Hilbert-space dimension to 577500, allowing us to obtain the four lowest eigenstates and their energies with the Lanczos algorithm.

For Yb, the electronic ground-state configuration is $(n_{f,1},n_{p,A},n_{p,B},n_{f,2})=(13,6,6,13)$, with 38 electrons in total. We instead work in the hole representation, where the ground-state configuration of holes is $(1,0,0,1)$. The full Hilbert space then has dimension 780, so no truncation is required to obtain the four lowest eigenstates and their energies.

\subsection{Extraction of the exchange interaction from exact diagonalization}

We extract the exchange interaction from the four lowest eigenstates as follows. First, we construct the ground-state manifold of the atomic Hamiltonian $\mathcal{H}_0$. In the absence of hopping and intersite interactions, this manifold is the direct product of the local ground states. Denoting the rare-earth doublet states by $\ket{\sigma_i}$ ($\sigma_i=\uparrow,\downarrow$) and the closed-shell ligand state by $\ket{p^6}$, the product states are
\begin{align}
\ket{\sigma_1 \sigma_2} = \ket{\sigma_1} \otimes \ket{p^6}_A \otimes \ket{p^6}_B \otimes \ket{\sigma_2}.
\end{align} 
We label these four states by $\ket{\phi_a}$ ($a=1,2,3,4$). We then diagonalize the full Hamiltonian $\mathcal{H}=\mathcal{H}_0+\mathcal{V}$ and obtain its four lowest eigenstates $\ket{\psi_n}$ and energies $E_n$ ($n=0,1,2,3$). Because the exact eigenstates contain small admixtures of charge-transfer configurations, their projections onto the unperturbed doublet subspace are not exactly orthonormal. We therefore construct the effective Hamiltonian in the four-dimensional product space using canonical L\"owdin orthonormalization.
By defining the overlap matrix $\mathsf{S}$ with the elements
\begin{align}
S_{an}=\braket{\phi_a|\psi_n},
\end{align}
and the corresponding projected metric
\begin{align}
\mathsf{G}=\mathsf{S}\mathsf{S}^{\dagger},
\end{align}
the L\"owdin-orthonormalized representation of the exact eigenstates in the doublet product basis is
\begin{equation}
\mathsf{U}=\mathsf{G}^{-1/2}\mathsf{S} .
\end{equation}
By construction, $\mathsf{U}\mathsf{U}^{\dagger}=1$ within the four-dimensional doublet subspace. The effective Hamiltonian obtained from exact diagonalization is defined as
\begin{align}
\mathcal{H}_{\rm eff}^{\rm ED}= \mathbf{F}^\dag \mathbb{K}_{\rm ED} \mathbf{F},
\end{align}
with
\begin{equation}
\mathbb{K}_{\rm ED}
=
\mathsf{G}^{-1/2}\mathsf{S}
\begin{pmatrix}
E_0-\bar{E} & & & \\
& E_1-\bar{E} & & \\
& & E_2-\bar{E} & \\
& & & E_3-\bar{E}
\end{pmatrix}
\mathsf{S}^{\dagger}\mathsf{G}^{-1/2},
\label{eq:K_ED_lowdin}
\end{equation}
where
\begin{equation}
\bar{E}=\frac{1}{4}\sum_{n=0}^{3}E_n .
\end{equation}
The subtraction of $\bar{E}$ removes the irrelevant constant energy shift. 
To quantify how well the exact low-energy subspace is represented by the target doublet product space, we define the subspace fidelity
\begin{align}
F_{\rm sub} = \frac{1}{4}\mathrm{Tr}(\mathsf{G}) = \frac{1}{4}\sum_{n=0}^{3}\sum_{a=1}^{4}|\braket{\phi_a|\psi_n}|^2.
\end{align}
The fidelity is unity when the four lowest exact eigenstates lie entirely within the target doublet product space and decreases when charge-transfer configurations admix into the low-energy subspace.

The exchange matrix $\mathbb{I}_{\rm ED}$ is obtained by expanding $\mathbb{K}_{\rm ED}$ in Pauli matrices acting on the two Kramers doublets. We compare it with the perturbative result $\mathbb{I}_{\rm pert}$ using the Frobenius norm
\begin{equation}
\delta I
=
\left[
\sum_{\alpha,\beta}
\left|
\mathbb{I}^{\alpha\beta}_{\rm ED} - \mathbb{I}^{\alpha\beta}_{\rm pert}
\right|^2
\right]^{1/2}.
\label{eq:delta_I_ED_pert}
\end{equation}
Since the leading superexchange interaction appears at fourth order in $t_{pf\sigma}$, agreement of the fourth-order perturbative expression implies $\delta I=O(t_{pf\sigma}^6)$.

\subsection{Comparison between exact diagonalization and perturbation theory}

\begin{figure}[t]
\centering
\includegraphics[width=1.0\textwidth]{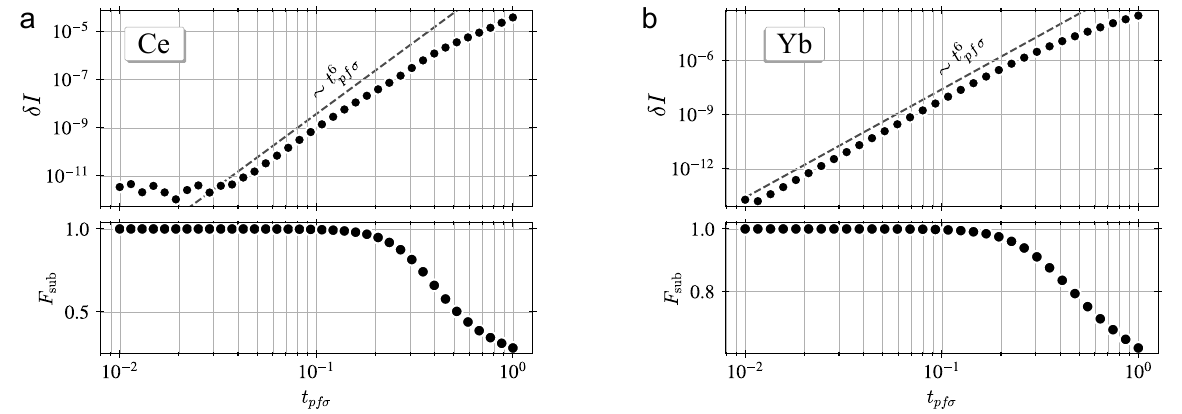}
\caption{\textbf{Comparison between the exchange interaction obtained from exact diagonalization and perturbation theory.} 
The difference between the two exchange matrices, $\delta I$ defined in Eq.~\eqref{eq:delta_I_ED_pert}, is plotted as a function of the hopping amplitude $t_{pf\sigma}$ for the Ce system (\textbf{a}) and the Yb system (\textbf{b}). The dashed lines represent the expected scaling $\delta I \propto t_{pf\sigma}^6$ for the leading order of the perturbation theory. The subspace fidelity $F_{\rm sub}$ is also plotted in the bottom panels.
}
\label{fig:delta_I_ED_pert}
\end{figure}

We now compare the exchange matrices obtained from exact diagonalization and perturbation theory, beginning with Ce. 
For the ED calculation, we employ the parameters for the Ce atomic Hamiltonian as follows (all parameters in eV): SOC $\lambda=0.067$ in Eq.~\eqref{eq:H_SOC}, on-site energy $\epsilon_f=-4.87$ in Eq.~\eqref{eq:Hf_ion_electron}, and Slater integrals $(F^0,F^2,F^4,F^6)=(10.92,6.80,4.38,2.61)$. In addition, we set the small energy penalty $\epsilon=0.01$ in Eq.~\eqref{eq:H_epsilon} to stabilize the ground-state doublet. For the ligand Hamiltonian, we set the on-site energy $\epsilon_p=-20.0$ and the Coulomb interaction $U_p=3.0$. The Slater--Koster parameter ratio is set to be $\rho=t_{pf\pi}/t_{pf\sigma}=-0.3$.
For the perturbative calculation, we use the expressions derived in \suppref{sm:perturbation}, with $U_f^{\pm}$ and $\Delta$ extracted from the ED parameters.
Figure~\ref{fig:delta_I_ED_pert}\textbf{a} shows the $t_{pf\sigma}$ dependence of $\delta I$. We find that $\delta I$ scales as $t_{pf\sigma}^6$ for intermediate $t_{pf\sigma}$, which is consistent with the leading order of the perturbation theory. For small $t_{pf\sigma}$, $\delta I$ stays $\sim 10^{-12}$ and does not decrease further due to the numerical precision of the Lanczos calculation. For larger $t_{pf\sigma}$, $\delta I$ deviates from the $t_{pf\sigma}^6$ scaling, since the eigenstates obtained from the exact diagonalization no longer have a large overlap with the target doublet subspace and the perturbation theory is invalid. 
Consistently, $F_{\rm sub}$ remains close to unity at small $t_{pf\sigma}$ but decreases rapidly for $t_{pf\sigma}\gtrsim0.3$, indicating that the exact low-energy subspace is no longer well represented by the target doublet product space.

For Yb, we use the following atomic parameters (all parameters in eV): SOC $\lambda=0.38$ in Eq.~\eqref{eq:H_SOC}, on-site hole energy $\epsilon_h=-4.31$ in Eq.~\eqref{eq:Hf_ion_hole}, and Slater integrals $(F^0,F^2,F^4,F^6)=(11.7895,14.184,9.846,6.89)$. For the ligand Hamiltonian, we set the on-site energy $\epsilon_q=4.0$ and the Coulomb interaction $U_p=3.0$. The Slater--Koster parameter ratio is set to be $\rho=t_{pf\pi}/t_{pf\sigma}=-0.3$. 
As in Ce, $\delta I$ follows the expected $t_{pf\sigma}^6$ scaling for $t_{pf\sigma}\lesssim0.2$. Because the Yb Hilbert space is small enough for full diagonalization, this scaling persists down to much smaller $t_{pf\sigma}$. At larger hopping amplitudes, the deviation from sixth-order scaling coincides with the decrease in $F_{\rm sub}$.

\section{Decomposition of the superexchange interaction into different channels}
\label{sm:exchange_decomposition}

\begin{figure}[t]
\centering
\includegraphics[width=1.0\textwidth]{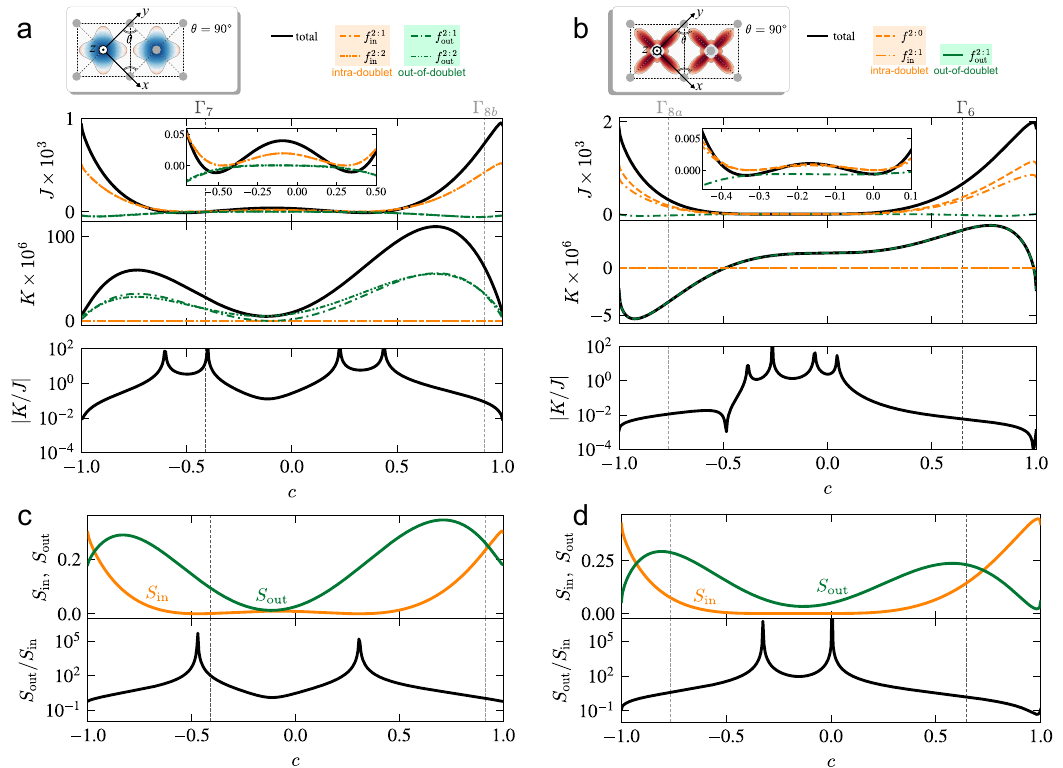}
\caption{\textbf{Decomposition of the superexchange interaction into the different channels and their relations to the intra- and out-of-doublet hopping.} 
The superexchange interaction in the Ce system (\textbf{a}) and the Yb system (\textbf{b}) is decomposed into contributions from different channels. The top and middle panels show the isotropic Heisenberg interaction $J$ and the anisotropic interaction $K$, respectively, and the bottom panels show their ratio $|K/J|$. 
In each panel, the black solid lines represent the total contribution. The orange and green lines represent the intra- and out-of-doublet contributions, respectively, and they are further decomposed into the contributions from different channels, as indicated in the legends. The parameters are the same as those used in Fig.~\ref{fig:wfn_J} of the main text. 
The squared amplitudes of the intra- and out-of-doublet hopping processes, $S_{\rm in}$ and $S_{\rm out}$ defined in Eqs.~\eqref{eq:S_in} and \eqref{eq:S_out}, for the Ce system (\textbf{c}) and the Yb system (\textbf{d}) are also plotted.
}
\label{fig:exchange_decomposition}
\end{figure}

As discussed in \suppref{sm:perturbation}, the perturbative expansion separates the superexchange into distinct virtual-charge channels. In Ce, the $f^{2:1}$ and $f^{2:2}$ channels each contain intra- and out-of-doublet contributions. In Yb, the $f^0$ channel is purely intra-doublet, whereas the $f^{2:1}$ channel contains both contributions.

Figures~\ref{fig:wfn_J}\textbf{c} and \textbf{d} of the main text show the total Ce and Yb superexchange interactions and their decomposition into intra- and out-of-doublet contributions. The former predominantly generates the isotropic Heisenberg exchange $J$, whereas the latter generates the anisotropic exchange $K$. Figure~\ref{fig:exchange_decomposition} further resolves these contributions into the $f^{2:1}$ and $f^{2:2}$ channels for Ce and the $f^0$ and $f^{2:1}$ channels for Yb. To connect this decomposition with the underlying hopping processes, we also evaluate the amplitude of each hopping channel. For the given bond geometry, we construct the ligand-mediated effective hopping matrix in Eq.~\eqref{eq:T_total} and define the squared amplitude of the intra- and out-of-doublet hopping processes as
\begin{align}
S_{\rm in} &= \sum_{\sigma, \sigma'=\uparrow, \downarrow} |T^{\sigma\sigma'}|^2, 
\label{eq:S_in} \\
S_{\rm out} &= \sum_{\substack{\sigma =\uparrow, \downarrow \\ \mu \neq \uparrow, \downarrow}} |T^{\sigma\mu}|^2+|T^{\mu\sigma}|^2,
\label{eq:S_out}
\end{align}
where $T^{\sigma\sigma'}$ and $T^{\sigma\mu}$ are the intra- and out-of-doublet hopping matrix elements, respectively.

Figure~\ref{fig:exchange_decomposition}\textbf{a} shows the decomposition of the superexchange interaction in the Ce system for the $\Gamma_7$ ground state in Eq.~\eqref{eq:Gamma7} of the main text. 
The isotropic Heisenberg exchange $J$ is dominated by intra-doublet processes, with comparable contributions from the $f^{2:1}$ and $f^{2:2}$ channels. As discussed in the main text, increasing $|c|$ enhances the $M=\pm5/2$ weight of the GS doublet and, consequently, the intra-doublet contribution to $J$. The Heisenberg exchange is strongly suppressed at intermediate and small $|c|$.
By contrast, $K$ is dominated by out-of-doublet processes, again with comparable $f^{2:1}$ and $f^{2:2}$ contributions. It is strongly suppressed at both large and small $|c|$ and reaches its maximum at intermediate $|c|$. The intra-doublet contribution to $K$ vanishes because of the special symmetry at $\theta=90^\circ$, as discussed below.
The relative anisotropy $|K/J|$ is strongly suppressed for large $|c|$ and becomes large at intermediate $|c|$, where the Heisenberg exchange is almost zero. 

To discuss the relationship between the hopping processes and the superexchange interaction, we plot the squared amplitudes of the intra- and out-of-doublet hopping processes, $S_{\rm in}$ and $S_{\rm out}$, in Fig.~\ref{fig:exchange_decomposition}\textbf{c}. We find that $S_{\rm in}$ is strongly enhanced for large $|c|$ and suppressed for intermediate and small $|c|$.
Although the squared hopping amplitudes do not reproduce the fine structure of $J$, $S_{\rm in}$ captures its overall variation.
The out-of-doublet hopping $S_{\rm out}$ is suppressed for both large and small $|c|$, leaving the maximum at the intermediate $|c|$, which is consistent with the behavior of the anisotropic exchange $K$. 
The relative anisotropy is then qualitatively captured by the ratio of the hopping amplitudes $S_{\rm out}/S_{\rm in}$, which has peaks at the intermediate $|c|$. 

Next, we discuss the decomposition of the superexchange interaction in the Yb system for the $\Gamma_6$ ground state in Eq.~\eqref{eq:Gamma6} of the main text. Figure~\ref{fig:exchange_decomposition}\textbf{b} shows the decomposition of the superexchange interaction into the contributions from the $f^{0}$ and $f^{2:1}$ channels. 
The isotropic Heisenberg exchange $J$ is again dominated by intra-doublet processes, with comparable contributions from the $f^0$ and $f^{2:1}$ channels.
The anisotropic exchange $K$ arises only from the out-of-doublet $f^{2:1}$ channel, and the $f^{0}$ channel does not contribute to the anisotropic exchange due to the special symmetry for $\theta=90^\circ$. 
The relative anisotropy $|K/J|$ is then strongly suppressed for large $|c|$ and becomes large for small $|c|$, where the Heisenberg exchange is almost zero. 

The $c$ dependences of $J$, $K$, and $|K/J|$ are qualitatively captured by $S_{\rm in}$, $S_{\rm out}$, and $S_{\rm out}/S_{\rm in}$, respectively, as shown in Fig.~\ref{fig:exchange_decomposition}\textbf{d}.  
As noted in the main text, $J$ is suppressed at small $|c|$ in Yb, as in Ce, but is less sensitive to variations in $|c|$. The same distinction appears in $S_{\rm out}/S_{\rm in}$, whose large values are confined to a narrow range of small $|c|$. The origin of this different sensitivity is discussed in \suppref{sm:angular_character}.

\section{Bond angle dependence of the superexchange interaction}
\label{sm:exchange_bond_angle}

\begin{figure}[t]
\centering
\includegraphics[width=1.0\textwidth]{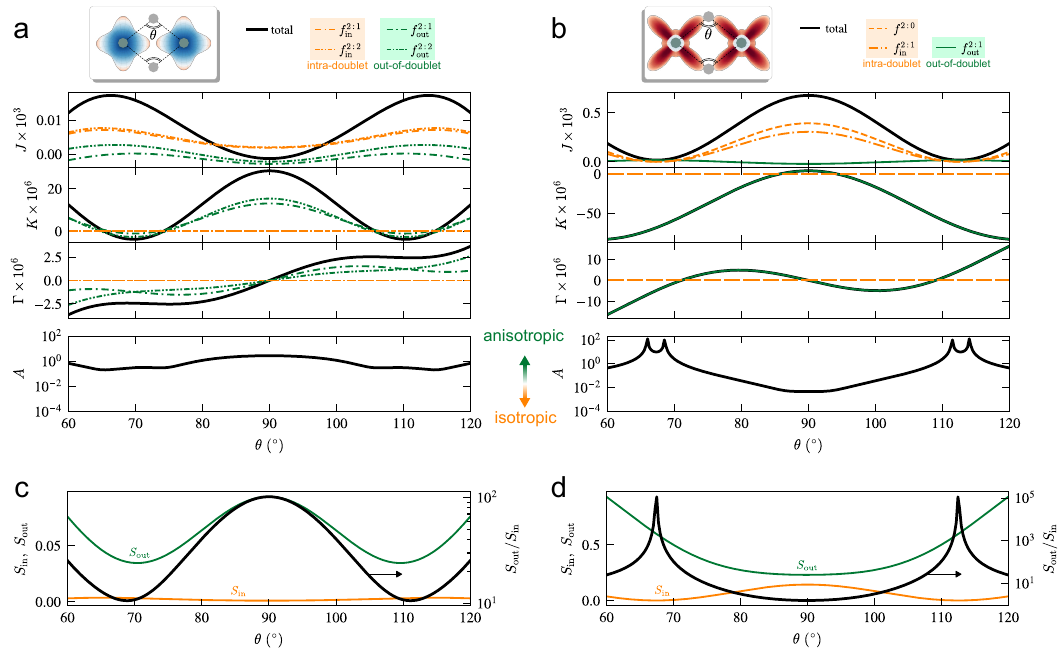}
\caption{\textbf{Bond angle dependence of the superexchange interaction and its decomposition.}
The superexchange interaction in the Ce system with the $\Gamma_7$ doublet (\textbf{a}) and the Yb system with the $\Gamma_6$ doublet (\textbf{b}) is plotted as a function of the bond angle $\theta$. 
The exchange interaction is represented by the three parameters $J$, $K$, and $\Gamma$ defined in Eq.~\eqref{eq:JKG}, and the measure of the relative anisotropy $A$ is also plotted. The parameters are the same as those used in Fig.~\ref{fig:wfn_J} of the main text.
The bond angle dependence of the intra- and out-of-doublet hopping amplitudes, $S_{\rm in}$ and $S_{\rm out}$, and their ratio $S_{\rm out}/S_{\rm in}$ are shown in \textbf{c} and \textbf{d}, respectively. 
}
\label{fig:exchange_bondangle}
\end{figure}

In this section, we discuss the bond angle dependence of the superexchange interaction. 
Figures~\ref{fig:exchange_bondangle}\textbf{a} and \textbf{b} show the bond-angle dependence of the Ce and Yb superexchange interactions, respectively. In real materials, both the bond angle and the ground-state wavefunction depend on the local crystal field. Here, we isolate the geometrical effect by fixing the ground-state doublet to $\Gamma_7$ for Ce and $\Gamma_6$ for Yb while varying $\theta$. %
For the bond angle $\theta \neq 90^\circ$, the exchange matrix is given by
\begin{align}
\mathsf{J}=
\begin{pmatrix}
    J & \Gamma & 0\\
    \Gamma & J & 0\\
    0 & 0 & J+K
\end{pmatrix}.
\label{eq:JKG}
\end{align}
In the present model, the vanishing of $\Gamma$ in \suppref{sm:exchange_decomposition} requires the combined conditions of a $90^\circ$ geometry, two equivalent ligand-mediated paths related by the local cubic rotation, and an exact cubic ground doublet. More generally, the bond geometry permits a finite $\Gamma$ term even at $\theta=90^\circ$~\cite{Rau2018}.
To account for the anisotropy of the exchange interaction, we also compute the measure of the relative anisotropy defined in Eq.~\eqref{eq:anisotropy_measure} of the main text. 

For Ce [Fig.~\ref{fig:exchange_bondangle}\textbf{a}], the intra- and out-of-doublet contributions to $J$ have opposite signs at $\theta=90^\circ$ and nearly cancel, strongly suppressing the net Heisenberg exchange. As $\theta$ moves away from $90^\circ$, both contributions increase moderately and the cancellation is lifted, enhancing $J$ near $\theta\sim(90\pm20)^\circ$. Within the parameter range considered, this angle dependence is symmetric about $90^\circ$, and the intra- and out-of-doublet contributions are of similar magnitude.
In contrast, the anisotropic diagonal exchange $K$ shows a large value at $\theta=90^\circ$ and is suppressed for $\theta \sim (90\pm20)^\circ$ in a symmetric manner. 
The off-diagonal exchange $\Gamma$ vanishes at $\theta=90^\circ$ and grows in magnitude away from this angle, with opposite signs for the two directions of deviation. It is therefore finite near $\theta\sim(90\pm20)^\circ$. As a result, the anisotropy measure $A$ stays relatively large in the whole range of $\theta$ and is maximized at $\theta=90^\circ$.

For Yb [Fig.~\ref{fig:exchange_bondangle}\textbf{b}], the intra-doublet-dominated Heisenberg exchange $J$ is maximal at $\theta=90^\circ$ and decreases approximately symmetrically away from this angle, becoming strongly suppressed near $\theta\sim(90\pm20)^\circ$. By contrast, the out-of-doublet-dominated diagonal anisotropy $K$ is minimized at $90^\circ$ and increases as the bond is distorted. The off-diagonal term $\Gamma$ also vanishes at $90^\circ$ but varies asymmetrically: for increasing $\theta$, it is initially small and negative, then increases and becomes positive for $\theta\gtrsim110^\circ$.
As a result of these behaviors, the anisotropy measure $A$ is significantly suppressed at $\theta=90^\circ$ but is enhanced for $\theta \sim (90\pm20)^\circ$. This trend is opposite to the Ce case, but the anisotropy in the Yb case is more sensitive to the bond angle $\theta$ than the Ce case.

Figures~\ref{fig:exchange_bondangle}\textbf{c} and \textbf{d} show the bond-angle dependence of the intra- and out-of-doublet hopping amplitudes $S_{\rm in}$ and $S_{\rm out}$, defined in Eqs.~\eqref{eq:S_in} and \eqref{eq:S_out}, together with their ratio $S_{\rm out}/S_{\rm in}$ for Ce and Yb, respectively.
The behavior of $S_{\rm in}$ and $S_{\rm out}$ is consistent with the behavior of $J$ and $K$, respectively, and the ratio $S_{\rm out}/S_{\rm in}$ qualitatively captures the behavior of the anisotropy measure $A$. For Ce, $S_{\rm in}$ remains small and varies only weakly, consistent with the small $J$. Consequently, $S_{\rm out}/S_{\rm in}$ and the anisotropy measure $A$ depend only weakly on $\theta$.
Such a weak bond angle dependence is observed in Fig.~\ref{fig:angle}\textbf{a} of the main text, where we show the bond angle dependence of $A$ for the Ce system while varying the $M=\pm5/2$ weight of the GS wavefunction, $c$; $A$ is less sensitive to $\theta$ for the Ce system compared to the Yb case.
For Yb [Fig.~\ref{fig:exchange_bondangle}\textbf{d}], $S_{\rm in}$ is strongly suppressed near $\theta\sim(90\pm20)^\circ$, whereas $S_{\rm out}$ increases monotonically away from $90^\circ$. Their ratio therefore grows by several orders of magnitude, mirroring the strong enhancement of the relative anisotropy.

\section{Effect of the complex mixing in the ground-state doublet}
\label{sm:exchange_complex_mixing}

\begin{figure}[t]
\centering
\includegraphics[width=1.0\textwidth]{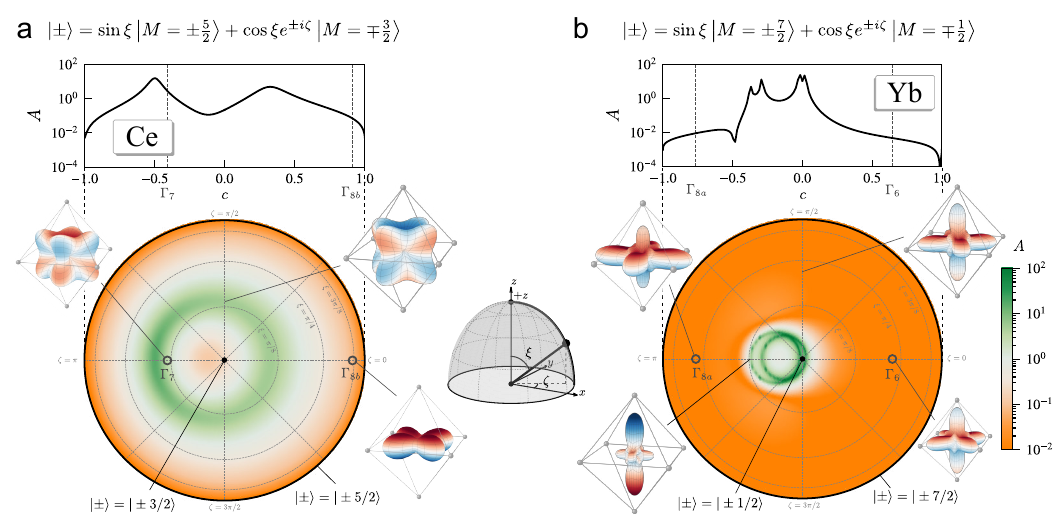}
\caption{\textbf{Measure of relative anisotropy $A$ for the complex mixing in the ground-state doublet.}
The measure of the relative anisotropy $A$ defined in Eq.~\eqref{eq:anisotropy_measure} of the main text is plotted as a function of the mixing parameters $(\xi, \zeta)$ for the Ce system (\textbf{a}) and the Yb system (\textbf{b}). The parameters are the same as those used in Fig.~\ref{fig:wfn_J} of the main text. The parameter space defined with $(\xi, \zeta)$ can be mapped on the hemisphere of the unit sphere (see the inset) and the view from the north pole is shown. 
The top panel represents the cut along the $\zeta=0,\pi$ line, which corresponds to the real mixing discussed in Fig.~\ref{fig:wfn_J} of the main text with $c=\sin\xi,~-\sin\xi$ ($\zeta=0,~\pi$). 
The shape of the wavefunction and the local ligand environment for some representative points are also shown.
}
\label{fig:exchange_complex_mixing}
\end{figure}

In the main text and \suppref{sm:exchange_decomposition}, we varied the ground-state wavefunction while keeping the relative phase between different $M$ components real. Here, we examine how a complex relative phase affects the superexchange interaction.
For Ce, we generalize the ground-state wavefunction to
\begin{align}
\ket{\pm} = \sin \xi \Ket{\pm 5/2} + \cos\xi\,e^{i\zeta} \Ket{\mp 3/2},
\label{eq:Ce_wfn_complex}
\end{align}
and for Yb to
\begin{align}
\ket{\pm} = \sin \xi \Ket{\pm 7/2} + \cos\xi\,e^{i\zeta} \Ket{\mp 1/2},
\label{eq:Yb_wfn_complex}
\end{align}
where $\xi \in [0, \pi/2]$ represents the mixing amplitude between the two $M$ components and $\zeta \in [0, 2\pi)$ is the relative phase. 
Using the same parameters as in Fig.~\ref{fig:wfn_J} of the main text, we evaluate the relative-anisotropy measure $A$ defined in Eq.~\eqref{eq:anisotropy_measure}.
The doublets in Eqs.~\eqref{eq:Ce_wfn_complex} and \eqref{eq:Yb_wfn_complex} do not, in general, diagonalize the $g$ tensor, making the relation between pseudospin operators and the physical magnetic moment less transparent. We therefore perform a unitary transformation to the basis that diagonalizes the $g$ tensor and evaluate the exchange interaction in that basis. This changes the matrix representation of the exchange tensor but not the underlying physics.

Figures~\ref{fig:exchange_complex_mixing}\textbf{a} and \textbf{b} show $A$ for Ce and Yb, respectively, as a function of $(\xi,\zeta)$. This parameter space maps onto a hemisphere of the unit sphere, shown from the north-pole direction (inset).
For Ce, $A$ is suppressed near $\xi\sim\pi/2$, where the GS wavefunction is dominated by the $M=\pm5/2$ components, and rises rapidly as the $M=\mp3/2$ component is admixed. The quasi-isotropic region, shown in orange, is therefore localized near the equator of the hemisphere. With stronger mixing, $A$ reaches a maximum near $\xi\sim\pi/8$, indicated by the green ring. The nearly concentric pattern shows that the $\zeta$ dependence is weak, although the ring is shifted slightly towards $\zeta\sim\pi$ and the anisotropy is somewhat larger on the $c<0$ side than on the $c>0$ side.
For Yb, the quasi-isotropic region persists over a broad range of $\xi$, reflecting the robustness discussed in the main text. Its dependence on $\zeta$ is weak within this regime.
The anisotropic regime is localized near the north pole ($\xi = 0$) but on the $\zeta \sim \pi$ side, where the intra-doublet hopping is strongly suppressed; near the north pole, $\zeta$ dependence is strong. 

Although a complex relative phase can have nontrivial effects, the overall behavior of $A$ is captured qualitatively by the real-mixing limit. In the present geometry, the two ligands lie in the $xy$ plane, and the planar localization of the orbital lobes is controlled primarily by the maximal-angular-momentum weight $\sin^2\xi$. The phase $\zeta$ changes their azimuthal orientation. Because the two components differ by $\Delta M=\pm4$, introducing $\zeta$ rotates the $\zeta=0$ orbital by $\zeta/4$ within the $xy$ plane. Thus, quasi-isotropy is governed mainly by $\xi$, while the effect of $\zeta$ is generally small.
The effect of $\zeta$ becomes more pronounced in the anisotropic regime, where the orbital lobes extend less strongly within the hopping plane and the intra-doublet hopping is suppressed. In this regime, the hopping is more sensitive to the azimuthal orientation controlled by $\zeta$. For instance, in the Ce case, the electron density is directed along the ligand direction for $\zeta = 0$ (see $\Gamma_{8b}$ wavefunction), while it is directed along the bond direction for $\zeta = \pi$ (see $\Gamma_{7}$ wavefunction). The former case allows stronger intra-doublet hopping than the latter case, which results in the stronger anisotropy for $\zeta = \pi$ than for $\zeta = 0$. Similarly, in the Yb case, $\zeta\sim0$ allows stronger intra-doublet hopping than $\zeta\sim\pi$.

\section{Angular character of the crystal-field wavefunctions}
\label{sm:angular_character}

In this section, we analyze the angular structure of the crystal-field wavefunctions used in the main text. We focus on the angular dependence of the $4f$ states to clarify why the $J=7/2$ manifold more readily supports orbitals with substantial weight in the exchange plane than the $J=5/2$ manifold.

\subsection{Planar character of the $4f$ wavefunction}

We first consider a general state in a fixed spin-orbit multiplet,
\begin{align}
\ket{\psi}= \sum_{M=-J}^{J} c_M \ket{J,M},
\end{align}
with $\sum_M |c_M|^2=1$.
We first expand $\ket{J,M}$ in the $\ket{m_\ell,m_s}$ basis.
For $J=L-1/2=5/2$, relevant for Ce$^{3+}$, we take
\begin{align}
\Ket{\frac{5}{2},M}
=\sqrt{\frac{7/2-M}{7}}\Ket{m_\ell=M-\frac12, m_s=\frac12}
-\sqrt{\frac{7/2+M}{7}}\Ket{m_\ell=M+\frac12, m_s=-\frac12},
\label{eq:CG_Ce}
\end{align}
For $J=L+1/2=7/2$, relevant for the $4f^{13}$ hole of Yb$^{3+}$, we use
\begin{align}
\Ket{\frac{7}{2},M}
=
\sqrt{\frac{M+7/2}{7}}
\Ket{m_\ell=M-\frac12, m_s=\frac12}
+
\sqrt{\frac{7/2-M}{7}}
\Ket{m_\ell=M+\frac12, m_s=-\frac12}.
\label{eq:CG_Yb}
\end{align}
The corresponding charge density is then expressed by the spherical harmonics $Y_{\ell,m}(\theta,\phi)$ as
\begin{align}
\rho(\theta,\phi)
=
\sum_{s=\pm1/2}
\left|
\sum_{m=-3}^{3} A_{m s} Y_{3,m}(\theta,\phi)
\right|^2 ,
\label{eq:rho_general}
\end{align}
where \(A_{m s}\) is obtained from Eqs.~\eqref{eq:CG_Ce} or \eqref{eq:CG_Yb}, as
\begin{align}
A_{m s}
=
(-1)^{2s+1}\sqrt{\frac{3-(-1)^{2s+1}m}{7}}\,c_{m+s},
\label{eq:Ams_Ce}
\end{align}
for the Ce case, and
\begin{align}
A_{m s}
=
\sqrt{\frac{4+(-1)^{2s+1}m}{7}}\,c_{m+s},
\label{eq:Ams_Yb}
\end{align}
for the Yb case.

As a global measure of orbital planarity, we consider the average of $\sin^2\theta$ over the charge density, 
\begin{align}
P_{\parallel}
=
\braket{\sin^2\theta}
=
\int d\Omega\, \rho(\theta,\phi)\sin^2\theta.
\label{eq:P_parallel_def}
\end{align}
This quantity is insensitive to the azimuthal orientation and measures the concentration of orbital weight in the $xy$ plane.
By using the relation
\begin{align}
\int d\Omega |Y_{\ell,m}(\theta,\phi)|^2 \sin^2\theta
=\frac{2(\ell (\ell +1) + m^2-1)}{(2\ell -1)(2\ell +3)},
\label{eq:sin2_values}
\end{align}
we obtain, for the $J=5/2$ multiplet,
\begin{align}
P_{\parallel}^{(5/2)}
=
\frac{1}{2}
+
\frac{2}{35}
\sum_M |c_M|^2 M^2 ,
\label{eq:P_parallel_Ce}
\end{align}
and for the $J=7/2$ multiplet,
\begin{align}
P_{\parallel}^{(7/2)}
=
\frac{1}{2}
+
\frac{2}{63}
\sum_M |c_M|^2 M^2.
\label{eq:P_parallel_Yb}
\end{align}
Thus, the most planar states are those with maximal angular-momentum projection, $\ket{J,\pm J}$.
Importantly, the $J=|M|=7/2$ state ($P_{\parallel}^{(7/2)}=8/9$) is slightly more planar than the $J=|M|=5/2$ state ($P_{\parallel}^{(5/2)}=6/7$) because
\begin{align}
\Ket{J=\frac{7}{2},M=\frac{7}{2}}
=
\Ket{m_\ell=3, m_s=\frac{1}{2}}
\end{align}
is a pure $m_\ell=3$ orbital, whereas
\begin{align}
\Ket{J=\frac{5}{2},M=\frac{5}{2}}
=
\sqrt{\frac{1}{7}}\Ket{m_\ell=2, m_s=\frac{1}{2}}
-
\sqrt{\frac{6}{7}}\Ket{m_\ell=3, m_s=-\frac{1}{2}}
\end{align}
already contains an $m_\ell=2$ component. 
The maximal-$|M|$ states of the $J=7/2$ multiplet are therefore more planar than their $J=5/2$ counterparts, favoring more nearly isotropic superexchange in Yb than in Ce.

\subsection{Mirror parity and reduced sensitivity of the exchange anisotropy in Yb}

We next examine the orbital shape locally near the exchange plane, beyond the global planarity measure.
To examine the density near $\theta=\pi/2$, we introduce $x=\cos\theta$ and expand the spherical harmonics in powers of $x$. 
For $\ell=3$, $Y_{3,m}(\theta,\phi)$ can be expanded as
\begin{align}
    Y_{3, \pm3} &\propto e^{\pm 3i\phi}
    \left(1-\frac{3}{2}x^2+\cdots\right),\\
    Y_{3, \pm2} &\propto x e^{\pm 2i\phi}+\cdots,\\
    Y_{3, \pm1} &\propto e^{\pm i\phi}
    \left(1-\frac{11}{2}x^2+\cdots\right),\\
    Y_{3, 0} &\propto x+\cdots.
\end{align}
The important distinction arises from the mirror parity of the spherical harmonics under $z \to -z$. 
For $m=\pm3,\pm1$, the leading term is a constant and the subleading term is quadratic in $x$, so that the density is maximal in the exchange plane and slowly decreases away from the plane for sufficiently small $|x|$. In contrast, for $m=0,\pm2$, the leading term is linear in $x$, so that the density vanishes in the exchange plane and grows rapidly away from the plane.  
Thus $m=\pm3,\pm1$ form equatorial components, whereas $m=0,\pm2$ are components that rise away from the exchange plane.

Mirror parity provides a further distinction between the $J=5/2$ and $J=7/2$ multiplets when the system is symmetric under reflection through the exchange plane. To express this distinction in the $J,M$ basis, we use the expansions in Eqs.~\eqref{eq:CG_Ce} and \eqref{eq:CG_Yb}.
First, we consider the $J=5/2$ multiplet relevant for Ce. 
Table~\ref{tab:J_5half_expansion_parity} summarizes the expansion of the $J=5/2$ states in the $\Ket{m_\ell,m_s}$ basis. Blue and orange cells denote even and odd parity, respectively, under $z\to-z$.
As noted above, the $M=\pm5/2$ states contain both the strongly planar $m_\ell=\pm3$ components and the less planar $m_\ell=\pm2$ components.
More importantly, the $M=\pm5/2$ states already contain mirror-even and mirror-odd orbital components in opposite spin sectors. Both parity sectors are therefore active even in the pure $|M|=5/2$ limit. Because hopping conserves spin and mirror parity in the present geometry, each component hybridizes only within its corresponding parity--spin sector.
The additional $M=\mp3/2$ component considered in the main text modifies the hopping amplitudes in both active sectors. It can therefore alter several orbital--spin channels simultaneously, allowing the out-of-doublet contribution and the exchange anisotropy to grow rapidly away from the pure $|M|=5/2$ limit.

\begin{table}[t]
\centering
\caption{
Expansion of the $J=5/2$ states in the $\Ket{m_\ell,m_s}$ basis. Each cell represents the Clebsch--Gordan coefficient with the corresponding $\Ket{m_\ell}$ state; the first and second rows correspond to $m_s=+1/2$ and $-1/2$, respectively.
The blue and orange cells denote the even and odd mirror parity under the mirror operation $z\to-z$, respectively.
}
\label{tab:J_5half_expansion_parity}
\scriptsize
\setlength{\tabcolsep}{2.5pt}
\renewcommand{\arraystretch}{1.25}
\begin{tabular}{c|cccccc}
    \hline
    \(\ket{J=5/2,M}\)& \(\Ket{\frac{5}{2}}\)
    & \(\Ket{\frac{3}{2}}\)
    & \(\Ket{\frac{1}{2}}\)
    & \(\Ket{-\frac{1}{2}}\)
    & \(\Ket{-\frac{3}{2}}\)
    & \(\Ket{-\frac{5}{2}}\) \\
    \hline
    \(m_s=+\frac{1}{2}\)
    & \cellcolor{mirrorOdd}\(\sqrt{\frac{1}{7}}\Ket{2}\)
    & \cellcolor{mirrorEven}\(\sqrt{\frac{2}{7}}\Ket{1}\)
    & \cellcolor{mirrorOdd}\(\sqrt{\frac{3}{7}}\Ket{0}\)
    & \cellcolor{mirrorEven}\(\sqrt{\frac{4}{7}}\Ket{-1}\)
    & \cellcolor{mirrorOdd}\(\sqrt{\frac{5}{7}}\Ket{-2}\)
    & \cellcolor{mirrorEven}\(\sqrt{\frac{6}{7}}\Ket{-3}\) \\
    \(m_s=-\frac{1}{2}\)
    & \cellcolor{mirrorEven}\(-\sqrt{\frac{6}{7}}\Ket{3}\)
    & \cellcolor{mirrorOdd}\(-\sqrt{\frac{5}{7}}\Ket{2}\)
    & \cellcolor{mirrorEven}\(-\sqrt{\frac{4}{7}}\Ket{1}\)
    & \cellcolor{mirrorOdd}\(-\sqrt{\frac{3}{7}}\Ket{0}\)
    & \cellcolor{mirrorEven}\(-\sqrt{\frac{2}{7}}\Ket{-1}\)
    & \cellcolor{mirrorOdd}\(-\sqrt{\frac{1}{7}}\Ket{-2}\) \\
    \hline
\end{tabular}
\end{table}

In contrast, the $M=\pm7/2$ states of the $J=7/2$ multiplet consist of a single $m_\ell=\pm3$ component. They therefore occupy only one mirror-even parity--spin sector in the maximal-$|M|$ limit. The expansion of the $J=7/2$ states is summarized in Table~\ref{tab:J_7half_expansion_parity}.
For example, the Yb wavefunction considered in the main text is
\begin{align}
\ket{\psi}
=
a\Ket{\frac{7}{2},\frac{7}{2}}
+
b\Ket{\frac{7}{2},-\frac{1}{2}},
\end{align}
with $|a|^2+|b|^2=1$. In the $\Ket{m_\ell,m_s}$ basis, it takes the form
\begin{align}
\ket{\psi}
=
a\Ket{3,\frac{1}{2}}
+
b\left(
\sqrt{\frac{3}{7}}\Ket{-1,\frac{1}{2}}
+
\sqrt{\frac{4}{7}}\Ket{0,-\frac{1}{2}}
\right).
\end{align}
The pure $|M|=7/2$ state only contains the $m_\ell=3$ component, which is mirror-even and belongs to the $m_s=+1/2$ spin sector. Due to the mirror symmetry, the $m_\ell=3$ component can only hybridize with orbitals in the same mirror-even and spin sector, which restricts the available out-of-doublet hopping channels. 
The $m_\ell=-1$ component introduced by the admixture belongs to the same mirror-even and spin sector as the dominant $m_\ell=3$ component. By contrast, the mirror-odd $m_\ell=0$ component belongs to the opposite spin sector and opens a new hopping channel that was not available in the pure $|M|=7/2$ limit. 

\begin{table}[t]
\centering
\caption{
Expansion of the $J=7/2$ states in the $\Ket{m_\ell,m_s}$ basis, using the same conventions as Table~\ref{tab:J_5half_expansion_parity}.
}
\label{tab:J_7half_expansion_parity}
\scriptsize
\setlength{\tabcolsep}{2.5pt}
\renewcommand{\arraystretch}{1.25}
\begin{tabular}{c|cccccccc}
    \hline
    \(\ket{J=7/2,M}\)& \(\Ket{\frac{7}{2}}\)
    & \(\Ket{\frac{5}{2}}\)
    & \(\Ket{\frac{3}{2}}\)
    & \(\Ket{\frac{1}{2}}\)
    & \(\Ket{-\frac{1}{2}}\)
    & \(\Ket{-\frac{3}{2}}\)
    & \(\Ket{-\frac{5}{2}}\)
    & \(\Ket{-\frac{7}{2}}\) \\
    \hline
    \(m_s=+\frac{1}{2}\)
    & \cellcolor{mirrorEven}\(\Ket{3}\)
    & \cellcolor{mirrorOdd}\(\sqrt{\frac{6}{7}}\Ket{2}\)
    & \cellcolor{mirrorEven}\(\sqrt{\frac{5}{7}}\Ket{1}\)
    & \cellcolor{mirrorOdd}\(\sqrt{\frac{4}{7}}\Ket{0}\)
    & \cellcolor{mirrorEven}\(\sqrt{\frac{3}{7}}\Ket{-1}\)
    & \cellcolor{mirrorOdd}\(\sqrt{\frac{2}{7}}\Ket{-2}\)
    & \cellcolor{mirrorEven}\(\sqrt{\frac{1}{7}}\Ket{-3}\)
    & -- \\
    \(m_s=-\frac{1}{2}\)
    & --
    & \cellcolor{mirrorEven}\(\sqrt{\frac{1}{7}}\Ket{3}\)
    & \cellcolor{mirrorOdd}\(\sqrt{\frac{2}{7}}\Ket{2}\)
    & \cellcolor{mirrorEven}\(\sqrt{\frac{3}{7}}\Ket{1}\)
    & \cellcolor{mirrorOdd}\(\sqrt{\frac{4}{7}}\Ket{0}\)
    & \cellcolor{mirrorEven}\(\sqrt{\frac{5}{7}}\Ket{-1}\)
    & \cellcolor{mirrorOdd}\(\sqrt{\frac{6}{7}}\Ket{-2}\)
    & \cellcolor{mirrorEven}\(\Ket{-3}\) \\
    \hline
\end{tabular}
\end{table}

The different sets of available parity--spin channels provide a qualitative explanation for the contrasting sensitivity of Ce and Yb to wavefunction admixture. In Ce, both mirror-parity sectors are active already in the maximal-$|M|$ state, so admixture modifies the corresponding out-of-doublet amplitudes through linear interference within each sector. In Yb, each maximal-$|M|$ Kramers partner is confined to a single parity--spin sector, and the opposite-parity channel appears only through admixture. Because the anisotropic exchange is governed by out-of-doublet virtual processes, this restricted channel structure slows the growth of anisotropy in Yb. Together with the greater planarity of the $J=7/2$ states, it makes quasi-isotropic exchange more robust against admixture of other $M$ components.

\begingroup
\def\refname{Supplementary References}
\def\bibname{Supplementary References}
\putbib[ref-insulators]
\endgroup
\end{bibunit}

\end{document}